\documentclass[a4paper,11pt]{article}
\usepackage{jheppub} %
\usepackage{tcolorbox}

\usepackage{package-notes}
\usepackage{bbold}

\usepackage{float} % for nuking figure placement with [H]

\usepackage{tikz}
\usetikzlibrary{positioning, arrows.meta, calc}
\usetikzlibrary{decorations.pathreplacing}

\usepackage{amsthm}
\theoremstyle{plain}

\theoremstyle{remark}

\usepackage{subfigure}
\usepackage{comment}
\usepackage{cancel}

\newcommand{\im}{\mathrm{Im}}
\newcommand{\re}{\mathrm{Re}}
\newcommand{\ba}{\begin{aligned}} 
\newcommand{\ea}{\end{aligned}}

\def\eps{{\epsilon}}
\def\be#1\ee{\begin{align}#1\end{align}}

\newcommand{\MP}{\text{MP}}
\newcommand{\el}{\text{el}}

\newcommand{\mmax}{\mathrm{max}}

\usepackage[normalem]{ulem}

\DeclareMathOperator\arccosh{arccosh}

\usepackage[notcite,notref]{}%{showkeys}

\preprint{LAPTH-048/25,~CERN-TH-2026-010}

\title{
Neural S-matrix bootstrap II: \\
solvable 4d amplitudes with particle production
}

\author[a]{Mehmet Asım Gümüş,}
\author[a]{Damien Leflot,}
\author[a]{Piotr Tourkine,}
\author[b]{Alexander Zhiboedov}

\affiliation[a]{LAPTh, CNRS et Universit\'{e} Savoie Mont-Blanc 
9 Chemin de Bellevue, F-74941 Annecy, France}
\affiliation[b]{CERN, Theoretical Physics Department, CH-1211 Geneva 23, Switzerland}

\abstract{
We study a model for nonperturbative unitarization of the four-point contact scalar amplitude in four dimensions. It is defined through an infinite sum of planar
diagrams, constructed using two-particle unitarity and crossing symmetry. We reformulate the problem in terms of a set of nonlinear integral equations obeyed by the single and double discontinuities of the amplitude. We then solve them using a neural-network ansatz trained by minimizing a physics-informed loss functional. We obtain a one-parameter family of amplitudes, which exhibit rich structure: sizeable particle production, nontrivial emergent Regge behavior, Landau curves, a logarithmic decay at high energy and fixed angle. 
Finally, we go beyond the two-particle-reducible setup by treating the multi-particle data---supported above the multi-particle Landau curves due to multi-particle unitarity---as a dynamical variable. We demonstrate that it can be tuned to suppress low-spin particle production---a phenomenon we call Aks screening---at the cost of generating larger and oscillatory double spectral density in the multi-particle region.
}

\begin{document}
\setcounter{tocdepth}{2} % % numbering up to subsection
\maketitle
%\tableofcontents

\newpage

\section{Introduction}

\begin{figure}[H]
    \centering
    \includegraphics[]{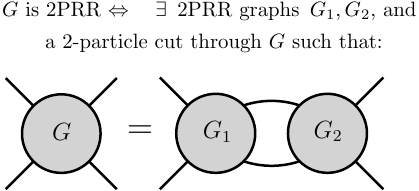}
    
    \vspace{16pt}
    
    \includegraphics[scale=1]{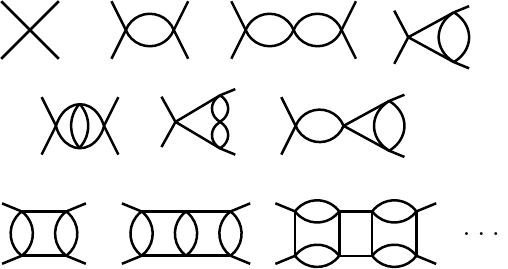}
    \caption{A series of \emph{two-particle-recursively-reducible} (2PRR) graphs that impose two-particle unitarity. Each of the graphs can be reduced to a single vertex by performing two-particle cuts. Each graph comes with its crossing images. We call the amplitudes defined via this series \emph{2PRR scattering amplitudes}.}
    \label{fig:series}
\end{figure}

In this paper we propose a solvable model for a nonperturbative $2\to 2$ scattering amplitude with particle production in four dimensions.  The model is defined as follows:

\paragraph{Graph definition.}
Consider the series of diagrams shown in Fig.~\ref{fig:series}.
Each diagram can be reduced to a single vertex by repeatedly cutting two internal lines.
Equivalently, a graph $G$ in this class admits a two-particle cut such that it splits into
two smaller graphs in the same class,
\begin{equation}
G\ \text{is 2PRR} \quad\Longleftrightarrow\quad
\exists\, G_1,G_2\ \text{2PRR and a two-particle cut with}\ G = G_1\,G_2 \, .
\end{equation}
We call these graphs \emph{two-particle-recursively-reducible} (2PRR). We also implicitly include their crossing-symmetry images. 
We next associate to each graph the corresponding Feynman diagram in ${\lambda \over 4!} \phi^4$ theory. 
The number of such graphs grows polynomially, see e.g. \cite{Tourkine:2023xtu}, and therefore we expect that such series has a finite radius of convergence and defines a well-defined function of the coupling $\lambda$. We call the resulting $2\to2$ scattering amplitudes \emph{2PRR amplitudes}.

We do not attempt to solve this problem by explicitly computing and resumming the relevant Feynman diagrams. Instead, we reformulate it in terms of nonperturbative unitarity equations for the amplitude.

\paragraph{Atkinson--Mandelstam approach.}
In this approach, we adopt a once-subtracted Mandelstam representation for the amplitude
\begin{equation}
  \label{eq:mandelstamRep1}
  \begin{aligned}
T(s,t) &= c_0+ B(s,t) + B(s,u) + B(t,u) ,  \\
B(s,t) &={1 \over 2} \int_{4 m^2}^\infty {d s' \over \pi} {\rho(s') \over s'-s_0} \Big( {s-s_0 \over s'-s} + {t-s_0 \over s'-t} \Big) \\
&+ \int_{4 m^2}^\infty {d s' d t' \over \pi^2}
{(s-s_0)(t-t_0) \rho(s',t') \over (s'-s)(t'-t)(s'-s_0)(t'-t_0)} \, , ~~~ s_0 = t_0 = u_0 = 4m^2/3 \, . 
\end{aligned}
\end{equation}
In this representation the amplitude is characterized by its value at the crossing-symmetric point  ${c_0} \equiv T(4m^2/3,4m^2/3)$, and a pair of spectral densities: single spectral density $\rho(s)$ and double spectral density $\rho(s,t)$.
The representation \eqref{eq:mandelstamRep1} makes crossing manifest, but unitarity has to be imposed. 

From the unitarity point of view, the set of 2PRR graphs is a minimal set of graphs that is necessary to solve \emph{two-particle unitarity}. A convenient way to impose two-particle unitarity is via the so-called Mandelstam equation \cite{Mandelstam:1958xc} (obeyed by the double spectral density), which imposes elastic unitarity for ${\rm Re}[J]>0$, and a separate $J=0$ unitarity condition (for the  single spectral density), which we also review in the main text. 

We conjecture that the 2PRR amplitudes can be equivalently defined as follows.\footnote{Our proposal involves a certain assumption about the Regge amplitude of the 2PRR amplitude, which we check a posteriori.} Their double spectral density is given by
\be
\rho_{\text{2PRR}}(s,t) = \rho_{\text{el}}(s,t) + \rho_{\text{el}}(t,s) ,
\ee
where $\rho_{\text{el}}(s,t)$ is the solution to the Mandelstam equation that we review in the main text of the paper. Here $\rho_{\text{el}}(s,t)$ guarantees that elastic unitarity $|S_J(s)|=1$ is obeyed for $16 m^2 > s \geq 4 m^2$. Crossing is imposed by adding the $s \leftrightarrow t$ image $\rho_{\text{el}}(t,s)$, which is responsible for particle production. The single spectral density $\rho_{\text{2PRR}}(s)$ is fixed by imposing $J=0$ elastic unitarity.

To connect to the perturbative definition above let us note that\footnote{The $O(\lambda^2)$ term  depends on the choice of the renormalization scheme in perturbative ${\lambda \over 4!} \phi^4$.}
\be
c_0 = - \lambda + O(\lambda^2) \ . 
\ee
The usual $\frac{\lambda}{4!} \phi^4$ theory with a repulsive potential that is bounded from below corresponds to $\lambda > 0$ and  $c_0 < 0$. In this regime, the theory is known to possess a Landau pole at large energies. In this paper, we will only consider the amplitudes with $c_0>0$.

\begin{figure}[H]
    \centering
    \includegraphics[width=0.5\linewidth]{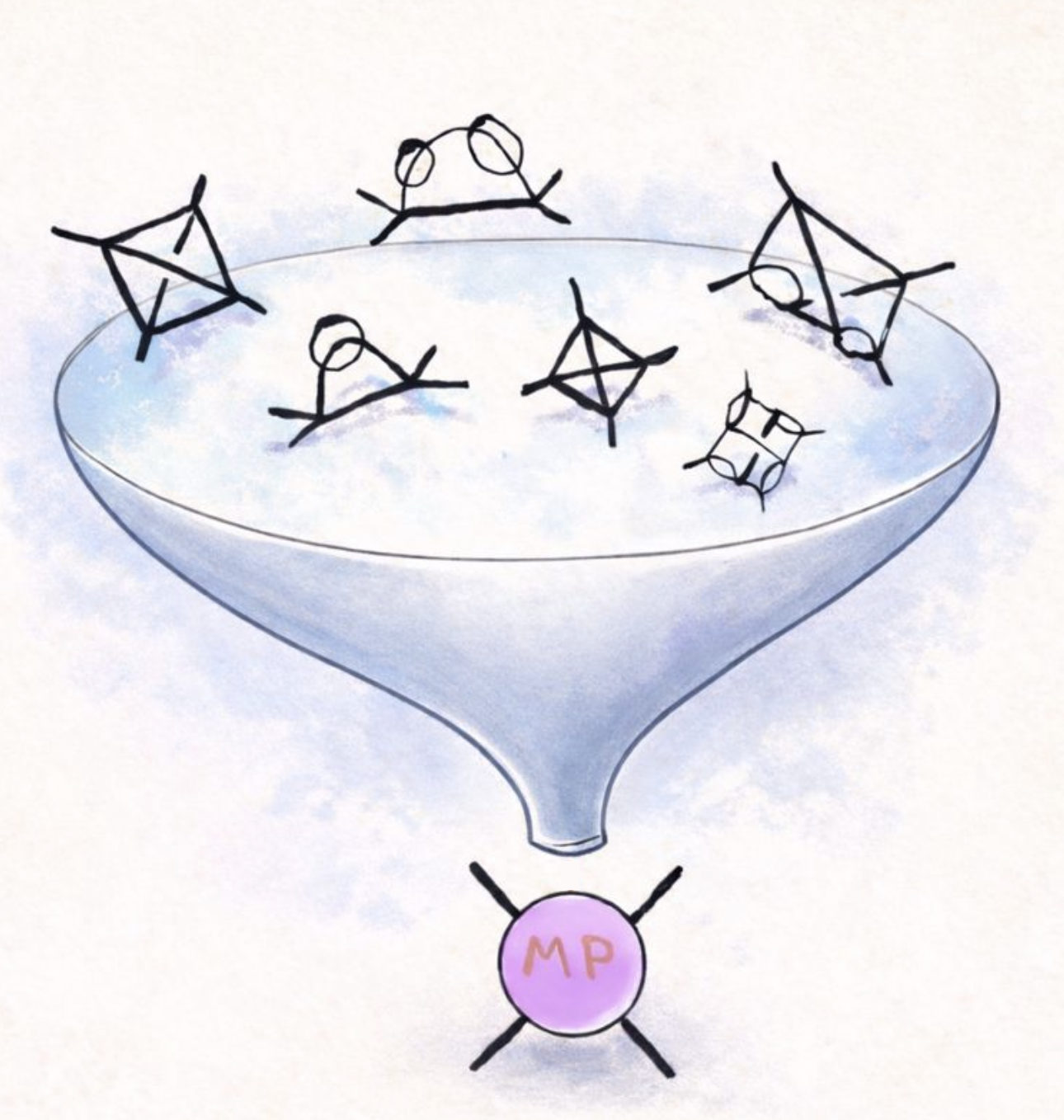}
    \caption{We characterize the space of amplitudes using \emph{the multi-particle data}. In perturbation theory, it is defined by the sum of all two-particle-irreducible graphs.}
    \label{fig:MP}
\end{figure}

In general, the double spectral density of the $2 \to 2$ scattering amplitude can instead be written as
\be
\rho(s,t) = \rho_{\text{el}}(s,t) + \rho_{\text{el}}(t,s) +   \rho_{\text{MP}}(s,t) ,
\ee
where $\rho_{\text{MP}}(s,t)$ encodes \emph{the multi-particle data} that corresponds to the graphs which are not 2PRR, see Figure \ref{fig:MP}.
\noindent In this language 2PRR amplitudes correspond to setting $\rho_{\text{MP}}(s,t) = 0$ (and similarly the multi-particle part of the single spectral density).

\paragraph{Solving the 2PRR model.} At this point we have reformulated the problem of interest in terms of non-linear unitarity equations satisfied by the spectral densities, and our next task is to solve them. We solve this problem numerically with the help of neural optimizers.\footnote{Our previous work on the subject \cite{Gumus:2024lmj} (Part I) studied a much simpler solvable model with
\emph{zero} double discontinuity. In that model, $J>0$ unitarity was not imposed.}

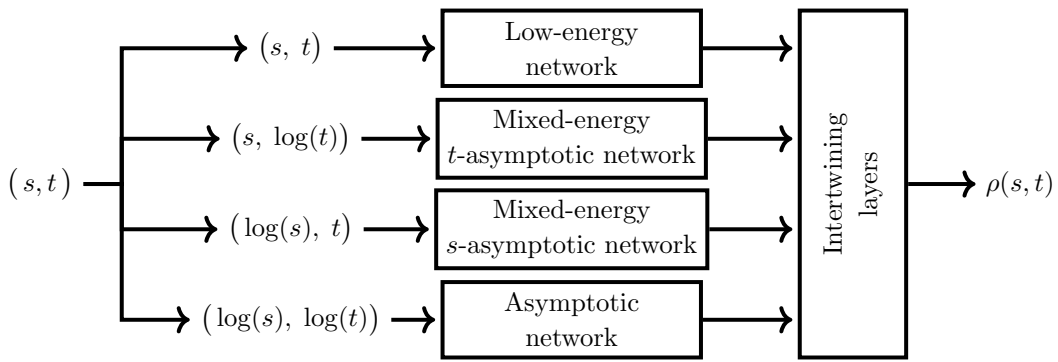
\begin{figure}[H]
    \centering
    \begin{tikzpicture}[font=\small]

\tikzset{
  block/.style = {draw, line width=1.5pt,
                  minimum width=3.4cm, minimum height=1.0cm,
                  align=center},
  % tallblock/.style = {draw, line width=1.5pt,
  %                     minimum width=1.4cm, minimum height=4.6cm,
  %                     align=center},
  tallblock/.style = {draw, line width=1.5pt,
                      minimum width=4.6cm, minimum height=1.4cm,
                      align=center},
  conn/.style = {->, thick},
}
%--------------------------------------------------------
% Input coordinate
%--------------------------------------------------------
\node (input) at (-4.5,0) {$
% \left(
% \begin{aligned}
% % x &= \frac{4m^2}{s} \\
% % y &= \frac{4m^2}{t}
% \,s\, \\
% \,t\,
% \end{aligned}\right)
\big(\,s,t\,\big)
$
};

%--------------------------------------------------------
% Input splits
%--------------------------------------------------------
% \node (x)    at (-1.2,  1.8) {$\big(x,\;y\big)$};
% \node (xlogy) at (-1.2,  0.6) {$\big(x,\;\log(y)\big)$};
% \node (logxy) at (-1.2, -0.6) {$\big(\log(x),\;y\big)$};
% \node (logxlogy) at (-1.2, -1.8) {$\big(\log(x),\;\log(y)\big)$};

\node (x)    at (-1.2,  1.8) {$\big(s,\;t\big)$};
\node (xlogy) at (-1.2,  0.6) {$\big(s,\;\log(t)\big)$};
\node (logxy) at (-1.2, -0.6) {$\big(\log(s),\;t\big)$};
\node (logxlogy) at (-1.2, -1.8) {$\big(\log(s),\;\log(t)\big)$};

%\draw[conn] (input.east) -- (x.west);
\draw[conn,line width=1.5pt] (input.east) -- ++(0.5,0) |- (x.west);
\draw[conn,line width=1.5pt] (input.east) -- ++(0.5,0) |- (xlogy.west);
\draw[conn,line width=1.5pt] (input.east) -- ++(0.5,0) |- (logxy.west);
\draw[conn,line width=1.5pt] (input.east) -- ++(0.5,0) |- (logxlogy.west);

%--------------------------------------------------------
% Sub-networks
%--------------------------------------------------------
\node[block] (NNlow) at (2.5,  1.8) {Low-energy\\network};
\node[block] (NNt)   at (2.5,  0.6) {Mixed-energy\\$t$-asymptotic network};
\node[block] (NNs)   at (2.5, -0.6) {Mixed-energy\\$s$-asymptotic network};
\node[block] (NNas)  at (2.5, -1.8) {Asymptotic\\network};

\draw[conn, shorten >=2pt,line width=1.5pt] (x.east) -- (NNlow.west);
\draw[conn, shorten >=2pt,line width=1.5pt] (xlogy.east) -- (NNt.west);
\draw[conn, shorten >=2pt,line width=1.5pt] (logxy.east) -- (NNs.west);
\draw[conn, shorten >=2pt,line width=1.5pt] (logxlogy.east) -- (NNas.west);

%--------------------------------------------------------
% Intertwining layers
%--------------------------------------------------------
%\node[tallblock] (merge) at (6.0,0) {Intertwining\\layers};
\node[tallblock,rotate=90] (merge) at (6.2,0) {Intertwining\\layers};

\draw[conn, shorten >=2pt,line width=1.5pt] (NNlow.east) -- (merge.north |- NNlow.east);
\draw[conn, shorten >=2pt,line width=1.5pt] (NNt.east)   -- (merge.north |- NNt.east);
\draw[conn, shorten >=2pt,line width=1.5pt] (NNs.east)   -- (merge.north |- NNs.east);
\draw[conn, shorten >=2pt,line width=1.5pt] (NNas.east)  -- (merge.north |- NNas.east);

%--------------------------------------------------------
% Output
%--------------------------------------------------------
\node (out) at (8.4,0) {$\rho(s,t)$};
%\draw[conn] (merge.east) -- (out.west);
\draw[conn,line width=1.5pt] (merge.south) -- (out.west);

\end{tikzpicture}
    \caption{Typical neural network used to parametrize the amplitude. For better expressivity, we design the network using connected subnetworks that deal with different regions.}
    \label{fig:NN_archi}
\end{figure}

We parameterize the spectral densities using neural networks, and we train them by imposing the two-particle unitarity. The amplitudes of interest effectively exhibit an emergent nonperturbative Regge limit, and to capture it faithfully we adopt a particular architecture which is suitable for the description of the Regge limit, see Figure \ref{fig:NN_archi}.
We reformulate the solving of the unitarity equations as an optimization problem over the space of spectral densities. For each unitarity condition to be imposed, we introduce a non-negative functional that vanishes if and only if the corresponding constraint is exactly satisfied. The total loss is defined as a weighted sum of these contributions,
\begin{equation}
\label{eq:schematic_loss}
\mathcal{L}\left(\rho(s),\rho(s,t)\right) = \sum_k \lambda_k\mathcal{L}_k\,,
\end{equation}
where each term $\mathcal{L}_k$ enforces a specific physical constraint and the coefficients $\lambda_k>0$ control their relative numerical weight. In an exact solution, all $\mathcal{L}_k$ vanish simultaneously.
The minimization is performed using standard gradient-based optimization methods, which rely on evaluating the gradient of the loss with respect to the neural-network parameters.

\subsection{Main results}

One of the main results of the paper is the numerical construction of the 2PRR amplitude for 
\be
0 \leq c_0 \lesssim 44 \pi ,
\ee
using neural networks and characterization of their physical properties. By construction, they exhibit \emph{fine analyticity properties} familiar from the study of the Feynman diagrams: the double discontinuity of the amplitude is only nonzero above the leading Landau curves. The physical properties include:
\begin{itemize}
\item \textbf{Emergent Regge and fixed-angle behavior.}
In \emph{the Regge limit},\footnote{Recall that the Regge limit corresponds to $s\to \infty$ with $t$ held fixed.} the amplitude approaches a $t$-dependent constant, consistent with an
emergent $J=0$ Regge-pole asymptotic. By contrast, the discontinuity of the amplitude exhibits a more intricate behavior, consistent with the presence of Regge cuts.
In the high-energy \emph{fixed-angle} regime,\footnote{In this limit, we have $s \to \infty$ with $t/s$ held fixed.} the amplitude decays as ${1 \over \log s}$ consistent with an effective logarithmic
running similar to $\phi^4$ theory at negative coupling~\cite{Symanzik:1961}.
\item \textbf{Particle production and transparency.} The partial waves $S_J(s)$ of the amplitude exhibit sizeable inelasticity about the four-particle threshold. Particle production has a clear origin at the level of the 2PRR graphs, see Fig. \ref{fig:series}. In the high-energy limit, they exhibit \emph{transparency}, i.e. $\lim_{s \to \infty} S_J(s) = 1$. 
\end{itemize}
Let us emphasize that the UV properties of the amplitude are highly nontrivial and are not visible at fixed-order perturbative analysis, or other approaches to the S-matrix bootstrap, which focus on the low-energy observables.

Finally, we go beyond the 2PRR setup and consider the problem with \emph{dynamical multi-particle data}:
\begin{itemize}
\item \textbf{Dynamical multi-particle data and Aks screening.}
The Aks theorem states that nontrivial scattering in spacetime dimensions $d>2$ implies particle production
\cite{Aks:1965qga}.
At the same time, extremal solutions of semidefinite $S$-matrix bootstrap studies often
appear close to elastic~\cite{Paulos:2017fhb}.
To probe this tension further, we go beyond the 2PRR model and allow a nonzero $\rho_{\rm MP}(s,t)$. Specifically, we train the network by solving two-particle unitarity and in addition \emph{minimizing particle production} in low-spin partial waves, in order to mimic the outcome of semidefinite bootstrap. In this setup $\rho_{\rm MP}(s,t)$ is generated dynamically during the training process and conceptually allows one to explore the full space of scattering amplitudes. We find that multi-particle input can be used to screen low-spin particle production, a phenomenon which we refer to as \textit{Aks screening}.

We find that suppressing inelasticity by tuning the multi-particle data drives the amplitude towards extremality, while at the same time necessarily leading to larger and more oscillatory $\rho_{\rm MP}(s,t)$. This phenomenon emphasizes the importance of a deeper
understanding of the physical principles that constrain the multi-particle data for further
progress in the S-matrix bootstrap studies. 
Furthermore, the use of variational multi-particle data opens new perspectives for exploring the space of consistent $2 \to 2$ scattering amplitudes with machine learning.

\end{itemize}

\paragraph{Plan of the paper}
The paper is organized as follows.
\begin{itemize}
    \item In Section~2, we introduce the Atkinson–Mandelstam framework and summarize several relevant results from the literature. We also provide a self-contained overview of our machine-learning approach, sufficient to follow the logic of the results.
    \item Section~3 contains the technical details of the numerical implementation, including the discretization strategy, the neural-network architecture, the loss functions, and the training procedure.
    \item In Section~4, we present our results for 2PRR amplitudes, with a focus on their Regge behavior and inelasticity. We also examine how the solutions evolve as the coupling $c_0$ is varied.
    \item Section~5 is devoted to inelastic effects. Within our Atkinson–Mandelstam approach, we treat the multi-particle double discontinuity $\rho_\mathrm{MP}$ as a dynamical degree of freedom and use it to screen inelasticity, a procedure we refer to as \emph{Aks screening}. Separately, we take the inelasticity profiles extracted from the 2PRR amplitudes and use them as input to the semidefinite bootstrap, showing that it consistently reproduces the same 2PRR solutions, including their Regge behavior.
    \item Finally, in Section~6 we briefly review key aspects of the Atkinson–Mandelstam approach and discuss several directions for future work.
\end{itemize}
In the appendices, we collect our definitions and conventions, as well as technical details of the semidefinite bootstrap analysis.

\subsection{Comparison to existing literature}

Modern scattering-amplitude and S-matrix bootstrap programs aim to constrain and characterize the space of physical S-matrices using the principles of causality, Lorentz invariance, and quantum mechanics.
In perturbative scattering (relevant for collider physics), amplitudes exhibit a rich analytic structure of singularities, reflecting perturbative unitarity; understanding this structure has driven many recent developments (see e.g. \cite{Hannesdottir:2021LandauIteratedIntegrals,Hannesdottir:2022SequentialDiscontinuities,Hannesdottir:2024LandauBootstrap,Caron-Huot:2024RecursiveLandau,Correia:2025SOFIA,Mizera:2021LandauDiscriminants,Fevola:2023LandauSingularitiesRevisited,Fevola:2023PrincipalLandauDeterminants}).
Modern nonperturbative approaches (see e.g. \cite{Kruczenski:2022lot}), building on earlier work \cite{Lukaszuk:1967zz,Lopez:1974cq,Lopez:1976zs,Bonnier:1975jz,Lopez:1975ca,Lopez:1975wf,Auberson:1977ss}, focus on constraining low-energy observables---such as Wilson coefficients and scattering lengths---using maximal (or axiomatic \cite{Guerrieri:2021tak}) analyticity, partial-wave unitarity $|S_J|\le 1$, and crossing symmetry.

For example, in the \emph{primal approach} of \cite{Paulos:2016fap,Paulos:2016but,Paulos:2017fhb}, one scans the space of low-energy observables by parametrizing the amplitude with an ansatz.\footnote{See also~\cite{Bhat:2023puy,deRham:2025vaq} for alternative primal formulations using partial waves as their basis.}
The ansatz is constructed to satisfy maximal analyticity and crossing symmetry, while unitarity is enforced by formulating a convex optimization problem over semidefinite matrices.
This algorithm yields four-point scattering amplitudes that numerically saturate the unitarity constraint $|S_J(s)|\leq 1$ (at all available energies and spins)
These methods were applied to study the scattering of  scalars~\cite{He:2021eqn,Chen:2022nym,EliasMiro:2022xaa,Acanfora:2023axz,Correia:2025uvc,Gumus:2025hwq}, or spinning particles~\cite{Hebbar:2020ukp,Haring:2022sdp}, and in the context of gauge theories such as QCD~\cite{Guerrieri:2018uew,Guerrieri:2020bto,Guerrieri:2023qbg,Guerrieri:2024jkn,He:2023lyy,He:2024nwd,He:2025gws,Bose:2020cod,Bose:2020shm,Li:2023qzs}, and string theory~\cite{Guerrieri:2021ivu,Guerrieri:2022sod}. 

In this work, we blend features of both perturbative and nonperturbative approaches. In the spirit of perturbative analyses, we impose the detailed analytic structure of the amplitude by implementing the corresponding support properties of the double spectral density, as dictated by the elastic \cite{Correia:2020xtr} and multi-particle \cite{Correia:2021etg} Landau curves.

This approach was pioneered  by Mandelstam and Atkinson
\cite{Mandelstam:1958xc,Mandelstam:1959bc,Atkinson:1968hza,Atkinson:1968exe,Atkinson:1969eh,Atkinson:1970pe,Atkinson:1970zza}. It allows to cleanly separate the elastic (two-particle) regime, where $|S_J|=1$, from the inelastic (multi-particle) regime, where $|S_J|\le 1$, while maintaining crossing symmetry.  We then use a neural-network parameterization of the amplitude to explore the space of low-energy observables in the spirit of the nonperturbative $S$-matrix bootstrap.

Using the Mandelstam representation~\eqref{eq:mandelstamRep1}, which is manifestly crossing-symmetric and respects analyticity, two-particle unitarity becomes a non-linear integral equation for the double spectral density.\footnote{
See also crossing-symmetric dispersion relations for an alternative representation~\cite{Auberson:1972prg,Mahoux:1974ej,Sinha:2020win,EliasMiro:2025rqo}.
} The original papers proposed an iterative strategy for constructing solutions, pursued in \cite{Tourkine:2021fqh,Tourkine:2023xtu}, in which the multi-particle data enters as a source term. In this work, following the approach of~\cite{Gumus:2024lmj}, we not only obtain numerically solutions for the 2PRR amplitudes, but also promote the multi-particle data from an external input to a dynamical quantity.

Relative to \cite{Tourkine:2023xtu}, where  2PRR amplitudes were initially studied, the present paper:
\begin{itemize}
    \item achieves numerical Reggeization, 
    \item reaches genuinely strong-coupling solutions,
    \item produces qualitatively similar, but significantly larger, multi-particle contributions,
    \item introduces the conceptual step of treating the multi-particle data as a dynamical degree of freedom.
\end{itemize}

The approach of~\cite{Gumus:2024lmj} uses physics-informed neural networks (PINNs)~\cite{karniadakis2021physics} to find solutions to the nonlinear unitarity equations.

See \cite{Dersy:2023job,Mizera:2023bsw,Niarchos:2024onf,Bhat:2024agd} for applications of neural networks to the S-matrix bootstrap, and \cite{Guerrieri:2024jkn} for a gradient-free approach to nonconvex optimization.

The amplitudes we obtain are \emph{analytically complete}, in the sense that we can access the full physical sheet.
This contrasts with other approaches that fail to converge in the double-spectral-density region $s,t\ge 4m^2$.
This completeness lies at the heart of the two-particle versus multi-particle separation of our method.
\section{Review and brief Machine-learning implementation}
\label{sec:AMreview}
In this section we review  our setup and write down the set of equations that we then solve numerically using the neural optimizer. The section is intended to be self-contained, but we refer to \cite{Tourkine:2023xtu} and \cite{Correia:2020xtr} for further technical details.

\subsection{The Atkinson--Mandelstam bootstrap}
\label{sec:AMbootstrap}
In~\cite{Mandelstam:1958xc,Mandelstam:1959bc,Mandelstam:1963iyb}, Mandelstam introduced a dispersive framework based on maximal analyticity to solve the non-perturbative $2\to2$ unitarity equation in a crossing-symmetric manner. To this end, he proposed a double-dispersive representation called the Mandelstam representation~\eqref{eq:mandelstamRep1}.
Of course unitarity cannot close with only $2\to2$ processes, as even this involves $2\to n$ processes in cuts via the optical theorem, which themselves can only close with the whole tower of $n\to m$ processes. 
The original idea of Mandelstam was therefore to use the genuinely multi-particle contributions as a source in the $2\to2$ unitarity to calculate the scattering amplitude from this input as a response by means of iterations. In this way, starting from a seed, possibly perturbative, one constructs a sum of integrals which can be rewritten as sums of Feynman diagrams.

This separation is particularly clean in the context of scattering of massive particles in gapped theories. In the paper, we work in a $\mathbb{Z}_2$-invariant theory with no bound-state below threshold:
in this framework, multi-particle physics enters for energies $s,t \geq 16 m^2$, whereas scattering for $s,t < 16 m^2$ is purely two-to-two. By exploiting analyticity and dispersion relations, Mandelstam, assuming maximal analyticity, in \cite{Mandelstam:1958xc} wrote down a set of equations that the two-to-two amplitude must satisfy. Multi-particle physics in this set of equations enters through the single- and double-discontinuity of the amplitude supported in the multi-particle region $s,t \geq 16 m^2$. The purpose of the present paper is to explore the space of solutions to these equations as a function of the multi-particle data.

It was already appreciated by him to some extent that some source of multi-particle physics is encoded in these non-linear equations, intuitively any process that is reducible to two-body physics in the $s$, $t$, or $u$ channels. In the language of cuts, what we made precise in~\cite{Tourkine:2023xtu} is that all graphs which are \textit{two-particle-recursively-reducible} are computed \textit{from first principles} by the formalism. This includes an infinite class of diagrams and processes with clear multi-particle production. See~Fig.~\ref{fig:series} for a graphical definition of this notion.

Let us explain the details of the formalism. We start from the double spectral function, which coincides with the double discontinuity of the amplitude $\rho(s,t)$.
We remind that it is defined as
\begin{equation}
\label{eq:ddisc-def}
\begin{aligned}
\rho(s,t) &=\frac{1}{(2i)^2}\,
\mathrm{Disc}_s\,\mathrm{Disc}_t\,T(s,t)\\
 &\equiv  \frac{1}{(2i)^2} \lim_{\eps \to 0}\Big[T(s{+}i\epsilon,t{+}i\epsilon)-T(s{-}i\epsilon,t{+}i\epsilon)-T(s{+}i\epsilon,t{-}i\epsilon)+T(s{-}i\epsilon,t{-}i\epsilon)\Big],
\end{aligned}
\end{equation}
with an analogous definition for single discontinuities in~\eqref{eq:single_disc-def}. %\pinote{added def, updated text a bit}
Following Mandelstam and Atkinson, we decompose $\rho(s,t)$ in three terms: an elastic piece, its crossing-symmetric counter-part and the genuinely multi-particle double-discontinuity
\begin{equation}
\label{eq:ddiscsum}
      \rho(s,t) = \rho_{\text{el}}(s,t) + \rho_{\text{el}}(t,s) +   \rho_{\text{MP}}(s,t) , ~~~ \rho_{\text{MP}}(s,t) = \rho_{\text{MP}}(t,s) ,
\end{equation}
where $\rho_{\text{el}}(s,t)$ is the solution of \emph{the Mandelstam equation}
%where $\rho_\mathrm{el}$ is given by
\begin{equation}
\label{eq:mandelstam-eqn}
%\label{eq:rho_el_def}
  \rho_{\text{el}}(s,t) = {(s - 4 m^2)^{1 \over2} \over  (4\pi)^{2} \sqrt{s} }\int\limits_{z_1}^{\infty} d \eta'  \int\limits_{z_1}^{\infty} d \eta'' \theta(z- \eta_+) %
 {T_t(s+i \eps, t(\eta') ) T_t(s- i \eps, t(\eta'') ) \over \sqrt{(z - \eta_-)(z - \eta_+)}} , 
\end{equation}
where 
\begin{equation}
\label{eq:eta-t-z1-def}
\begin{aligned}
    &\eta_{\pm} = \eta' \eta'' \pm \sqrt{\eta'^2 -1} \sqrt{\eta''^2 - 1}\,,\\
    &t(\eta) = {s - 4 m^2 \over 2} (\eta - 1)\,,\\
    &z_1=1+{8m^2 \over s-4m^2}\,,
\end{aligned}
\end{equation}
and 
\begin{equation}
    T_t(s,t) \equiv \lim_{\eps \to 0} {T(s,t+i \eps) - T(s, t-i \eps) \over 2 i}\,,
    \label{eq:single_disc-def}
\end{equation}
is the $t$-channel discontinuity of the amplitude; using the parametrization given in Eq.~\eqref{eq:mandelstamRep1}, this definition yields
\begin{equation}
\label{eq:single_disc}
    T_t(s,t)=\rho(t)+\!\!\int_{4m^2}^\infty\!\!\dfrac{\mathrm{ds'}}{\pi}\dfrac{(s-s_0)\rho(s',t)}{(s'-s)(s'-s_0)}+\!\int_{4m^2}^\infty\!\!\dfrac{\mathrm{du'}}{\pi}\dfrac{(u-u_0)\rho(u',t)}{(u'-u)(u'-u_0)}\,.
\end{equation}
In the elastic strip, for $4m^2\leq s\leq 16m^2$, $\rho_\el(s,t) = \rho(s,t)$ and  \eqref{eq:mandelstam-eqn} is a closed equation that enfores \textit{elastic} unitarity
\begin{equation}
    \text{Elastic unitarity:}\qquad |S_J(s)|=1,\quad 4m^2 \leq s < 16 m^2\,,
\end{equation}
analytically continued in spin $J$ for all $J$ such that ${\rm Re}[J]>0$. The partial waves $S_J(s)$ are defined in appendix~\ref{app:definitions}.
In this work we \emph{define} $\rho_{\rm el}(s,t)$ for all $s,t\ge 4m^2$, in particular beyond $16m^2$, by the right-hand side of
\eqref{eq:mandelstam-eqn}, i.e. we impose \eqref{eq:mandelstam-eqn} as a functional identity beyond the elastic strip. This continuation is essential for describing the 2PRR class and is checked a posteriori through the resulting partial waves and discontinuities.

The physical support of the elastic double discontinuities $\rho_\mathrm{el}$ can be derived from Eq.~\eqref{eq:mandelstam-eqn} (see~\cite{Correia:2020xtr}) and is constrained by the following Landau curves:
\begin{subequations}
    \begin{align}
    t &\geq \dfrac{16s}{s-4m^2}, \quad (s\text{-channel}),\label{eq:LeadingLC-s}\\
    t &\geq \dfrac{4s}{s-16m^2}, \quad (t\text{-channel}),\label{eq:LeadingLC-t}
    \end{align}
    \label{eq:LeadingLC}
\end{subequations}
which determine the regions of the $(s,t)$ plane in which the elastic double discontinuities $\rho_{\text{el}}(s,t)$ and $\rho_{\text{el}}(t,s)$ are non-vanishing,
respectively.
Intuitively the fact that the double-discontinuity should have support in the region $s,t\geq4m^2$ is obvious: for a process to go on-shell simultaneously in the $s$ and $t$ channels, both $s$ and $t$ must exceed the amount of energy needed to send at least two particles on-shell in \textit{each} channel.

Finally, \(\rho_{\text{MP}}(s,t)\) encodes genuinely multi-particle contributions that cannot be reduced to iterated two-body interactions.\footnote{In terms of Feynman diagrams, these are graphs with more than two-particle cuts in all channels, see Fig.~\ref{fig:MP}.}
Whenever we include \(\rho_{\text{MP}}\) in this work, we \emph{assume} that its support is bounded by the \emph{planar-cross} Landau curve,
\begin{equation}
\label{eq:PlanarCrossLandauCurve}
s^3(t-16m^2)+t^3(s-16m^2)+24st(s+t-18m^2)m^2-2s^2t^2=0,
\end{equation}
following~\cite{Correia:2021etg}, where it was conjectured that no other graphs generate singularities prior to this Landau curve. This assumption about the support of \(\rho_{\text{MP}}\) is expected to follow from multi-particle unitarity. 

\begin{figure}[h!]
    \centering
    \includegraphics[scale=1]{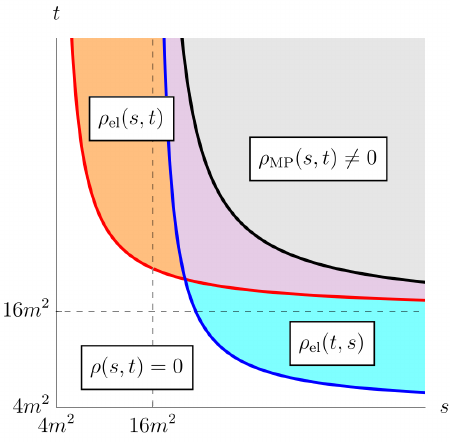}
    \caption{Structure of the double discontinuity $\rho(s,t)$ in the $(s,t)$-plane.
    {\color{red}\textbf{Red}:} leading $s$-channel Landau curve (eq.\eqref{eq:LeadingLC-s}), {\color{blue} \textbf{Blue}:} leading $t$-channel Landau curve (Eq.~\eqref{eq:LeadingLC-t})
\textbf{Black}: conjectured leading Landau curve for the onset of the inelastic contribution $\rho_\MP$~\cite{Correia:2021etg} (see Eq.~\eqref{eq:PlanarCrossLandauCurve}).
The purple region indicates the domain where both elastic contributions overlap. In the $(x=4/s,y=4/t)$-plane, the red and blue LC become straight lines, see e.g.\cite[Fig.~4]{Tourkine:2023xtu}. }
    \label{fig:DDiscStructure}
\end{figure}

Importantly, with the definitions above, $\rho_{\mathrm{MP}}$ is not the only source of inelasticity. Since \eqref{eq:mandelstam-eqn} enforces elastic scattering, if we had $\rho(s,t)=\rho_\mathrm{el}(s,t)$ for all $t$, we would have a purely elastic amplitude. But in our ansatz, Eq.~\eqref{eq:ddiscsum}, crossing symmetry forces $\rho_\el(t,s)$ to be present, therefore this term has to generate multi-particle contributions (see Fig.~\ref{fig:DDiscStructure}). Such inelasticity is unavoidable in any interacting quantum field theory in spacetime dimensions greater than two~\cite{Aks:1965qga}. We summarize the support in the $(s,t)$-plane of the various pieces of the double discontinuity in Fig.~\ref{fig:DDiscStructure}.

To impose the $J=0$ unitarity, which is not captured by~\eqref{eq:mandelstam-eqn}, we write 
\begin{equation}
\begin{aligned}
    \label{eq:single-disc-unitarity}
  &2\,\im\,f_0(s)=\sqrt{s-4 m^2 \over s} \left|  f_0(s) \right|^2 +
    \sqrt{s \over {s - 4 m^2} }\; \eta_{0}(s) \, ,\\
    &\text{where}\quad f_0(s)=\dfrac{1}{16\pi}\int_0^{2\pi}\!\!\!\mathrm{d}\theta\;T\!\left(s,-\dfrac{s-4m^2}{2}(1-\cos\theta)\right)\,,
\end{aligned}
\end{equation}
and $\eta_0$ is the inelasticity in the spin-0 partial wave
\begin{equation}
\label{eq:partialwavezero}
    \eta_0(s)=1-|S_0(s)|^2\,.
\end{equation}
The explicit expression of $f_0$ in terms of the spectral densities is provided in Appendix~\ref{app:definitions}, Eq.~\eqref{eq:f0def}.
At this point, $\eta_0$ is an unconstrained function for which only the support is known.
A priori, Eq.~\eqref{eq:single-disc-unitarity} allows for two possible threshold behaviors
\begin{align}
    \im\,f_0(s)&=c_\mathrm{Th}\dfrac{\sqrt{s-4m^2}}{4m}+\mathcal{O}\big((s-4m^2)^{3/2}\big)\,,\label{eq:regular}\\
    \im\,f_0(s)&=\dfrac{4m}{\sqrt{s-4m^2}}+\mathcal{O}\big(\sqrt{s-4m^2}\big)\,.\label{eq:singular}
\end{align}
In this work we will only consider the regular threshold behavior Eq.~\eqref{eq:regular}.

To summarize our formalism, we start from the Mandelstam representation~\eqref{eq:mandelstamRep1} and we impose unitarity using the Mandelstam equation~\eqref{eq:mandelstam-eqn} and and the spin-0 unitarity condition~\eqref{eq:single-disc-unitarity}.
We have three unconstrained quantities ($c_0$, $\eta_0$, $\rho_\MP$) which, in the original spirit of Mandelstam, can serve as input.

In the following section, we review and explain how to actually compute $\eta_0$ in terms of the double spectral density $\rho(s,t)$ and we motivate why we believe choice together with $\rho_\MP$ precisely select the 2PRR amplitudes mentioned in the introduction.

\paragraph{Spin-0 particle production and analyticity in spin.} %Froissart-Gribov at $J=0$.}
We now come to a subtle and important point, which was already discussed in~\cite[eq 2.25]{Tourkine:2023xtu}. Let us first assume that we set by hand $\rho_\MP$ to zero.
On its own, the Mandelstam equation, together with~\eqref{eq:ddiscsum}, and its extension beyond $s,t=16m^2$ to $\infty$ solves a crossing-symmetric 2-to-2 unitarity for graphs with a nonzero double discontinuity. From this alone, the diagrammatics of 2PRR is already clear, as it captures in a crossing-symmetric manner the 2-particle unitarity cuts.
The Mandelstam equation~\eqref{eq:mandelstam-eqn} imposes unitarity for all spins  $J>0$ but does not capture the $J=0$ sector.

Therefore, the Mandelstam equation firstly obviously does not capture graphs with a vanishing double-discontinuity, such as bubble chains, and other graphs with only a single cut (see examples in the first two lines of Fig.~\ref{fig:series}), which are pure $J=0$. A separate treatment of the $J=0$ sector is therefore required to capture them. Note that none of these graphs generate particle production.%\pinote{check}

Secondly, about graphs with a double-discontinuity, the Mandelstam equation on its own does only reconstruct their $J\neq 0$ part. If we set by hand $|S_0|=1$ ($\eta_0=0$), we obtain a class of amplitudes which we called \emph{quasi-elastic} in \cite{Tourkine:2023xtu}, where multi-particle production is projected out from the 2PRR graphs. These amplitudes are simpler to construct but are severely non-analytic in spin at $J=0$.
To obtain the equations describing the full 2PRR graphs, we need to add a new equation in order to backreact into the unitarity relation for $J=0$ the scalar sector of these graphs.

We start with the definition of inelasticity in partial waves $\eta_J$ for $J\geq0$:
\begin{equation}
\label{eq:partialwaveJ}
1 - |S_{J}(s)|^2 = \eta_{J}(s) . 
%\eta_{\text{MP}}(s)   .
\end{equation}
We are interested in how the Mandelstam equation generates an $\eta_0(s)$: while it imposes \textit{elastic unitarity} for $J>0$, it nevertheless \textit{generates} some $J=0$ contribution $\eta_0$, which we want to impose as a unitary condition in the $J=0$ sector, as was explained in~\cite{Tourkine:2023xtu}.

We recall that for $J>0$, the Froissart-Gribov formula for the imaginary part of $f_J$ gives
\begin{equation}
    \im f_J(s) \equiv \dfrac{1}{8\pi^2}\int_{z1}^\infty\mathrm{d}z\;Q_J^{(4)}(z)\rho(s,t(z)).
\end{equation}
Since $\rho_\el(s,t)$ generates elastic scattering, via Froissart-Gribov, Eq.~\eqref{eq:partialwavezero} gives the explicit formula for inelasiticities in higher spins:
\begin{equation}
\label{eq:eta-J-Q-J}
    \eta_J(s)=\dfrac{1}{4\pi^2}\sqrt{\dfrac{s-4m^2}{s}}  \int_{4}^{\infty} Q_J\!\left(1+\frac{2t}{s-4m^2}\right)\big(\rho_{\el}(t,s)+\rho_{\MP}(t,s)\big)\,dt
\end{equation}

Pretending at first that it can be extended to $J=0$, and using the fact that $\rho_\text{el}(s,t)$, we introduce the following quantities:
\begin{subequations}
\begin{align}
    \eta_0(s)&\equiv\eta_\text{2PRR}(s)+\bar \eta_\text{MP}(s) \ , 
    \label{eq:eta_0_el_mp} \\
    \eta_\text{2PRR}(s) &\equiv \dfrac{1}{4\pi^2}\sqrt{\dfrac{s-4m^2}{s}}  \int_{z_1}^\infty\!\!\!\! \mathrm{d} z\, \operatorname{tanh^{-1}}\!\!\left(\dfrac{1}{z}\right) \rho_{\text{el}}(t(z),s) \ , 
    \label{eq:eta_el_def} \\
    \bar \eta_\text{MP}(s)&=  \eta_\text{MP}(s) + \dfrac{1}{4\pi^2}\sqrt{\dfrac{s-4m^2}{s}}  \int_{z_1}^\infty\!\!\!\! \mathrm{d} z\, \operatorname{tanh^{-1}}\!\!\left(\dfrac{1}{z}\right) \rho_{\text{MP}}(s,t(z)) \,,
    \label{eq:eta_MP_def}
\end{align}
\label{eq:inel_el_mp}
\end{subequations}
where $ \eta_\text{MP}(s) $ is a genuine spin-zero inelasticity that is not connected to the double spectral density and unrelated to the 2PRR graphs.
To be self-consistent, we then need to check \textit{a posteriori} that the various integrals are well-defined and convergent. In particular, Eq.~\eqref{eq:eta_el_def} requires
\begin{equation}
\label{eq:consistent2PRR}
    \lim_{t\to\infty}\rho_\mathrm{el}(t,s)=0\,,\quad\forall s\,.
\end{equation}
For the 2PRR class of amplitudes considered in this paper, all such integrals are observed to converge nicely.
These considerations complete Mandelstam's original construction and allow to describe physical systems where particle production is present in the $S$-wave.

\begin{tcolorbox}[title={Definition (2PRR amplitudes).}]
In this work, by \textit{2PRR amplitudes} we mean solutions of the Mandelstam equations \eqref{eq:mandelstam-eqn}, \eqref{eq:single-disc-unitarity} and  \eqref{eq:inel_el_mp} with subtraction constant $c_0$ and
\begin{equation}
\label{eq:2PRR_def}
\rho_{\rm MP}(s,t)=0,\qquad \eta_{\rm MP}(s)=0,
\end{equation}
and with the $S$--wave inelasticity determined self--consistently from the crossed elastic
double discontinuity via \eqref{eq:eta_el_def}.
Equivalently, the full double discontinuity takes the form
\begin{equation}
\rho(s,t) = \rho_{\rm el}(s,t) + \rho_{\rm el}(t,s).
\end{equation}
Furthermore, {2PRR} amplitudes are defined to have regular 2-particle threshold behavior (see below Eq.~\eqref{eq:regular}), and have no bound states. We interpret these solutions as the sum of two-particle-recursively-reducible graphs (see Fig.~\ref{fig:series}).
\end{tcolorbox}
We emphasize that this definition does not guarantee uniqueness: additional physical features,
such as the threshold behavior (Eqs.~\eqref{eq:regular} and~\eqref{eq:singular}) or the structure of resonances, are not fixed by Eq.~\eqref{eq:2PRR_def}.

\subsection{Collected positivity statements}
\label{sec:coll_pos}
We summarize now several useful results related to the sign properties of the double spectral function $\rho(s,t)$.

\subsubsection{Mahoux--Martin region}
\begin{figure}[h!]
    \centering
    \includegraphics[scale=1]{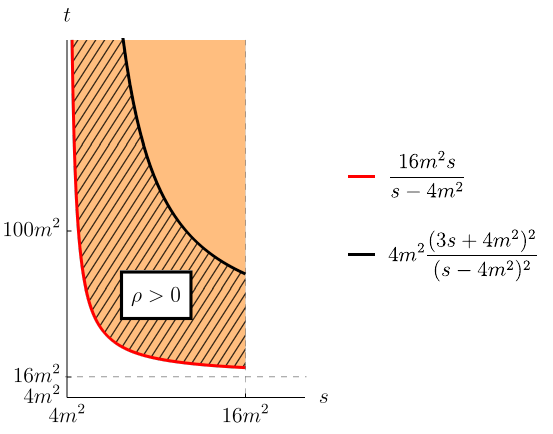}
    \caption{The Mahoux--Martin region in the $(s,t)$ plane, shown as a hatched area. In this region the double spectral function $\rho(s,t)$ is strictly positive.}
    \label{fig:mahoux-martin}
\end{figure}

As first established in Ref.~\cite{Mahoux:929517}, the double discontinuity is known to be strictly positive in the \emph{Mahoux--Martin region}
\begin{equation}
\label{eq:mm_region}
\rho(s,t) > 0\,,
\qquad
4m^2 < s < 16m^2\,,
\qquad
\frac{16m^2 s}{s - 4m^2} < t < 4m^2 \frac{(3s + 4m^2)^2}{(s - 4m^2)^2}\,,
\end{equation}
This follows from the Mandelstam equation~\eqref{eq:mandelstam-eqn}. In the Mahoux--Martin region, the single discontinuity $T_t(s,t)$ can be shown to be positive over the entire integration domain thanks to its partial wave expansion. As a result, the integrand of~\eqref{eq:mandelstam-eqn} is positive, which implies $\rho>0$ (see, for instance, Sec.~4.1 of Ref.~\cite{Correia:2020xtr} for a detailed derivation).
The support of the Mahoux--Martin region is illustrated in Fig.~\ref{fig:mahoux-martin}. In Sec.~\ref{sec:NNarchitecture} we incorporate this positivity constraint directly (hardwire) into the final layer of the neural parametrization of $\rho(s,t)$.

Importantly, this positivity result is restricted to the Mahoux–Martin region and does not extend to the full support of $\rho(s,t)$.
Nevertheless, \emph{if one assumes} that the double spectral function is positive everywhere on its support, a number of strong consequences follow.
In particular, Ref.~\cite{Martin:1969cq} claims that under this assumption the total cross section obeys the bound
\be
\sigma_\text{tot} < \frac{C}{\log s} \, ,
\ee
and that the Froissart–Gribov formula is valid for partial waves with spin $J>1$. This represents a much stronger constraint than the standard Froissart bound $\sigma_{\text{tot}} < C \log^2 s$.

Further implications were explored in Ref.~\cite{Yndurain:1969np}, where it was shown that the same positivity assumption implies the inequalities $\im f_{\ell}(s)/\im f_{\ell'}(s) > 1$ for $\ell' > \ell > 1$. These relations exclude elastic resonances except in the $S$-wave where saturating the partial wave bound $\im f_0(s)=2\sqrt{s}/\sqrt{s-4}$ is not forbidden.

Finally, Ref.~\cite{Martin:1969wv} argues that high-energy diffraction peaks are incompatible with a positive double spectral density.

\subsubsection{Mandelstam's  kernel pairing's sign-indefiniteness}

We now discuss briefly a simple but important property of the Mandelstam kernel in \eqref{eq:mandelstam-eqn}, 
\begin{equation}
K(z,\eta',\eta'') \equiv \Theta(z-\eta_+)\frac{1}{\sqrt{\bigl(z-\eta_+(\eta',\eta'')\bigr)\bigl(z-\eta_-(\eta',\eta'')\bigr)}} \, ,
\end{equation}
namely the fact that it does not define a positive semidefinite pairing. As we just explained, positivity of the double-discontinuity is only guaranteed in the Mahoux--Martin, and in general the double-discontinuity is allowed to change sign. The sign-indefiniteness of the Mandelstam kernel illustrates one aspect of this question.

\begin{figure}[h]
    \centering
    \includegraphics[scale=1.1]{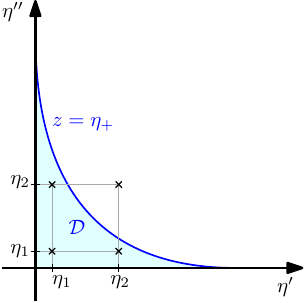}
    \caption{Integration domain of the Mandelstam equation and choice of pair $(\eta_1,\eta_2)$.}
    \label{fig:Mandelstam-region}
\end{figure}

Consider the right-hand-side of the Mandelstam equation at fixed $s,t$: take two functions $f_1,f_2$, we define their pairing by the quadratic form
\begin{equation}
    (f_1,f_2) = \iint  K(z,\eta',\eta'') f_1(\eta') f_2(\eta'')d \eta' d \eta''
\end{equation}
This pairing does not define a positive-definite product. It is straightforward to prove.
Firstly, note that a positive-definite product must obey the Cauchy--Schwarz inequality, i.e. for any $f_1,f_2$ we should have:
\begin{equation}
    (f_1,f_2)^2\leq (f_1,f_1) (f_2,f_2)
    \label{eq:CS-K}
\end{equation}
It is immediate to see that our pairing does not obey \eqref{eq:CS-K}. Because of the Heaviside theta-function support $\Theta(z-\eta_+)$, the integration is defined over a domain which is not the square, but a hyperbola-like-bounded region, of which an example is shown in blue in Fig.~\ref{fig:Mandelstam-region}, see also \cite[Fig.~3]{Tourkine:2023xtu}.
(Remember that we work at fixed $s,t$ and do not display the dependence on these variables in $z$ and $z_1$.)

Now choose a pair of points $\eta_1$ and $\eta_2$ such that $(\eta_2,\eta_2)$ is outside of the region $z-\eta_+\geq0$, while  $(\eta_1,\eta_2)$,  $(\eta_2,\eta_1)$ and  $(\eta_1,\eta_1)$ are strictly inside. This can always be done. An example is shown in Fig.~\ref{fig:Mandelstam-region}.
If we finally choose $f_i(\eta) = \delta(\eta-\eta-i)$ for $i=1,2$, we see that the pairing $(f_2,f_2)$ must vanish, as $(\eta_2,\eta_2)$ lies outside of the integration range, while the other pairings are non-zero.
This implies that the Mandelstam kernel pairing does not respect the CS inequality. The reasoning can be trivially extended to actual functions with compact supports around $\eta_1,\eta_2$. 

We finally also tested by direct inspection that upon discretization, the pairing possesses indeed large negative eigenvalues.
Overall, this implies that the Mandelstam kernel pairing is not sign-definite and, hence, it does not guarantee the positivity of $\rho(s,t)$ in general.

In addition, this criterion sheds light on a more general aspect of the Atkinson--Mandelstam bootstrap: its non-convexity. Consider the discretized Mandelstam equation, schematically it is a quadratic equation of the form $\rho_{i} = \sum K_{i,j,k} \rho_j\rho_k$. Such an equality is not a convex equation on its own, but the inequality $\rho_{i} \geq \sum K_{i,j,k} \rho_j\rho_k$ \textit{could} be.\footnote{This is obvious when considering the definition of convexity and using linear combinations of solutions. This is the same statement as saying that the \textit{curve} $y=x^2$ is not a convex region, while the area $y\geq x^2$ is.} It happens that since the quadratic form $K$ is not semi-positive definite, that is still not enough to define a convex problem, which motivates the use of general non-convex solver methods, such as iterative fixed-points in Mandelstam's original framework, or gradient based as we use in this paper, to which we turn now.

\subsection{Solving unitarity equations through loss minimization}
\label{sec:integral_opt}

\begin{tcolorbox}[title={The Atkinson-Mandelstam problem.}]
{Starting from the Mandelstam representation of the scattering amplitude~\eqref{eq:mandelstamRep1}, and multi-particle data $(\eta_{\MP},\rho_\MP)$, our objective is to impose unitarity.
To this end, we solve the following equations, introduced in the previous section:
\begin{itemize}
    \item Mandelstam equation~\eqref{eq:mandelstam-eqn} (elastic unitarity for spin $J>0$),
    \item Spin-0 unitarity~\eqref{eq:single-disc-unitarity},
    \item Inelastic unitarity for spin $J>0$\,:
    \begin{equation}
    \label{eq:inel-unitarity-SJ}
        |S_J(s)|\leq 1\,,\quad\forall s>16m^2\,.
    \end{equation} %Eq.~\eqref{eq:inel-unitarity-SJ}\pinote{check}
\end{itemize}
}
\end{tcolorbox}

In order to solve this system of coupled non-linear functional equations, one must resort to \emph{numerical} %functional 
analysis methods.

\paragraph{Iterative solution.} One possible solution strategy is to reformulate this system as a fixed-point problem. After fixing the external inputs $(c_0,\eta_0,\rho_\mathrm{MP})$, one defines a sequence of spectral densities $(\rho^{(n)}(s),\rho^{(n)}(s,t))$ by computing the left-hand side of Eqs.~\eqref{eq:single-disc-unitarity} and~\eqref{eq:mandelstam-eqn} using $(\rho^{(n-1)}(s),\rho^{(n-1)}(s,t))$ on the right-hand side. Given an initial guess $(\rho^{(0)}(s),\rho^{(0)}(s,t))$, this procedure defines an iterative map on the space of spectral densities. When the sequence converges, its fixed point is a solution of the unitarity equations.
This approach was pursued in~\cite{Tourkine:2021fqh,Tourkine:2023xtu}, where solutions could be obtained whenever the fixed point was attractive. However, for sufficiently large coupling ($c_0 \gtrsim 5\pi$), the fixed point becomes repulsive and the iteration diverges, rendering this method ineffective.

\paragraph{Gradient descent}
In this work, we continue the approach initiated in~\cite{Gumus:2024lmj} and inspired from \cite{Dersy:2023job}, and adopt a more direct strategy. We reformulate the solving of the unitarity equations as an optimization problem over the space of spectral densities. For each physical condition to be imposed, such as spin-0 unitarity or the Mandelstam equation, we introduce a non-negative functional that vanishes if and only if the corresponding constraint is exactly satisfied. The total loss is defined as a weighted sum of these contributions,
\begin{equation}
\label{eq:schematic_loss}
\mathcal{L}\left(\rho(s),\rho(s,t)\right) = \sum_k \lambda_k\mathcal{L}_k\,,
\end{equation}
where each term $\mathcal{L}_k$ enforces a specific bootstrap constraint and the coefficients $\lambda_k>0$ control their relative numerical weight. In an exact solution, all $\mathcal{L}_k$ vanish simultaneously.

Since the constraints are functional equations defined over continuous kinematic variables, they must be enforced numerically on discrete sets of sampling points. For each constraint $\mathcal{L}_k$, we therefore choose a grid $\mathcal{G}_k$ in the relevant kinematic domain and penalize the deviation between the left-hand side (LHS) and right-hand side (RHS) of the corresponding equation at each grid point. Schematically\footnote{In Eq.~\eqref{eq:grid_loss} we use an $\ell_1$-type penalty based on absolute values. Other choices, such as an $\ell_2$ penalty proportional to $\big(\mathrm{LHS}-\mathrm{RHS}\big)^2$, are equally possible. We have verified that our numerical results are not qualitatively sensitive to this choice, and we adopt the absolute value for simplicity.},
\begin{equation}
\label{eq:grid_loss}
\mathcal{L}_k(\rho(s),\rho(s,t)) =
\sum_{s_i\in\mathcal{G}_k}
w_{k}(s_i)\,\big|\mathrm{LHS}_i(\rho(s),\rho(s,t))-\mathrm{RHS}_i(\rho(s),\rho(s,t))\big|,
\end{equation}
where $\mathrm{LHS}_i$ and $\mathrm{RHS}_i$ denote the left- and right-hand sides of the corresponding bootstrap equation evaluated at the sampling point $s_i$, and $w_{k}(s_i)>0$ are numerical weights accounting for the relative importance of grid points.
The role of these weights is to enhance the contribution of regions where both $\mathrm{LHS}_i$ and $\mathrm{RHS}_i$ are small. In such regions, the absolute deviation $\big|\mathrm{LHS}_i-\mathrm{RHS}_i\big|$ can be numerically suppressed even when the relative error is large. Without appropriate weighting, these kinematic regimes would contribute negligibly to the loss and could therefore be poorly constrained during training. 
The grids $\mathcal{G}_k$ share a common energy cutoff $s_{\max}$; unitarity constraints are not enforced for energies above this scale.

In this formulation, the bootstrap problem reduces to the minimization of a single scalar functional $\mathcal{L}$, whose global minimum is achieved by spectral densities that simultaneously satisfy the Mandelstam equation and spin-0 unitarity (in the limit of dense sampling grids, where the continuum equations are recovered). Additional physical constraints can be incorporated straightforwardly within the same framework by adding corresponding terms to the loss.
This formulation nevertheless comes at a numerical cost: different loss terms can partially screen each other, requiring a careful choice of weights and a suitable training strategy which is described in Sec.~\ref{sec:TrainingNN}.

In the following subsections we first explain our neural network parametrization of the spectral densities and then provide the explicit losses $\mathcal{L}_k$ used to solve the unitarity equations.

\subsubsection{Neural network parametrization}
%\paragraph{Neural networks parametrization.}
\label{sec:NN-param}
To implement the procedure described above, one must choose a flexible parametrization of the spectral densities entering the Mandelstam representation. In this work, following our previous studies~\cite{Gumus:2024lmj,Dersy:2023job}, we use neural networks for this purpose.
This choice is motivated by the non-linear structure of the unitarity equations~(\ref{eq:mandelstam-eqn},~\ref{eq:single_disc}, and \ref{eq:inel-unitarity-SJ}) in terms of the spectral densities, which leads to coupled constraints beyond the scope of standard linear or semidefinite bootstrap methods~\cite{Boyd:2004fnq,Simmons-Duffin:2015qma}.
Neural networks offer a convenient parametrization in this setting. When taken sufficiently large, they can approximate arbitrary continuous functions, as guaranteed by the universal approximation theorem~\cite{cybenko1989approximation,hornik1989multilayer}. Their trainable parameters therefore provide a practical set of coordinates on the space of spectral densities, well suited for gradient-based methods. In practice, modern machine-learning frameworks combine this flexibility with efficient optimization algorithms and automatic differentiation, making it possible to handle large systems of coupled non-linear equations in a scalable way.
A more detailed discussion of the motivation for using neural networks in this context, together with further references to the machine-learning literature, can be found in our previous work~\cite{Gumus:2024lmj}.

In our previous work~\cite{Gumus:2024lmj}, we worked on a class of toy-model amplitudes with only a single spectral function and no-double-discontinuity, thus requiring only one network. In the present work, we use two neural networks, denoted $\mathrm{NN}_{\rho}$ and $\mathrm{NN}_{\mathrm{DDisc}}$, to parametrize the single and double spectral densities respectively. 
As usual in such approaches, it is useful to build in as much information as possible so that the network does not have to learn the already known physical information. For instance, if we know that a target function $f(s)$ decays as $C^{ st}/s$ as $s\to\infty$, it is useful to parametrize it with a neural network $\rm{NN}(s)$ such that $f(s) = \rm{NN}(s)/s$, so that the solver only has to determine the constant $c$ and not the full functional decay.
In our construction, we employ the following parametrization:
\begin{subequations}
\begin{equation}
\begin{aligned}
    \rho(s)&=f_{\rho}(s)\times\mathrm{NN}_{\rho}(s)\,,\\
    \rho(s,t)&=\rho_\text{NN}(s,t)+\rho_\text{NN}(t,s)\,,\quad\text{where\,,}\\
    \rho_\text{NN}(s,t)&=f_\mathrm{DDisc}(s,t)\times\mathrm{NN}_{\mathrm{DDisc}}(s,t)\,,
    \label{eq:spectralParametrization}
\end{aligned}
\end{equation}
and
\begin{equation}
\begin{aligned}
\label{eq:funcrion_params}
    f_{\rho}(s)&=\sqrt{\dfrac{s-4m^2}{s}}\dfrac{1}{\big(1+\log(\frac{s}{4m^2})\big)^3},\\
    f_\mathrm{DDisc}(s,t)&=\Theta\left(t-\dfrac{16m^2s}{s-4m^2}\right)\dfrac{8m^2}{s}\dfrac{1}{\big(1+\log(\frac{t}{16m^2})\big)^2}\sqrt{\dfrac{s-4m^2}{16m^2s}-\dfrac{1}{t}}.\\
\end{aligned}
\end{equation}
\end{subequations}
With this choice, the single spectral density $\rho(s)$ automatically reproduces the known threshold behavior Eq.~\eqref{eq:regular}, while the double spectral density $\rho(s,t)$ has support only above the leading Landau curves in Eq.~\eqref{eq:LeadingLC} and is explicitly symmetric under exchange of its arguments. The remaining prefactors act as mild regulators.
Importantly, this parametrization does not impose the asymptotic behavior of the solution. The neural networks are free to compensate the prefactors and to grow at large energies. Indeed, as shown in Sec.~\ref{sec:2PRR_amplitude}, the trained spectral densities do not decay according to the naive asymptotics suggested by $f_{\rho}$ and $f_{\mathrm{DDisc}}$, demonstrating that the parametrization does not artificially restrict the space of solutions.

We further designed our neural networks to be able to exhibit qualitatively different behavior in different kinematic regimes. For this purpose we employ neural-network architectures composed of several parallel subnetworks, each receiving a different representation of the kinematic variables as input as shown in Fig.~\ref{fig:NN_architecture} for $\text{NN}_\text{DDisc}$. Concretely, separate subnetworks are fed with combinations of linear and logarithmic variables, such as $(4m^2/s,4m^2/t)$ and $\big(\log(s/m^2),\,\log(t/m^2)\big)$, allowing different branches to specialize in distinct regions of kinematic space. The outputs of these subnetworks are then merged by the intertwining network to produce the final spectral density (Eq.~\eqref{eq:spectralParametrization} is used to obtain $\rho(s,t)$ from the neural network). We used a similar architecture for $\text{NN}_\rho$ which only depends on one Mandelstam variable. 

\subsubsection{Spin-0 unitarity and the Mandelstam equation}

We now illustrate this general construction by explicitly defining the loss terms associated with the two central constraints of the bootstrap problem: spin-0 unitarity and the Mandelstam equation.

\paragraph{Spin-0 unitarity.}
For instance, to impose the spin-0 unitarity equation~\eqref{eq:partialwavezero}, we define the loss
\begin{equation}
    \mathcal{L}_0
    = \sum_{i\in\mathcal{G}_0}
    w_0(s_i)\,
    \big|\,1-\eta_0(s_i)-|S_0(s_i)|^2\,\big|,
    \label{eq:s0_loss}
\end{equation}
where $S_0(s)$ is computed from the spectral densities (using the Mandelstam representation), and $\eta_0(s)$ can either be fixed or given by Eq.~\eqref{eq:inel_el_mp}. The exact weights $w_0(s_i)>0$ are provided in Sec.~\ref{sec:NumericalImplementation}

\paragraph{Mandelstam equation}
\label{sec:multi-particle-input}

In Sections~\ref{sec:2PRR_amplitude}~and~\ref{sec:Aks_screening} we consider two different setups, corresponding respectively to turning off and turning on the multi-particle input $\rho_\text{MP}$. In this subsection, we explain how these constraints are implemented within our optimization framework.

The 2PRR amplitudes are defined by the conditions listed in Eq.~\eqref{eq:2PRR_def}. While setting $\eta_\text{MP}=0$ is straightforward in our framework, the implementation of $\rho_\text{MP}=0$ is more subtle, and we now describe our approach.
Setting the multi-particle input to zero implies that the double discontinuity takes the form
\begin{equation}
    \rho(s,t)=\rho_\text{el}(s,t)+\rho_\text{el}(t,s).
    \label{eq:2PRR_ddisc}
\end{equation}
Using our parametrization of the double discontinuity, Eq.~\eqref{eq:spectralParametrization}, this condition is satisfied provided that
\begin{equation}
   \rho_\text{NN}(s,t)=\rho_\text{el}(s,t)\quad \forall s,t\,,
\end{equation}
where $\rho_\text{NN}(s,t)$ denotes the neural-network parametrization of the double discontinuity.
We enforce this constraint thanks to the following contribution in the loss function:
\begin{equation}
\label{eq:Mandelstam_loss}
    \mathcal{L}_M=\!\!\!\!\!\!\sum_{(s_i,t_i)\in\mathcal{G}_M}\!\!\!\!\left|\rho_\text{NN}(s_i,t_i)-\rho_\text{el}(s_i,t_i)\right|\,,
\end{equation}
where $\rho_\text{el}(s,t)$ is understood as the right-hand side of the Mandelstam equation~\eqref{eq:mandelstam-eqn}, and the grid $\mathcal{G}_M$ covers the entire region in which $\rho_\text{el}(s,t)$ is supported.

In the Aks screening setup discussed in Section~\ref{sec:Aks_screening}, we allow for a {dynamical} multi-particle contribution $\rho_\text{MP}$. This is achieved by modifying the grid used in the Mandelstam loss $\mathcal{L}_M$ defined in Eq.~\eqref{eq:Mandelstam_loss}: we introduce a reduced grid $\mathcal{G}_M^{*}$ that excludes the region where $\rho_\text{MP}(s,t)$ is supported, while keeping the same loss functional. With this choice, $\rho_\text{NN}(s,t)$ remains constrained to coincide with $\rho_\text{el}(s,t)$ outside the multi-particle region, but is left unconstrained inside it.
As a result, a symmetric multi-particle contribution $\rho_\text{MP}(s,t)$ is dynamically generated during training. In this setup, $\rho_\text{MP}$ is not treated as a fixed external input but rather as an emergent degree of freedom of the optimization problem, which can be exploited to satisfy additional physical constraints, as we demonstrate in Section~\ref{sec:Aks_screening}.

$\mathcal{L}_0$ and $\mathcal{L}_M$ are not the only loss terms. For instance, imposing inelastic unitarity, $|S_J(s)|\leq 1$ for all $s>16m^2$ and spins $J>0$, requires an additional contribution to the loss of the schematic form $\mathcal{L}\sim\sum_s \Theta\big(|S_J(s)|^2-1\big)\big(|S_J(s)|^2-1\big)$. All loss terms used in our implementation are presented in Sec.~\ref{sec:TrainingNN}.

\subsubsection{Training}

We now describe the general logic of the training procedure by which unitarity is enforced. The discussion here is intended as a conceptual overview; full implementation details are given in Sec.~\ref{sec:NumericalImplementation}.

With the parametrization just explained, the neural-network parameters are optimized by minimizing a loss function $\mathcal{L}$ similar to what was explained in the previous section. 
The minimization is performed using standard gradient-based optimization methods, which rely on evaluating the gradient of the loss with respect to the neural-network parameters. One last aspect, technical at first sight but essential in practice is  "automatic-differentiation", a method which implements the \textit{backpropagation} algorithm. It relies on a conceptually trivial mathematical fact (Leibniz's rule for derivatives of composed functions), but it allows to evaluate gradients at no computational cost during the optimization. This requirement is what forces a discretization of every-single integral present in the formalism (by comparison, in the fixed-point approach used in~\cite{Tourkine:2023xtu}, all integrals were performed numerically at each iterations).

Once training is completed, the loss is typically small but non-zero, reflecting residual violations due to discretization and finite numerical precision. The trained neural networks then provide explicit representations of the spectral densities, from which the scattering amplitude and partial-waves are computed. These quantities can be used to verify a posteriori that the physical constraints have been successfully imposed within the desired numerical accuracy. For instance, higher-spin $J>0$ partial waves indicate how well the Mandelstam equation is satisfied.

We conclude this subsection with a note: even though our solving strategy technically belongs to a class of "machine learning" problems, no training data are used: the training is said to be "unsupervised". The loss function itself fully encodes the physical constraints. This approach is inspired by physics-informed neural networks (PINNs)~\cite{RAISSI2019686}. However, the goal here is not to learn unknown dynamics from data, but rather to solve known functional equations dictated by fundamental physical principles.

In the next section, we present a detailed account of the numerical implementation, including the discretization strategy, the neural-network parametrizations, the weights used in the loss, and the training procedure. 
Readers primarily interested in the results may skip Section~\ref{sec:NumericalImplementation} without loss of continuity.

\section{Numerical implementation}
\label{sec:NumericalImplementation}
This section provides a detailed description of the numerical realization of the bootstrap problem outlined in Section~\ref{sec:AMreview}. We discuss the discretization of the kinematic domains, the neural-network architectures used to parametrize the spectral densities, and the training strategy employed to enforce the physical constraints.

While these details are not required to follow the results presented in the remainder of the paper, they clarify how the constraints are implemented in practice and shed light on several nontrivial physical features, such as the inelastic unitarity in higher partial waves and the positivity of double spectral density in the Mahoux--Martin region~\cite{Mahoux:929517}.

For numerical convenience, we work with the dimensionless variables
\begin{equation}
    x=\dfrac{4m^2}{s}\,,\quad\quad y=\dfrac{4m^2}{t}\,,
\end{equation}
which map the physical energy range $s,t\in[4m^2,\infty)$ onto the compact interval $(0,1]$.

\subsection{Discretization}
\label{sec:discretization}

The physical constraints defining the bootstrap problem are formulated as integral equations over continuous kinematic variables. In order to solve them numerically, we discretize the relevant domains and enforce the constraints on finite sets of sampling points.

We introduce three grids:
\begin{itemize}
\item a one-dimensional grid $\mathcal{G}_0$ for the single spectral density $\rho(s)$,
\item a two-dimensional grid $\mathcal{G}_{\mathrm{DD}}$ for the double spectral density $\rho(s,t)$,
\item and another two-dimensional grid $\mathcal{G}_S$ for the real part of the $t$-discontinuity of the amplitude $\re\,T_t(s,t)$.
\end{itemize}
All loss terms are expressed entirely in terms of the values of the spectral densities evaluated on these grids meaning that the neural networks are evaluated only at grid points
\begin{equation}
\rho(s)\quad \text{with } s\in \mathcal{G}_0 \, ,
\qquad
\rho(s,t)\quad \text{with } (s,t)\in\mathcal{G}_{\mathrm{DD}}\,.
\end{equation}
Whenever a constraint requires values of the spectral densities away from the grid points, these values are obtained by linear interpolation from the corresponding grid, rather than by direct evaluation of the neural networks.

For the two-dimensional grids, we use linear interpolation 
within a triangle formed by grid points containing the evaluation point.

The third grid $\mathcal{G}_S$ serves as a convenient intermediate representation, since $\re\,T_t$ enters both the Mandelstam equation~\eqref{eq:mandelstam-eqn} and the Froissart–Gribov formula~\eqref{eq:FG}.

\paragraph{Numerical evaluation of the loss and GPU implementation.}

All integrals entering the loss function are evaluated numerically. A key feature of our implementation is that, once the grids are fixed, all integrals can be rewritten as matrix or tensor operations acting on the values of the spectral densities at the grid points. This makes the full evaluation of the loss function compatible with efficient GPU execution.

We illustrate the procedure with a one-dimensional integral of the form
\begin{equation}
\int_0^1 \mathrm{d}x'\,\rho(x')\,K(x',x)\,,
\end{equation}
where $K(x,x')$ is a known kernel. Let $x_i,\,i=0,\cdots N$ denote a discretization of the integration domain. The integral is approximated by decomposing it into intervals $[x_{i-1},x_i]$ and linearly interpolating the spectral density within each interval,
\begin{subequations}
    \begin{equation}
\label{eq:NumericalIntegration}
\int_0^1 \mathrm{d}x'\,\rho(x')K(x',x)
\;\longrightarrow\;
\sum_{i=1}^N \int_{x_{i-1}}^{x_i} \mathrm{d}x'\,\rho_i(x'),K(x',x)\,,
\end{equation}
with
\begin{equation}
\rho_i(x') = \rho(x_{i-1})+\bigl(\rho(x_i)-\rho(x_{i-1})\bigr)
\frac{x'-x_{i-1}}{x_i-x_{i-1}}\,.
\end{equation}
\end{subequations}
In this representation, the integral depends only on the values $\rho(x_i)$ at the grid points. After discretizing the external variable $x$ as well, the integrals over $x'$ can be precomputed once, and Eq.~\eqref{eq:NumericalIntegration} reduces to a matrix multiplication acting on the vector of spectral densities evaluated on the grid.

The same strategy is used for all integrals appearing in the Mandelstam equation, the Froissart–Gribov formula, and the spin-0 unitarity constraint. When off-grid values of the spectral densities are required, they are obtained exclusively by interpolation from the grid, an operation which, again, can be seen as a tensor product. Once the grids $\mathcal{G}0$, $\mathcal{G}_{\mathrm{DD}}$, and $\mathcal{G}_S$ are fixed, all tensors required to perform the numerical integrations can be precomputed and reused throughout training on the same grid.

\paragraph{Grid coverage and memory complexity.}
The grids are chosen so as to adequately cover all kinematic regions relevant to the constraints. In particular, the two-dimensional grid $\mathcal{G}_{\mathrm{DD}}$ samples configurations corresponding to
\begin{equation}
s \sim t \sim \mathcal{O}(4m^2), \qquad
s \gg t \sim \mathcal{O}(4m^2), \qquad
t \gg s \sim \mathcal{O}(4m^2), \qquad
s,t \gg 4m^2 .
\end{equation}
This ensures that the double spectral density is accurately resolved both near threshold and in asymptotic regimes.

In practice, the grid $\mathcal{G}_{\mathrm{DD}}$ is not chosen as a Cartesian grid in $(s,t)$ (or $(x,y)$), but is instead organized along lines of constant complex scattering angle
\begin{equation}
z = 1 + \frac{2t}{s-4m^2}
= 1 + \frac{2x}{y(1-x)}\, .
\end{equation}
This choice is motivated by the structure of the Mandelstam equation~\eqref{eq:mandelstam-eqn}, whose kernel
\begin{equation}
\label{eq:MandelstamKernel}
\frac{1}{\sqrt{\bigl(z-\eta_+(\eta',\eta'')\bigr)\bigl(z-\eta_-(\eta',\eta'')\bigr)}},
\end{equation}
depends on the kinematic variables $(x,y)$ only through the combination $z$.
Naively discretizing the variables $(s,t,\eta',\eta'')$ with $N$ points each would require storing $\mathcal{O}(N^4)$ kernel values.
By organizing $\mathcal{G}_{\mathrm{DD}}$ along constant-$z$ lines, the same kernel values can be reused for all grid points $(x,y)$ sharing the same $z$, reducing the memory scaling to $\mathcal{O}(N^3)$.
However, for a fixed value of $z$, the discretization in $\eta'$ and $\eta''$ does not always resolve the infrared region of $T_t\big(s,t(\eta')\big)$ adequately. To remedy this, we supplement the $\eta$-grid with an additional set of $N_{\mathrm{IR}}$ grid points concentrated at low energies, $t(\eta)\sim \mathcal{O}(4m^2)$.
In practice, the additional memory cost remains manageable: we find that $N_{\mathrm{IR}}=25$ is sufficient, while typical dense grids use $N\simeq150$ points, so the overall cost is still well below $\mathcal{O}(N^4)$.

\begin{figure}[h!]
    \centering
    \includegraphics[scale=1.4]{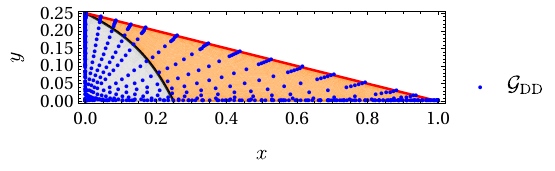}
\caption{
Example of a sparse, constant-$z$ grid $\mathcal{G}_\mathrm{DD}$, shown as blue dots.
The grid consists of $26$ lines of constant complex scattering angle $z=\cos\theta$, each sampled at $26$ points (with an accumulation near $x=0$ and $y=0$, not visible by eye on the figure).
The red curve denotes the $s$-channel–leaning Landau curve, which delimits the support of $\rho_\mathrm{NN}$ (orange and gray region). The gray region shows the domain where the multi-particle data $\rho_\mathrm{MP}$ is supported; it is bounded by the planar cross-graph Landau curve shown in black.
}
\label{fig:DDiscGrid}
\end{figure}

An explicit example of a constant-$z$ grid is shown in Fig.~\ref{fig:DDiscGrid}. The figure illustrates the smallest grid which we refer to as \textit{sparse} used at the initial stage of training, consisting of $26$ constant-$z$ lines, each sampled with $26$ points. This coarse discretization enables a rapid and computationally inexpensive convergence toward a broad approximation of the spectral density. Once this initial stage is completed, progressively denser constant-$z$ grids are employed to refine the solution.

For the densest grid used in this work, $\mathcal{G}_{\mathrm{DD}}$ contains $170$ constant-$z$ lines, each sampled with $165$ points. Of these, $60$ points are distributed logarithmically between the ultraviolet cutoff $x_{\min}$ and the boundary of the double discontinuity support.
The constant-$z$ lines are chosen to cover the regions all relevant kinematic regions. 
The one-dimensional grid $\mathcal{G}_0$ used for the single spectral density $\rho(s)$ contains $299$ points, about half of them are linearly-spaced while the others are distributed logarithmically near both $x=0$ and $x=1$ to resolve the infrared and threshold regions.
The largest tensor required for the evaluation of the loss function occupies approximately $2\,\mathrm{GB}$ of memory when stored as a double-precision array in .npy format.

\paragraph{Physical versus unphysical degrees of freedom.}

An important point is that the physical degrees of freedom of the problem are entirely encoded in the values of the spectral densities on the grids $\mathcal{G}_0$ and $\mathcal{G}_{\mathrm{DD}}$. Two parametrizations that agree on all grid points define the same scattering amplitude, even if their underlying neural-network parameters differ.

By contrast, the parameters of the neural networks (weights and biases) are unphysical. They merely provide a coordinate system for navigating a high-dimensional parameter space in which the optimization is carried out. Increasing the number of neural-network parameters does not introduce additional physical degrees of freedom; instead, it enlarges the optimization manifold on which gradient-based methods operate. Empirically, we find that such overparameterization significantly improves convergence and reduces the likelihood that the optimization becomes trapped in poor local minima. This behavior is consistent with general observations in the machine-learning literature, where highly overparameterized models are known to exhibit smoother loss landscapes and more favorable optimization properties.

\subsection{Neural network architecture}
\label{sec:NNarchitecture}

In this work, neural networks are used as flexible parametrizations of the spectral densities entering the Mandelstam representation. Concretely, a feed-forward neural network can be viewed as an alternating sequence of affine transformations, parameterized by weights and biases, and nonlinear activation functions applied to an input vector. The width of a layer refers to the dimensionality of its output feature vector.

\paragraph{Activation function.}

We employ the Continuously Differentiable Exponential Linear Unit (CELU) as activation function throughout,
\begin{equation}
x\longmapsto\mathrm{CELU}(x)=
\begin{cases}
\alpha\big(e^{x/\alpha}-1\big), & x \leq 0, \\ 
x, & x > 0,
\end{cases}
\end{equation}
with $\alpha = 1/10$.
The CELU activation can be viewed as a differentiable version of the more commonly used ReLU,
\begin{equation}
x\longmapsto\mathrm{ReLU}(x)=
\begin{cases}
0, & x < 0, \\
x, & x \geq 0.
\end{cases}
\end{equation}
This choice is motivated by the expected smoothness of the target physical functions.

\paragraph{Double spectral density.}

\begin{figure}
    \centering
    \begin{tikzpicture}[>=Latex, font=\small]
\tikzset{
  block/.style = {draw, thick, rounded corners,
                  minimum width=2.6cm, minimum height=0.9cm,
                  fill=gray!15, align=center},
  smallblock/.style = {draw, thick, rounded corners,
                       minimum width=1.2cm, minimum height=0.5cm,
                       fill=gray!5, align=center, font=\scriptsize},
  invisibleblock/.style = {rounded corners,
                           minimum width=1.2cm, minimum height=0.5cm,
                           fill=none, draw=none, align=center, font=\scriptsize},
  conn/.style = {->, thick},
  every node/.style={inner sep=2pt}
}

%--------------------------------------------------------
% Coordinates (absolute placement)
%--------------------------------------------------------
\coordinate (A1) at (0,0);
\coordinate (A2) at (0,-1.5);
\coordinate (A3) at (0,-3.0);
\coordinate (A4) at (0,-4.5);

\coordinate (B1) at (1,0);
\coordinate (B2) at (1,-1.5);
\coordinate (B3) at (1,-3.0);
\coordinate (B4) at (1,-4.5);

\coordinate (C)  at (6.5,-2.25); % center of intertwining block
\coordinate (D)  at (8.3,-2.25); % center of final block
\coordinate (OUT) at (10.5,-2.25);

%--------------------------------------------------------
% Inputs
%--------------------------------------------------------
\node[left=0.1cm of A1] (in1) {$(x, y)$};
\node[left=0.1cm of A2] (in2) {$(x, \tfrac{\log_{10}y}{10})$};
\node[left=0.1cm of A3] (in3) {$(\tfrac{\log_{10}x}{10}, y)$};
\node[left=0.1cm of A4] (in4) {$(\tfrac{\log_{10}x}{10}, \tfrac{\log_{10}y}{10})$};

%--------------------------------------------------------
% Sub-networks
%--------------------------------------------------------
\node[block, anchor=west, minimum width=4.0cm] (NN1) at (B1) {Low-energy\\network};
\node[block, anchor=west, minimum width=4.0cm] (NN2) at (B2) {Mixed-energy\\$t$-asymptotic network};
\node[block, anchor=west, minimum width=4.0cm] (NN3) at (B3) {Mixed-energy\\$s$-asymptotic network};
\node[block, anchor=west, minimum width=4.0cm] (NN4) at (B4) {Asymptotic\\network};

%--------------------------------------------------------
% Skip connections inside NN4
%--------------------------------------------------------
\node[smallblock, below=0.6cm of NN4, xshift=-1.8cm, rotate=90] (skip1) {Layer 1};
\node[smallblock, right=0.9cm of skip1.west, rotate=90] (skip2) {Layer 2};
\node[smallblock, right=0.9cm of skip2.west, rotate=90] (skip3) {Layer 3};
\node[invisibleblock, right=0.9cm of skip3.west, rotate=90] (skip4) {...};
\draw[conn] (skip1) -- (skip2);
\draw[conn] (skip2) -- (skip3);
\draw[conn] (skip3) -- (skip4);
\draw[conn] (skip1.west) |- ($(skip1.west)+(0,-0.45)$) -| (skip3.west);
\draw[conn] (skip1.west) |- ($(skip1.west)+(0,-0.45)$) -| (skip4.west);
\draw[conn] (skip2.west) |- ($(skip2.west)+(0,-0.65)$) -| (skip4.west);

%--------------------------------------------------------
% Final network
%--------------------------------------------------------
\node[block, anchor=center, fill=gray!25, minimum width=5.4cm, minimum height=1cm,rotate=90]
  (NNinter) at (C) {Intertwining Layers};
\node[block, anchor=center, fill=gray!25, minimum width=3.4cm, minimum height=1cm,rotate=90]
  (NNfinal) at (D) {Final Layer};
%\node[right=1.0cm of NNfinal] (out) {NN$_\sigma(x,y)$};
\node[] (output) at (OUT) {NN$_\text{DDisc}(x,y)$};

%--------------------------------------------------------
% Arrows from inputs → sub-NNs
%--------------------------------------------------------
\foreach \i/\n in {1/NN1, 2/NN2, 3/NN3, 4/NN4}
  \draw[conn] (in\i.east) -- (\n.west);

%--------------------------------------------------------
% Arrows from sub-NNs → NNfinal
%--------------------------------------------------------
\foreach \n in {NN1,NN2,NN3,NN4}
  \draw[conn] (\n.east) -- (NNinter.north |- \n.east);

\draw[conn] (NNinter.south) -- (NNfinal.north);

%--------------------------------------------------------
% Output
%--------------------------------------------------------
\draw[conn] (NNfinal.south) -- (output);
\end{tikzpicture}
    \caption{Neural-network architecture used to model the double discontinuity (see Eq.~\eqref{eq:spectralParametrization}).
The model consists of five subnetworks: four input networks receiving pairs of variables chosen from ${x,\,y,\,\log_{10}x,\,\log_{10}y}$, and a final network that concatenates and merges their outputs to produce $\mathrm{NN}_{\mathrm{DDisc}}(x,y)$.
The final layer of this network incorporates the positivity constraint in the Mahoux--Martin region~\eqref{eq:mm_region}.
This architecture allows the model to capture both sharp features near threshold and the asymptotic behavior at large energies.
Training is performed on grids with energy cutoffs $10^{8}m^{2}$, $10^{12}m^{2}$, and $10^{16}m^{2}$; a logarithmic rescaling by a factor $1/10$ ensures that all inputs remain within comparable numerical ranges (e.g. $[-1.6,1]$ for the largest cutoff).
Skip connections are implemented within each input network to mitigate vanishing-gradient effects and improve training stability; these are illustrated schematically below the asymptotic network.}
    \label{fig:NN_architecture}
\end{figure}
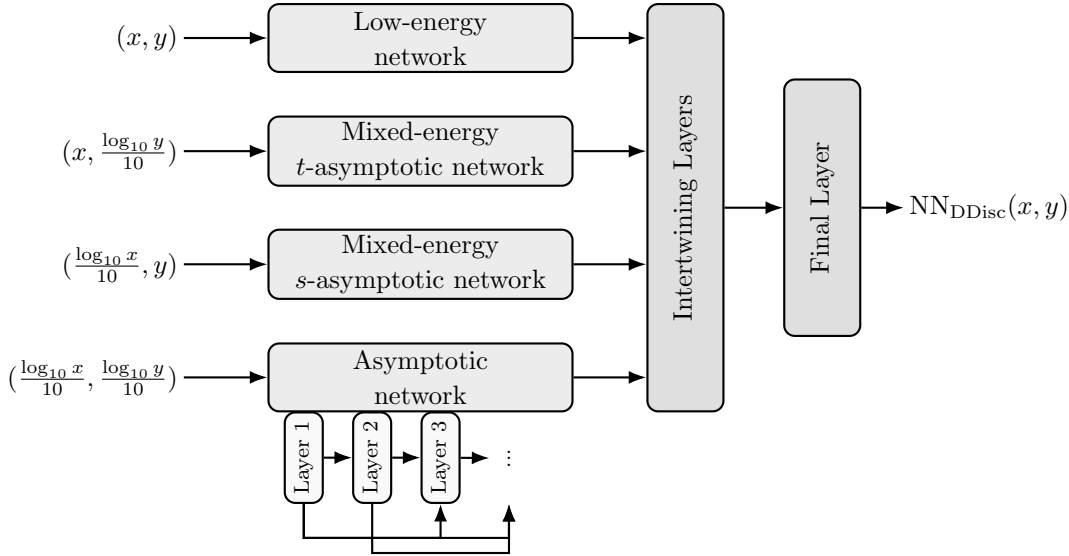

The architecture used to represent the double spectral density $\rho(x,y)$ is shown in Fig.~\ref{fig:NN_architecture}. It consists of five subnetworks. Four input subnetworks process different combinations of $(x,y)$ variables,
\begin{equation}
    (x,y)\,,\quad(x,\dfrac{\log_{10}y}{10})\,,\quad(\dfrac{\log_{10}x}{10},y)\,,\quad(\dfrac{\log_{10}x}{10},\dfrac{\log_{10}y}{10})\,.
\end{equation}
This design allows the network to remain expressive across a wide range of energy scales, simultaneously capturing near-threshold behavior ($x,y\sim \mathcal{O}(1)$) and asymptotic regimes ($x,y\lll1$).

Each input subnetwork is equipped with concatenating skip connections: the output of a given layer is concatenated with its input before being passed to the next layer. As a consequence, the feature width grows exponentially with depth, following the sequence
$$2 \rightarrow 4\rightarrow 8\rightarrow 16\rightarrow 32\rightarrow 64\rightarrow 128,$$ 
for the seven layers of each input branch. The outputs of the four input subnetworks are then concatenated and passed to a merging network composed of four layers of width $512$, followed by a final layer producing a single scalar output, denoted $\mathrm{NN}_\text{DDisc}(x,y)$.

The use of skip connections plays a crucial role in mitigating the vanishing-gradient problem, whereby gradients decay exponentially with depth and deeper layers receive negligible updates. By directly propagating early activations to deeper layers through concatenation, skip connections preserve gradient flow and ensure that all layers remain trainable throughout optimization.

{Finally, as discussed in Section~\ref{sec:coll_pos}, the double discontinuity is built to be non-negative in the Mahoux--Martin region~\eqref{eq:mm_region}.} We enforce this constraint directly at the level of the neural-network architecture by modifying the final activation layer of the network $\text{NN}_\text{DDisc}$. Concretely, the scalar output $x$ of the merging network is transformed according to
\begin{equation}
    (x,s,t)\mapsto\begin{cases}
\text{ReLU}\big(x-\tanh(x)\big)\,,  &\text{if }  4m^2<s<16m^2\,,\text{and } \dfrac{16m^2s}{s-4m^2}<t<4m^2\dfrac{(3s+4m^2)^2}{(s-4m^2)^2}\,, \\ 
x-\tanh(x)\,, & \text{otherwise}\,.
\end{cases}
\end{equation}
In the positivity region, the ReLU activation guarantees $\rho(s,t)\geq 0$ by construction. Outside this region, the smooth function $x-\tanh(x)$ allows for both positive and negative values.
The full architecture of the neural network used to parametrize the double discontinuity is summarized in Fig.~\ref{fig:NN_architecture}.

\paragraph{Single spectral density.}

The neural network used to parametrize the single spectral density $\rho(x)$ follows a similar logic, but with reduced complexity. It consists of two input subnetworks, taking as input $y$ and $\log_{10}(y)/10$, respectively, each equipped with skip connections. The feature widths follow the sequence
$$2 \rightarrow 4\rightarrow 8\rightarrow 16\rightarrow 32\rightarrow 64,$$ 
over six layers. The corresponding merging network contains three layers of width $128$, followed by a final scalar output $\mathrm{NN}_\rho(x)$.

\paragraph{Remarks on scale separation.}

The explicit use of both linear and logarithmic variables in the inputs, together with the multi-branch structure of the networks, is essential for capturing sharp features at low energies while simultaneously resolving slowly varying behavior at large energies. The logarithmic rescaling by a factor of $1/10$ ensures that all inputs remain within comparable numerical ranges, even when training on grids extending up to $s,t \sim 10^{16} m^2$.

\subsection{Loss function and training procedure}
\label{sec:TrainingNN}
Training proceeds over successive epochs, during which the network parameters (weights and biases) are updated using a gradient-based optimization algorithm.

At each epoch, the loss function encoding the physical constraints of the bootstrap problem is evaluated, and its gradient with respect to the network parameters is computed and used by an optimizer to update the network's parameters.

We now turn to the loss function.
Each physical constraint: spin-0 unitarity, the Mandelstam equation (elastic unitarity for $J>0$), and inelastic unitarity for higher spins is enforced by introducing a corresponding non-negative loss term $\mathcal{L}_k$.
\begin{subequations}
\label{eq:unitarity_losses}
\paragraph{Spin-0 unitarity.}
The spin-0 unitarity constraint is enforced through the loss
\begin{equation}
    \begin{aligned}
        \mathcal{L}_0=&\sum_{x\in\mathcal{G}_0}\Big(w_0(x)\Big||S_0(x)|^2-1+\eta_{0}(x)\Big| + w_0^{(\text{inel})}(x)\,\text{ReLU}\left(|S_0(x)|^2-1+\eta_{0}(x)\right)\Big)\,,\\
        &w_0(x)= 2 + \log^2(1-x) - \log(x)\,,\\
        &w_0^{(\text{inel})}(x)=2 + \log^2(1-x) + \log^2(x_\text{min})/2+ \log^2(x)/2\,,\\
    \end{aligned}
\end{equation}
The two weight functions $w_0(x)$ and $w_0^{(\mathrm{inel})}(x)$ are introduced to prevent the training from under-constraining regions in which the spin-0 partial wave is naturally small.

Near threshold, $S_0(s)\sim_{s\to4m^2}1+\mathcal{O}(\sqrt{s-4m^2})$ and therefore approaches unity. As a result, violations of unitarity produce only small absolute deviations and their numerical contribution to the loss is suppressed, even when the relative error is significant.
A similar issue arises at large energies. In this regime, we inferred from previous training runs that the spin-0 partial wave $f_0(s)$ decreases with energy (see Fig.~\ref{fig:S0_2PRR} in Sec.~\ref{sec:2PRR_amplitude}), which again weakens the absolute contribution to the loss in this regime.

The logarithmic factors in $w_0$ and $w_0^{(\mathrm{inel})}$ compensate for these effects by enhancing the loss both near threshold ($x\to1$) and at high energies ($x\to0$), ensuring that unitarity is enforced with comparable accuracy across the full kinematic range. A squared logarithm is used near threshold to counteract the rapid $\sqrt{s-4m^2}$ suppression, while the milder logarithmic enhancement at large $s$ reflects the slower, empirically observed decay of $f_0(s)$.

In $\mathcal{L}_0$, the term proportional to $w_0$ enforces spin-0 unitarity, while the second term, proportional to $w_0^{(\mathrm{inel})}$, further increases the penalty whenever $|S_0(x)| > 1 - \eta_0(x)$.

\paragraph{Consistency condition.}
Whenever $\eta_0$ is treated as a dynamical parameter (e.g. when it is defined through Eq.~\eqref{eq:inel_el_mp}), an additional loss term is introduced to enforce consistency condition $\eta_0\geq0$. This is implemented as
\begin{equation}
    \mathcal{L}_\eta=\sum_{x\in\mathcal{G}_0}w_0(x)\text{ReLU}\!\left(-\eta_0(x)\right)\,.
\end{equation}

\paragraph{Mandelstam equation (elastic unitarity for spin $J>0$).}
The Mandelstam equation is enforced through the loss
\begin{equation}
\begin{aligned}
    \mathcal{L}_\mathrm{M}=&\!\!\!\!\!\sum_{{(x,y)\in\mathcal{G}_\mathrm{DD}}}\!\!\!\!
    w_\mathrm{M}^{(e)}(x,y)\Big|\rho_\mathrm{NN}(x,y)-\mathrm{\rho_{el}}\!\left(x,y,\rho(x'),\rho(x',y')\,\right)\Big|\,,\\
    &w_\text{M}^{(e)}(x,y)=\beta\, w_\text{M}^{(e-1)}+(1-\beta)\,\dfrac{2}{\left|\rho_\text{NN}(x,y)+\rho_\text{el}(x,y)\right|+\dfrac{1}{w_\text{max}}}\\
    &\quad\quad\quad\quad\quad+(1-\beta)\,\min\!\left(w_\text{max},\,\exp\!\left(\dfrac{1}{L}\left|\rho_\text{NN}(x,y)-\rho_\text{el}(x,y)\right|\right)\right)\,.\\
\end{aligned}
\end{equation}
Let us unpack this function $w_{\mathrm{M}}^{(e)}(x,y)$ step by step.
The exponential term strongly penalizes large absolute deviations between $\rho_\mathrm{NN}$ and $\rho_\mathrm{el}$. We use $L=0.2$, and the weight is capped at $w_\mmax$ to avoid numerical instabilities.
By itself, this term does not sufficiently penalize large relative deviations when both $|\rho_\mathrm{NN}|$ and $|\rho_\mathrm{el}|$ are smaller than $L$. To address this, we include a second contribution proportional to $\left(|\rho_\mathrm{NN}+\rho_\mathrm{el}|+1/w_\text{max}\right)^{-1}$, which enhances the weight precisely in regions where the double discontinuity is small. Here again, $w_\text{max}$ acts as a regulator.
Since this weight depends explicitly on the neural-network output, it evolves during training. To prevent rapid oscillations and improve numerical stability, we update it using an exponential moving average: at epoch $e$, the weight 
$w_\mathrm{M}^{(e)}$ is obtained by combining its value from the previous epoch 
$w_\mathrm{M}^{(e-1)}$ with the newly evaluated contribution, with relative weights set by the parameter $\beta$. This procedure smooths the evolution of the weight while allowing it to adapt gradually during training. We use $\beta=0.9$ throughout.
The initial weight function is chosen as 
$$w_\text{M}^{(0)}=\dfrac{1}{\dfrac{x\sqrt{\frac{1-x}{4}-y}}{(1-\log y)^2}+\dfrac{1}{w_\text{max}}}\,.$$

Importantly, although the weights $w_\mathrm{M}^{(e)}$ depend on the network output, they are not treated as trainable quantities. During gradient descent, they are held fixed and act only as multiplicative prefactors in the loss. The weight-update rule is applied separately between epochs, so that no gradients propagate through the adaptive weighting mechanism itself.

\paragraph{Inelastic unitarity at spin $J>0$.}
Inelastic unitarity for higher-spin partial waves is enforced through the loss
\begin{equation}
    \begin{aligned}
        \mathcal{L}_J=&\sum_J\sum_x w_J(x)\,\mathrm{ReLU}\Big(|S_J(x)|^2-1\Big)\,,\\
        &w_J(x)=(\pi J)^{3/2}\chi\left(z_1(x)\right)^{J+1}\,(1-x)(1-\log x)^2\,,\\
        &\chi(z)=z+\sqrt{z^2-1}\,,\\
        &z_1(x)=1+\dfrac{2x}{1-x}\,.
        \end{aligned}
\end{equation}
The sum over $x$ runs over the grid the $x$-values in $\mathcal{G}_S$, where the partial waves $f_J$ are computed from $\mathrm{Re}\,T_t(x,y)$ and $\rho(x,y)$ using Froissart-Gribov formula~\eqref{eq:FG}. The sum over spins is truncated at $J=10$ which serves as a numerical cutoff.

At large spin, the partial waves $f_J$ exhibit a strong kinematic suppression governed by the Gegenbauer function $Q_J^{(d)}$ appearing in the Froissart–Gribov formula~\eqref{eq:FG}. The large-$J$ behavior of $Q_J^{(d)}$ was analyzed in~\cite{Correia:2020xtr}, and the dominant factors in the weight $w_J(x)$ are directly inspired by this asymptotic form. The additional factors $(1-x)(1-\log x)^2$ regulate the threshold region and enhance the penalty at large energies.
As a consistency check, we verify in Sec.~\ref{sec:2PRR_amplitude} that rescaling the inelasticities $\eta_J$ the large-$J$ behavior of $Q_J^{(d)}$ brings the different partial waves to comparable numerical scales.

For the Aks screening setup, the ReLU function is replaced by an absolute value for all $J\leq J_{\max}$. This simple modification enforces
\begin{equation}
    |S_J|^2=1,\quad\forall\,J\leq J_\mmax\,,
\end{equation}
thereby suppressing particle production in the corresponding low-spin partial waves.
\end{subequations}

\paragraph{Gradient computation and task separation.}
Before describing the gradient computation in detail, it is useful to clarify the roles played by the two neural networks. The Mandelstam equation primarily constrains the double spectral density $\rho(s,t)$, while $\text{spin-0}$ unitarity predominantly constrains the single spectral density $\rho(s)$. This motivates a training strategy in which each network is driven mainly by the loss terms most directly associated with its physical role.

\paragraph{Dynamical loss combination.}
The constraints entering the double-discontinuity network $\mathrm{NN}_\mathrm{DDisc}$ have very different numerical scales and convergence rates. Rather than fixing their relative weights by hand, we combine them using a standard uncertainty-based weighting scheme, commonly employed in multi-objective optimization and multi-task learning.
Concretely, we define the total loss
\begin{equation}
    \mathcal{L}_\text{DDisc}=\mathcal{L}_\text{M}\dfrac{e^{-2\sigma_\text{M}}}{2}+\mathcal{L}_J\dfrac{e^{-2\sigma_J}}{2}+\mathcal{L}_\eta\dfrac{e^{-2\sigma_\eta}}{2}+\sigma_\text{M}+\sigma_J+\sigma_\eta\,,
\end{equation}
where $\sigma_{\mathrm{M}},\sigma_J,\sigma_{\eta}$ are trainable scalar parameters, updated alongside the neural-network weights by gradient descent.

This construction can be interpreted as assigning a learnable scaling factor to each constraint: loss terms with larger numerical values are down-weighted, while smaller ones are relatively amplified, preventing any single constraint from dominating the training purely due to its scale.
The additive $\sigma$ terms prevent the trivial solution $\sigma\to\infty$ and ensure a well-defined minimum. During training, the relative importance of the different constraints is therefore adjusted dynamically, without the need for manual tuning.

This form of loss balancing is standard in the machine-learning literature (see~\cite{Kendall_2018_CVPR}) and has been shown to improve stability and convergence when optimizing competing objectives.

\paragraph{Gradient mixing strategy.}
The parameters of the two neural networks are updated using a gradient-based optimizer (Adam~\cite{kingma2017adammethodstochasticoptimization}).
At each training epoch, we explicitly construct two update directions,
$\mathbf{g}_{\rho}$ and $\mathbf{g}_{\mathrm{DDisc}}$,
which are applied to the parameters of $\mathrm{NN}_{\rho}$ and
$\mathrm{NN}_{\mathrm{DDisc}}$, respectively.
Before forming these update directions, all individual gradients
$\nabla\mathcal{L}$ are clipped to unit norm, ensuring numerical stability
throughout training.
\begin{subequations}
    \begin{equation}
    \begin{aligned}
        &\mathbf{g}_{\rho}=\,\,\nabla_{\!\!\rho}\,\mathcal{L}_0+\tilde{\varepsilon}_\rho^{(e)}\,\nabla_{\!\!\rho}\,\mathcal{L}_\text{DDisc}\,,\\
        &\mathbf{g}_{\mathrm{DDisc}}=\,\,\nabla_{\!\!\text{DDisc}}\,\mathcal{L}_\text{DDisc}+\tilde{\varepsilon}_\text{DDisc}^{(e)}\,\nabla_{\!\text{DDisc}}\,\mathcal{L}_0\,,\\
    \end{aligned}
\end{equation}
with
\begin{equation}
        \tilde{\varepsilon}_\rho^{(e)}=\beta\,\tilde{\varepsilon}_\rho^{(e-1)}+(1-\beta)\,\varepsilon\,\dfrac{\lVert\nabla_{\!\!\rho}\,\mathcal{L}_0\rVert}{1+\frac{\lVert\nabla_{\!\!\rho}\,\mathcal{L}_0\rVert}{\lVert\nabla_{\!\!\rho}\,\mathcal{L}_\text{DDisc}\rVert}}\,,
        %&\text{where}\quad\tilde{\varepsilon}_\text{DDisc}^{(e)}=\beta\,\tilde{\varepsilon}_\text{DDisc}^{(e-1)}+(1-\beta)\,\varepsilon\,\dfrac{\lVert\nabla_{\!\!\text{DDisc}}\,\mathcal{L}_\text{DDisc}\rVert}{1+\frac{\lVert\nabla_{\!\!\text{DDisc}}\,\mathcal{L}_\text{DDisc}\rVert}{\lVert\nabla_{\!\!\text{DDisc}}\,\mathcal{L}_0\rVert}}\,.
\end{equation}
and an analogous expression for $\tilde{\varepsilon}_{\mathrm{DDisc}}^{(e)}$. The mixing coefficients $\tilde{\varepsilon}_{\rho}^{(e)}$ and
$\tilde{\varepsilon}_{\mathrm{DDisc}}^{(e)}$ are updated dynamically at each
training epoch $e$.
Their definition involves an exponential moving average controlled by the
parameter $\beta$, which we fix to $\beta = 0.9$ throughout this work.
As a result, the value of $\tilde{\varepsilon}^{(e)}$ depends smoothly on its
value at the previous epoch, suppressing rapid oscillations that would
otherwise destabilize the optimization.
\end{subequations}

This construction ensures that each network is predominantly driven by
the loss associated with the physical quantity it parametrizes.
In particular, when updating $\mathrm{NN}_{\rho}$, the contribution of
$\mathcal{L}_{\mathrm{DDisc}}$ is bounded to be at most a fraction
$\varepsilon$ of the contribution from $\mathcal{L}_0$, and vice versa for
$\mathrm{NN}_{\mathrm{DDisc}}$.
Empirically, this gradient-mixing strategy leads to significantly more stable
and reliable convergence than minimizing a single combined loss
$\mathcal{L}_0+\mathcal{L}_{\mathrm{DDisc}}$ using an unconstrained shared gradient.

\subsection{Initialization, optimization, and convergence}

All neural networks are implemented in \texttt{PyTorch}, which provides automatic differentiation and backpropagation.
Network parameters are initialized using Kaiming normal initialization~\cite{he2015delvingdeeprectifierssurpassing}, appropriate for ReLU-type activations and empirically found to yield stable training in our setup.

Optimization is performed with the Adam optimizer using standard parameters $\beta_1=0.9$ and $\beta_2=0.999$.
To improve convergence and mitigate trapping in local minima, we use a \texttt{CosineAnnealingWarmRestarts} learning-rate scheduler, with the learning rate periodically annealed between $l_{\max}=10^{-5}$ and $l_{\min}=10^{-7}$.

At each epoch, all loss terms in Eq.~\eqref{eq:unitarity_losses} are evaluated on the current grids.
Their gradients are computed by automatic differentiation and combined into the effective update directions $\mathbf{g}_{\rho}$ and $\mathbf{g}_{\mathrm{DDisc}}$ according to the gradient-mixing strategy described in Sec.~\ref{sec:TrainingNN}, before being passed to the optimizer to update the network parameters.

Training is performed in stages.
We begin with a sparse grid to obtain a coarse approximation of the spectral densities at low cost, and then progressively switch to denser grids with increasing energy cutoffs,
$s_{\max}=10^{8}m^{2}$, $10^{12}m^{2}$, and $10^{16}m^{2}$.
With this procedure, a fully Reggeized amplitude is typically obtained within about $18$ hours of training on a single GPU, corresponding to roughly $3\times10^{5}$ epochs.

Convergence is not guaranteed for all random initializations; however, with our setup described above, convergence was achieved reliably.
For the eleven amplitudes studied with couplings $c_{0}\leq 44\pi$, three
initial runs failed to converge but successfully converged upon restarting the training with a different random initialization.

\section{Results for 2PRR amplitudes}
\label{sec:2PRR_amplitude}

In this section, we present our results for a class of 2PRR amplitudes, as defined in Eq.~\eqref{eq:2PRR_def}, and discuss some of their physical properties. Our analysis focuses on amplitudes with regular threshold behavior~\eqref{eq:regular}. Within this restricted class, we find a family of solutions that depends smoothly on the
remaining input parameter $c_0$. These solutions are characterized by the absence of bound states and resonances.\footnote{Here, ``resonance" refers to a sharp phase rotation in any partial wave. We cannot exclude the presence of zeros of $|S_J|$ located far from the real axis.}
Other solutions consistent with the same definition, such as amplitudes with different
threshold behavior or additional resonances, may exist but were not explored here.

Our main result is that, for couplings in the range $0 \leq c_0 \lesssim 44\pi$, we obtain amplitudes with numerically stable emergent non-perturbative Regge behavior.
For $44\pi \lesssim c_0 \lesssim 56\pi$, we obtain amplitudes that are fully consistent at low energies, but for which the non-perturbative Regge behavior cannot be firmly observed. 
In the range between $56\pi$ and the maximal coupling $c_0^\text{max}\simeq85.16\pi$~\cite{Paulos:2017fhb,Guerrieri:2021tak}, we were not able to produce converged amplitudes.

We remind the reader that we use the following variables%\pinote{didn't we use them in the previous section too?}
\begin{equation}
    x=\dfrac{4m^2}{s}\,,\quad y=\dfrac{4m^2}{t}\,,
\end{equation}
that compactify the interval $[4m^2,\infty)$ into $(0,1]$ and work in units
\begin{equation}
    m=1\,.
\end{equation}
%\dnote{since when are we using $m^2=1$ ?}

\begin{figure}[t!]
\begin{minipage}{1.1\textwidth}
    \centering
    \includegraphics[scale=1]{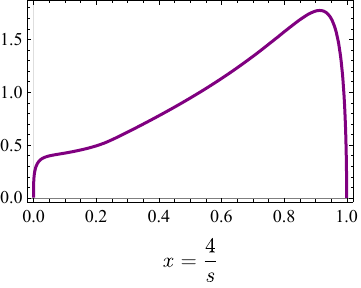}
    \quad
    \includegraphics[scale=1]{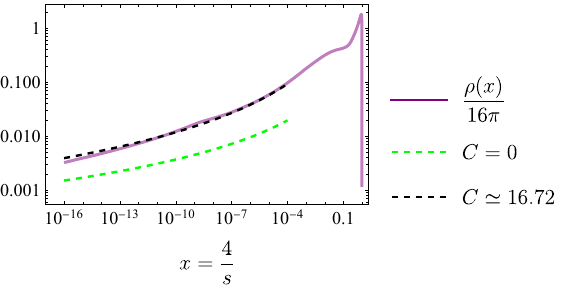}
    \caption{Single spectral density $\rho(x)$ of the 2PRR amplitude at $c_0=40\pi$ as a function of $x=4/s$. Right panel zooms in the large energy region $s > 10^4$ which can be well fitted by the logarithmic decay given by
    $\frac{32\pi^3}{9} \frac{1}{\log(s)^2} \left( 1 + C \,  \frac{\log(\log(s))}{\log(s)} \right)$ 
    and the best-fit is given by $C \simeq 16.72$.
    %\dnote{fitted funtion in legend or at leats in caption}
    }
    \label{fig:Sing_2PRR}
\end{minipage}
\begin{minipage}{\textwidth}
    \centering
    \raisebox{1.8cm}{\includegraphics[scale=0.95]{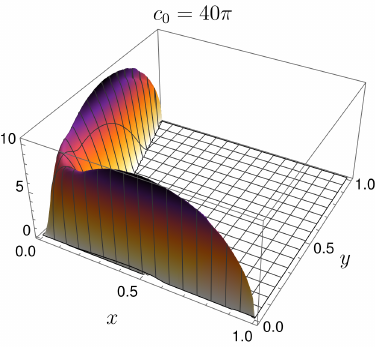}}\quad
    \includegraphics[scale=0.95]{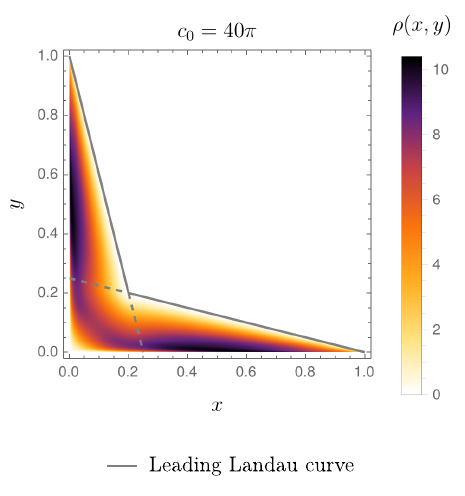}
    \vspace{-10pt}
    \caption{Double spectral density of the 2PRR amplitude at $c_0 = 40\pi$ in the $(x,y)=(4/s,4/t)$ plane, shown both as a 3D surface plot (left) and a density plot (right). In the right panel, leading Landau curves~\eqref{eq:LeadingLC} delineate the support of the double discontinuity, and are indicated in gray.}
    \label{fig:ddisc_2PRR}
    \end{minipage}
\end{figure}

\subsection{Selected amplitude: $c_0=40\pi$}

As the coupling constant is increased, the qualitative structure of the amplitudes remains largely unchanged. In order to make the various physical features more visible, we therefore focus on a representative case with relatively large coupling, $c_0=40\pi$, and present detailed results for this amplitude. We finally return to the dependence of various observables on the size of the coupling in Section~\ref{sec:couplingDependence}.

\begin{figure}[h!]
    \begin{minipage}{\textwidth}
    \centering
    \includegraphics[scale=1]{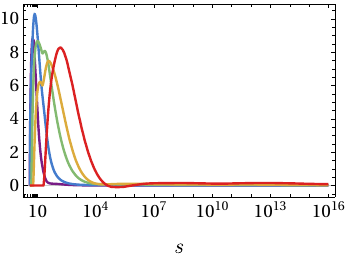}
    \includegraphics[scale=1]{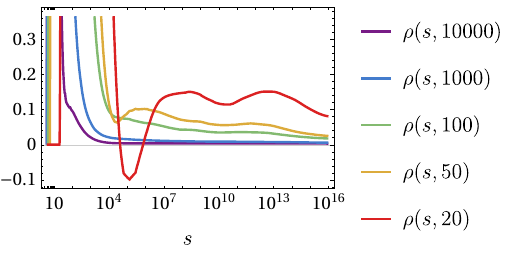}
    \vspace{-10pt}
    \caption{Fixed-$t$ slices of the double discontinuity $\rho(s,t)$ of the 2PRR amplitude at $c_0=40\pi$ on log-linear scale. The right panel provides a zoomed-in view of the $y$-range of the left panel, highlighting the sign changes of the $t=20$ slice (red) around $s\sim10^5$. In the limit $s\to\infty$, all fixed-$t$ slices slowly decay.}
    \label{fig:DDtotSlices_2PRR}
\end{minipage}
\begin{minipage}{\textwidth}
    \centering
    \includegraphics[scale=1]{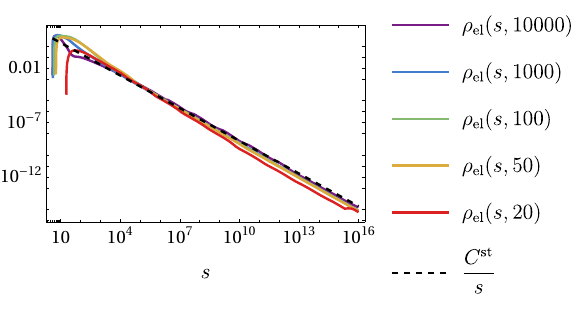}
    \vspace{-10pt}
    \caption{Fixed-$t$ slices of the elastic piece $\rho_{\mathrm{el}}(s,t)$ in the double discontinuity at $c_0=40\pi$, as defined in \eqref{eq:mandelstam-eqn}, on log-log scale. All slices exhibit a universal $1/s$ decay in the large-$s$ regime. The same $1/s$ decay is observed for all values of the coupling $c_0$ (for $c_0=40\pi$ the fit gives $C^\mathrm{st}=20$).
    }
    \label{fig:DDelDecay_2PRR}
    \end{minipage}
\end{figure}

\subsubsection{Spectral densities}
\label{sec:spectral_densities}
The outcome of the neural-network training consists of single and double spectral densities $\rho(s)$ and $\rho(s,t)$, which are shown in Figures~\ref{fig:Sing_2PRR} and~\ref{fig:ddisc_2PRR}, respectively, for $c_0=40\pi$.

The single spectral density $\rho(s)$, displayed in Figure~\ref{fig:Sing_2PRR}, is positive at all energies for the 2PRR amplitudes at all couplings. We emphasize that this positivity is not a consequence of unitarity, since unitarity only constrains $\mathrm{Im}\,f_0$ to be positive. The spectral density reaches a maximum near two-particle threshold and remains of order $\mathcal{O}(1)\times 16\pi$ up to energies of order $s\sim 10^{2}$, beyond which a decaying behavior sets in. At large energies, we observe a logarithmic falloff, similar to that found in a toy model with vanishing double discontinuity studied in~\cite[Appendix D]{Tourkine:2023xtu}
\be
\label{eq:sing_log_decay}
\frac{32\pi^3}{9}
\frac{1}{\log(s)^2} \left( 1 + C \,  \frac{\log(\log(s))}{\log(s)}% + D \, \frac{\log(\log(s))^2}{(\log(s))^2} \,
+ \cdots \right) \, .
\ee
In that simplified setting, the full asymptotic expansion in $\log s$ can be derived analytically. Extending such an analysis to the full problem would require a detailed study of the large-$s$ behavior of the Mandelstam equation~\eqref{eq:mandelstam-eqn}.

The double spectral density $\rho(s,t)$ is shown in Figure~\ref{fig:ddisc_2PRR}. By construction, its support is determined by the leading Landau curves defined in Eqs.~\eqref{eq:LeadingLC}. As discussed in Sec.~\ref{sec:coll_pos}, $\rho(s,t)$ is not required to be positive over its entire support, and we indeed find regions of the $(s,t)$ plane where it becomes negative. At large energies, $\rho(s,t)$ exhibits a slow decay. Both the sign changes and the large-$s$ behavior are more clearly illustrated in Fig.~\ref{fig:DDtotSlices_2PRR}, which shows several fixed-$t$ slices of $\rho(s,t)$ as functions of $s$.
Since $\rho(s,t)$ is symmetric under the interchange $s \leftrightarrow t$, analogous statements hold for fixed-$s$ slices.

A defining property of 2PRR amplitudes is the convergence of Eq.~\eqref{eq:eta_el_def}, which in turn requires the condition~\eqref{eq:consistent2PRR} to be satisfied by the elastic piece $\rho_\mathrm{el}(s,t)$ of the double discontinuity. Figure~\ref{fig:DDelDecay_2PRR} shows the large-$s$ behavior of $\rho_\mathrm{el}(s,t)$ for several fixed values of $t$. On each slice, we observe that $\rho_\mathrm{el}(s,t)$ decays as $s^{-1}$, ensuring convergence of the integral defining $\eta_\mathrm{2PRR}$. We remark that this $s^{-1}$ decay is observed\footnote{This decay is not enforced by the explicit $s^{-1}$ factor in the parametrization of Eq.~\eqref{eq:funcrion_params}. Rather, it was already present in preliminary numerical solutions and was subsequently built into the parametrization to improve stability and convergence.} for all couplings in the range $0 \leq c_0 \leq 44\pi$.
\begin{figure}[t!]
\begin{minipage}{\textwidth}
    \centering
    \includegraphics[scale=1]{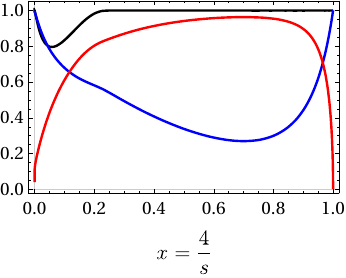}\qquad
    \includegraphics[scale=1]{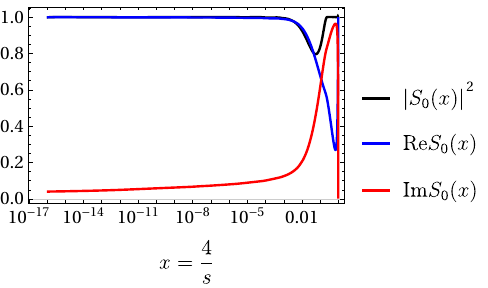}
    \vspace{-10pt}
    \caption{Spin-zero partial wave for the 2PRR amplitude at $c_0=40\pi$. It exhibits particle production beyond four-particle threshold $x<0.25$ and is transparent as $x\to0$.}
    \vspace{10pt}
    \label{fig:S0_2PRR}
\end{minipage}
\end{figure}

\begin{figure}
%\centering
\begin{minipage}{\textwidth}
\centering
\includegraphics[scale=1]{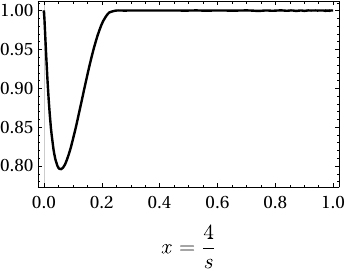}\quad
\begin{tikzpicture}
  \node[anchor=south west,inner sep=0] (img) at (0,0)
    {\includegraphics[scale=1]{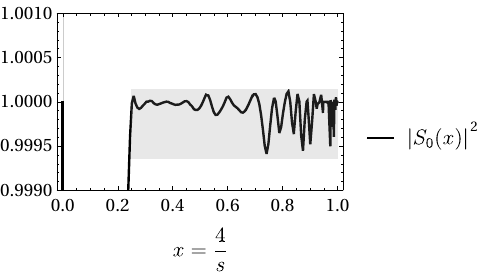}};

  % CRUCIAL: lock the bounding box to the image
  \path[use as bounding box] (img.south west) rectangle (img.north east);

  \begin{scope}[x={(img.south east)},y={(img.north west)}]
    \draw[|-|,very thick, gray] (0.22,0.42) -- (0.22,0.67);
    \node[anchor=south, gray] at (0.35,0.7) {error estimate};

    % this can now go outside without pushing layout
    \draw[thin] (-0.08,0.96) -- (0.20,0.67);
    \draw[thin] (-0.08,0.93) -- (0.20,0.42);
    %\draw[thin] (-0.1,1.2) -- (0.22,0.67);
  \end{scope}
\end{tikzpicture}

\vspace{-10pt}
\caption{\underline{Left:} Modulus squared of the spin-zero partial wave for the 2PRR amplitude at $c_0 = 40\pi$. \underline{Right:} Magnified plot: we estimate a posteriori the numerical error based on the size of the oscillations.
}
\label{fig:S0_2PRR_precision}
\end{minipage}
\end{figure}
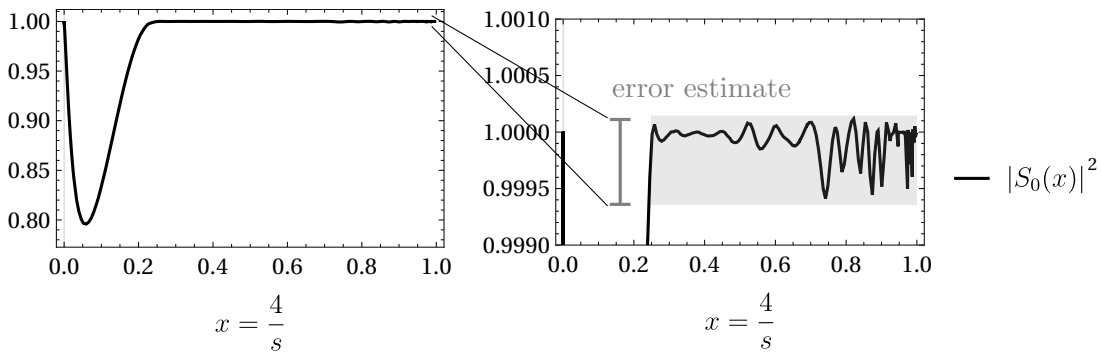

\subsubsection{Partial waves}

Given the spectral densities, we can use Mandelstam representation~\eqref{eq:mandelstam-eqn} to obtain the partial wave projections~\eqref{eq:partialwave_expansion} of the amplitude $T(s,t)$. Let us start by examining the $J=0$ partial wave. In Figure~\ref{fig:S0_2PRR}, we show the modulus squared, the real and imaginary parts of $S_0$.
%$|S_0|^2$, $\re f_0$, and $\im f_0$ as a function of $x=4/s$,
The modulus squared $|S_0|^2$ stays around unity up to four-particle production threshold $s=16$, as dictated by elastic unitarity, and particle production kicks in beyond, as expected since these constraints are explicitly enforced during training.
In Figure~\ref{fig:S0_2PRR_precision}, we provide a zoomed-in view of $|S_0|^2$. From the left panel, we see that the inelastic contribution reaches a $20\%$ effect at its peak. We comment on the relation between the amount of inelasticity and the coupling $c_0$ later in Section~\ref{sec:couplingDependence}. The right panel shows a further zoom around unity, revealing that elastic unitarity is satisfied up to small oscillations of order $5\times10^{-4}$. It would be nice to cross-check this number against an estimated amount of error due to the discretization choices in numerical integrals. \\

\begin{figure}[h!]
\begin{minipage}{\textwidth}
      \centering
  \subfigure[]{
    \includegraphics[scale=1]{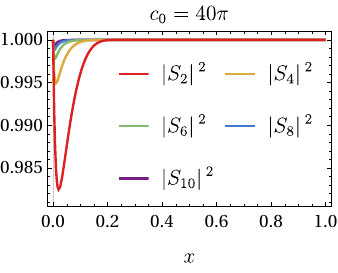}
    \label{fig:Sj_PartialWaves}
  }
  \hfill
  \subfigure[]{
    \includegraphics[scale=1]{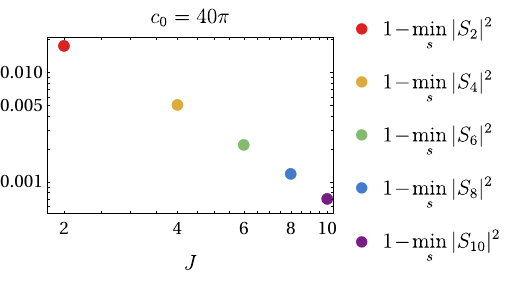}
    \label{fig:minSJ}
  }
  \vspace{-10pt}
  \caption{\underline{Left:} Modulus squared of higher partial waves ($|S_J|^2$ with $J = 2, 4, \dots, 10$) of the 2PRR amplitude at $c_0 = 40\pi$. \underline{Right:} Maximal inelasticity for each spin $J$, i.e.\ $1 - \min_s |S_J|^2$, plotted in log-log scale as a function of $J$. The plot highlights the kinematic decay of inelasticity at large spin. % can be understood kinematically from the Gegenbauer $Q$-functions appearing in the Froissart-Gribov representation of the partial waves.
  }
  \label{fig:higher_PW}
\end{minipage}
\begin{minipage}{\textwidth}
\vspace{5pt}
    \centering
    \includegraphics[scale=1]{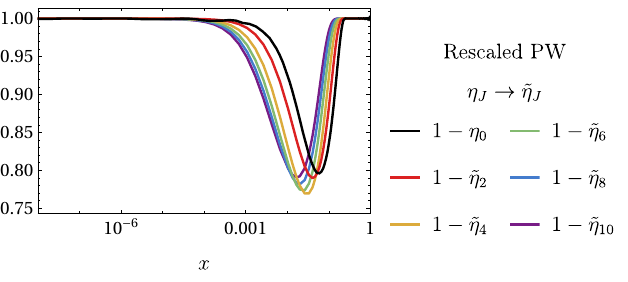}
    \vspace{-10pt}
    \caption{Rescaled inelasticities $\tilde{\eta}_J$ for $J = 2, \dots, 10$ (with $|S_J|^2 = 1 - \eta_J$) as a function of $x$ in log-linear scale, defined in Eq.~\eqref{eq:rescaled_J}, plotted alongside the spin-zero inelasticity as a function of $x = 4/s$.}
    \label{fig:SJ_2PRR}
\end{minipage}
\end{figure}

We now turn to partial waves with spin $J>0$. For these higher-spin partial waves, the double discontinuity plays a crucial role, as it fully determines their imaginary part via the Froissart-Gribov formula~\eqref{eq:FG}. Figure~\ref{fig:Sj_PartialWaves} displays the modulus squared $|S_J|^2$ for $J=2,4,\dots,10$. For all spins, elastic unitarity is exactly satisfied in the range $4 \leq s \leq 16$, while there is particle production $s \geq 16$. We also observe that the size of inelasticity decreases with increasing spin, as Figure~\ref{fig:minSJ} displays in terms of the maximal inelasticity that we define to be $1-\min_s |S_J|^2$. This decrease is purely kinematic and can be understood by analyzing the large-$J$ behavior of the Legendre functions $Q_J$ entering the Froissart-Gribov formula~\eqref{eq:FG}. A detailed analytic study in the large-$J$ limit was carried out in~\cite{Correia:2020xtr}.

To illustrate this scaling better, we present in Figure~\ref{fig:SJ_2PRR} the modulus of the first few partial waves after applying a rescaling procedure inspired by~\cite{Correia:2020xtr}. We first define the inelasticities in each partial wave, similar to~\eqref{eq:partialwavezero}
\begin{equation}
    \eta_J=1-|S_J|^2,
    \label{eq:eta_higher_waves}
\end{equation}
and then rescale them according to
\begin{equation}
\begin{aligned}
    &\tilde{\eta}_J(s)=C^\mathrm{st}\times J^{3/2}\chi(z_1(s))^{J+1}\eta_J(s),\\
        &\chi(z)=z+\sqrt{z^2-1}\,,\quad z_1(s)=1+\dfrac{8}{s-4}\,.\\
    \end{aligned}
    \label{eq:rescaled_J}
\end{equation}
The formula for $\tilde{\eta}_J$ is derived in the large-$J$ limit, and it is not applicable to $\eta_0$ which is hence excluded from the rescaling.
The constant $C^\mathrm{st}$ is then chosen to be $3/2$ in order to match the overall scale of the spin-zero inelasticity. 

All 2PRR amplitudes obtained for $0 \leq c_0 \leq 44\pi$ exhibit qualitatively similar inelastic profiles. In the high-energy limit $s \to \infty$, all partial waves become trivial $f_J(s) \to 0$ and we get that for \emph{fixed $J$}
\begin{equation}
\label{eq:pwtransp}
    \lim_{s\to\infty}S_J(s)=1,\quad J\geq0,
\end{equation}
which is clearly visible in Figure~\ref{fig:SJ_2PRR} by all $\tilde{\eta}_J \to 0$ as $x \to 0$. Using the terminology of \cite{Caron-Huot:2020ouj}, we can refer to the statement \eqref{eq:pwtransp} by saying that scattering becomes \emph{transparent}. Let us emphasize that \eqref{eq:pwtransp} \emph{does not} imply that the scattering amplitude itself becomes trivial, as we will explicitly demonstrate in the next section.

To fully characterize different physical regimes of the high-energy scattering amplitude we would also need to consider high-energy scattering at fixed impact parameter $b$, i.e. $\lim_{s\to\infty}S_{J={\sqrt{s} b \over 2}}(s)$. We do not pursue this in the present paper, but we expect that the result exhibits the standard Yukawa decay at large impact parameters $S(s,b) \sim {e^{-2 m b} \over s}$.

\subsubsection{Regge limit and fixed angle scattering}
\label{sec:2PRR_Regge}

In this section, we study the high-energy limit of our amplitudes. %
We check that our solutions do not depend on the energy cutoff $s_\mmax$ up to which unitarity is imposed, and we observe that the emergent non-perturbative Regge behavior is stable, a phenomenon which we call \textit{Reggeization}.
Then we study the Regge limit ($s \to \infty$, $t<0$ fixed) and the fixed-angle high-energy regime ($s \to \infty$, $s/t<0$ fixed) of these amplitudes.

In order to test the robustness of the numerical UV completion output by our solver, we successively trained the same networks on various grids with increasing cut-offs, so that unitarity is enforced to larger and larger energies. Concretely, we consider three different cut-offs: \{$10^{8}m^2$, $10^{12}m^2$, $10^{16}m^2$\}. This procedure allows us to test whether the amplitudes have Reggeized, i.e. have reached their true asymptotic behavior, and whether the results are insensitive to the choice of the ultraviolet cut-off.

\begin{figure}
    \centering
    \includegraphics[scale=1]{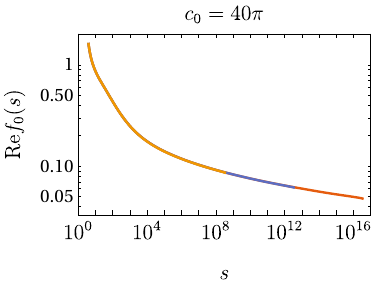}\quad
    \includegraphics[scale=1]{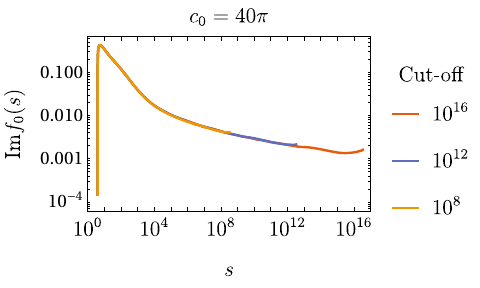}
    \vspace{-10pt}
    \caption{Regge behavior of the real (left panel) and imaginary (right panel) parts of spin-zero partial wave ($\im f_0$ and $\re f_0$) for different high-energy cutoffs, shown for the 2PRR amplitude at $c_0 = 40\pi$. Increasing the cutoff leads to a smooth extension of the slow Regge decay, indicating that the asymptotic regime has been reached. Using this criterion, we find asymptotic behavior for all amplitudes with $c_0 < 44\pi$.}
    \label{fig:Reggeization}
\end{figure}

The outcome of this procedure is illustrated in Fig.~\ref{fig:Reggeization}, where we display the spin-zero partial wave $f_0(s)$ obtained using the increasing cut-offs. We see that the partial wave smoothly extends its previously established behavior. This stability therefore strongly suggests that the amplitudes have indeed Reggeized.
We find that this consistency check is successfully passed only by the 2PRR amplitudes with coupling $c_0 \leq 44\pi$. For larger values of $c_0$, the procedure breaks down: while a solution can be obtained on grids with moderate energy cutoffs, the solver fails to converge once the cutoff is increased. Finally, for $c_0 > 56\pi$ we are unable to reach convergence on any grid.
At present, we interpret this result as an empirical observation rather than an exclusion: it may be related to convergence issues of the numerical optimization or to the specific choice of training parameters, such as grid resolution or loss weights. 
We display in Figure~\ref{fig:Ref0_regge} results at various couplings that illustrate further the (non)-Reggeization property.\footnote{\label{fn:qe} Finally, note that our numerical Reggeization fails for the quasi-elastic amplitudes, no matter the coupling. Numerically, we cannot catch the true asymptotic regime for these amplitudes with oiur solver.}

\begin{figure}[H]
    \begin{minipage}{\textwidth}
    \centering
    \includegraphics[scale=.9]{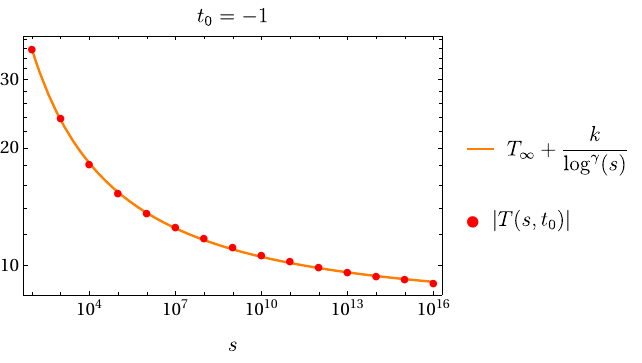}
    \vspace{-10pt}
    \caption{Regge behavior of the 2PRR scattering amplitude $|T(s,-1)|$ at the coupling $c_0=40\pi$ as a function of $s$.
    The orange curve corresponds to a fitting function of the form $T_\infty + k/\log^{\gamma}(s)$.
    For $t_0=-1$, we find that $T_\infty \simeq 7.5$, $k \simeq 240$, and $\gamma \simeq 1.4$. We also probed that at a different coupling, $c_0=20\pi$ for instance, the constant at infinity becomes $T_\infty = 4.5$.
    }

    \label{fig:Regge_behavior}
    \end{minipage}
    \begin{minipage}{\textwidth}
    \vspace{10pt}
    \centering
    \includegraphics[scale=.9]{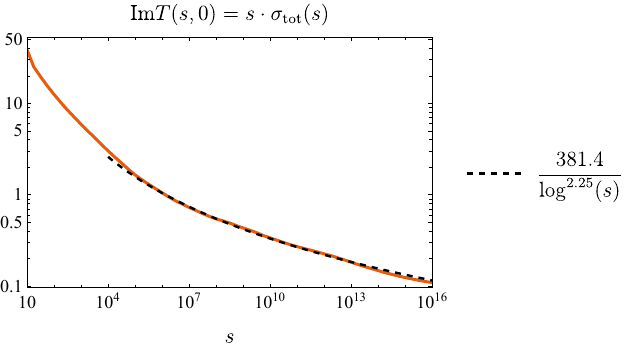}
    \vspace{-10pt}
    \caption{Imaginary part of the amplitude for the 2PRR amplitude at $c_0=40\pi$. It falls off logarithmically which is well-fitted by the black dashed line.}
    \label{fig:ImT_2PRR}
    \end{minipage}

    \begin{minipage}{\textwidth}
    \vspace{10pt}
    \centering
    \includegraphics[scale=.9]{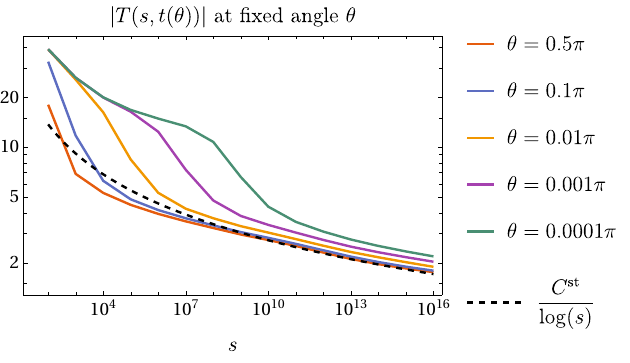}
    \vspace{-10pt}
    \caption{
    Fixed-angle high-energy behavior of the 2PRR scattering amplitude at $c_0=40\pi$.
    The modulus $|T(s,t(\theta))|$ is shown as a function $s$ for several fixed scattering angles~$\theta$.
    The black dashed curve corresponds to a fit of the large-$s$ behavior at $\theta=0.5\pi$ 
    of the form~\eqref{eq:amp_log_decay}, with best-fit coefficient $C^\text{st} \simeq 63$. Note the transition from Regge to fixed-angle regime at energies $s \sim \theta^{-2}$.
    }
    \label{fig:fixedAngle}
    \end{minipage}
\end{figure}

We now turn to the Regge limit of the selected 2PRR amplitude with $c_0=40\pi$. In Fig.~\ref{fig:Regge_behavior} we plot the modulus $|T(s,t_0)|$ at fixed momentum transfer $t_0=-1$ as a function of $s$. The amplitude approaches a constant value at large $s$
\begin{equation}
\label{eq:asymptbeh}
    T(s,t_0)\underset{s\to\infty}{\to} f(t_0)\times (s)^0\,,
\end{equation}
indicating the presence of a Regge pole at $J=0$.
For this value of $t_0$, we can extract the constant by a large-$s$ ``phenomenological'' fit:
\begin{equation}
\label{eq:Ttot_fit}
    |T(s,t_0)| \simeq T_\infty + \frac{k}{\log^{\gamma}(s)} \, ,
\end{equation}
where $T_\infty\simeq 7.5$ (and $\gamma\simeq 1.4$). We emphasize that, although our amplitude is transparent, i.e. individual partial waves $f_J(s)$ vanish in the Regge limit, their infinite sum does not, consistent with this observed behavior.

Next we consider the $s$-channel discontinuity of the amplitude, which connects to the total cross-section via the optical theorem,
\be
    \sigma^\text{tot}(s) = \frac{\im\,T(s,t=0)}{\sqrt{s(s-4)}}.
\ee
The total cross-section and the imaginary part of the amplitude are displayed in Figure~\ref{fig:ImT_2PRR} and they both vanish as $s \to \infty$. The imaginary part in particular exhibits a logarithmic decay at large-$s$ behavior, which is well fitted by
\begin{equation}
    |\im\, T(s,0)| \simeq \frac{k}{\log^{\gamma}(s)} \, ,
\end{equation}
where $k \simeq 381, \gamma\simeq9/4$. Given that the imaginary part decays at large-$s$ faster than $\log^{-2}s$, we can propose an independent sum rule to obtain $T_\infty$ by leveraging the unsubtracted Cauchy relation $T(s,t)=\frac{1}{2\pi i} \oint ds' \frac{M(s',t)}{s'-s}$. By deforming the contour and \emph{not} dropping the arc at infinite, we get to
\begin{equation}
T_\infty(t) = T(s,t) - \int_4^{\infty} \frac{ds'}{\pi}  \, \im T(s',t) \left(\frac{1}{s'-s}+\frac{1}{s'-u}\right) \, ,
\label{eq:sum_rule}
\end{equation}
where the left-hand side can be computed\footnote{This is true when the integrand does not exhibit a dependence on $\theta$ at large $R$ which we confirm in our amplitudes.} via the integral along a large arc
\begin{equation}
T_\infty(t) = \lim_{R\to\infty} \frac{1}{\pi} \int_0^{\pi} \!\!\!d\theta \, \re \, T(R e^{i\theta},t) \, .
\end{equation}
By integrating the fitted behavior for $\im \, T$, we evaluate \eqref{eq:sum_rule}, which agrees qualitatively with the $T_\infty$ fit in Figure~\ref{fig:Regge_behavior}. A robust computation of $T_\infty$ likely requires an enlargement of our grid to exponentially higher energies, or finding an analytic solution for the amplitude in the Regge limit.

Finally, we study the fixed-angle high-energy scattering regime. For physical kinematics, remember that $t(\theta) = -\frac{1}{2}(s-4)(1-\cos \theta)$. The modulus of the amplitude $|T(s,t(\theta))|$ at various fixed scattering angles $\theta$ is shown in Fig.~\ref{fig:fixedAngle}. 
The plots clearly exhibit the transition between Regge scattering and fixed-angle scattering, which occurs at energies of order $s \sim \theta^{-2}$.
In the fixed-angle regime, the decay of the amplitude is well fitted by
\begin{equation}
    |T(s,t(\theta))| \sim \frac{C^{\text{st}}}{\log s} \, ,
    \label{eq:amp_log_decay}
\end{equation}
where $C^{\text{st}} \simeq 63$ for our best-fit. This decay profile is compatible with the expected running behavior of the quartic coupling: The running of the coupling as a function of energy takes the form 
\begin{equation}
    %\lambda_p=\dfrac{\lambda}{1-\dfrac{3\lambda}{16\pi}\log\left(\dfrac{p}{m}\right)}.
    \lambda(s)=\dfrac{\lambda}{1-\dfrac{3\lambda}{16\pi^2}\log\left(\dfrac{\sqrt{s}}{m}\right)}.
    \label{eq:phi4running}
\end{equation}
The denominator is always strictly positive for $\lambda<0$, the effective coupling decreases logarithmically at large energies and the theory is asymptotically free, thus in qualitative agreement with~\eqref{eq:amp_log_decay}.

\newpage

\subsection{Coupling dependence}
\label{sec:couplingDependence}
In this section, we provide an overview of how our results vary as a function of the coupling $0 \leq c_0 \leq 44\pi$. These amplitudes all possess a smooth large-$s$ limit verified by our Reggeization procedure up to $s=10^{16}$, as explained in Sec.~\ref{sec:2PRR_Regge}. We also comment on the convergence issues regarding $c_0 \geq 44\pi$.
\begin{figure}[h!]
\begin{minipage}{1.\textwidth}
    \centering
    \includegraphics[scale=1]{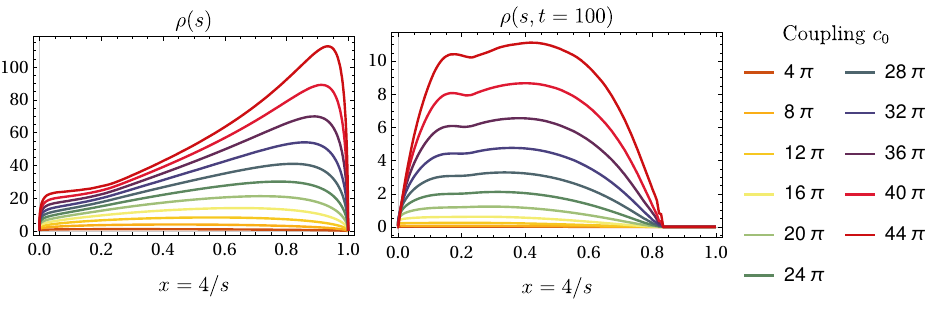}
    \vspace{-10pt}
    \caption{Single (left) and double (right) spectral densities at fixed-$t=100$ for 2PRR amplitudes at all available couplings $c_0$. 
    Their sizes grow with increasing $c_0$. The growth rate as a function of the coupling can be fitted by a power-law $(c_0)^\gamma$ with an exponent $2 \leq \gamma \leq 3$. 
    {The bump around $x=0.2$ coincide with the $t$-channel leading Landau curve. At $t=100$ both $\rho_{\rm el}(s,t)$ and $\rho_{\rm el}(s,t)$ contribute for $x\leq0.24$, but only $\rho_{\rm el}(s,t)$ remains for $x>0.24$.}
    }
    \label{fig:Sing_DDtot_coupling}
\end{minipage}
\begin{minipage}{1.\textwidth}
    \centering
    \includegraphics[scale=1]{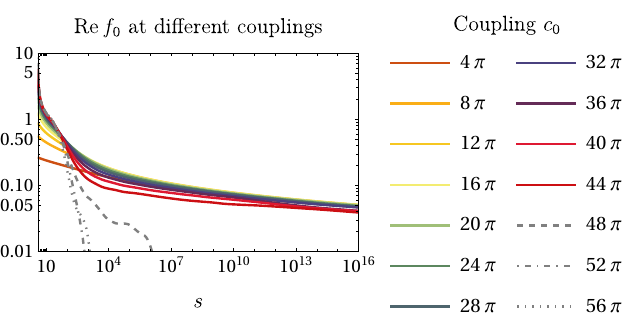}
    \vspace{-10pt}
    \caption{Real part of the spin-zero partial wave coefficient, $\re f_0(s)$, shown in log–log scale for couplings $c_0$ ranging from $4\pi$ to $56\pi$.
    Reggeized solutions are displayed as solid curves and color-coded, while non-Reggeized amplitudes are shown in gray with different dashed styles. 
    For the largest three couplings, Reggeization could not be achieved; these curves were therefore obtained using the lowest energy cutoff, $s_{\max}=10^8\,m^2$. 
    Unlike the Reggeized solutions, the real part $\re f_0(s)$ for these non-Reggeized amplitudes becomes negative at large energies. 
   The origin of this behavior, physical or numerical, cannot be firmly established at present.}
    \label{fig:Ref0_regge}
\end{minipage}
\end{figure}
\begin{figure}[h!]
    \centering
    \includegraphics[scale=1]{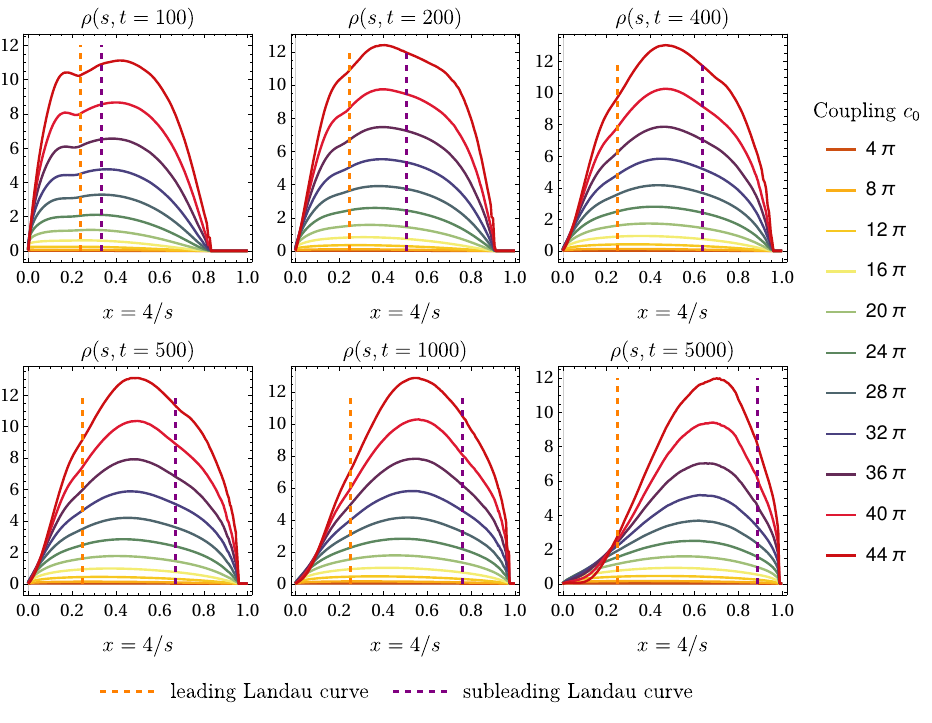}
    \vspace{-30pt}
    \caption{Double discontinuity $\rho(s,t)$ at fixed momentum transfer $t$ as a function of $x=4/s$, shown for several values of the coupling (colored solid curves).
The orange dashed line indicates the position of the leading $t$-channel Landau curve, while the purple dashed line marks the location of the subleading Landau curve at this value of $t$.
A mild change in convexity of $\rho(s,t)$ is visible near the latter for $t=200,\,400$ and $500$, although the effect is subtle.}
    \label{fig:subleadingLC}
\end{figure}

We present the single and double spectral densities for $c_0 \leq 44\pi$ in Figure~\ref{fig:Sing_DDtot_coupling}. Their overall magnitudes grow with increasing $c_0$.
We observe that double spectral densities at $t=100$ share a prominent feature around $x=0.2$. This feature is independent from the coupling, and has a kinematical origin, since its position clearly correlates with the location of the leading $t$-channel Landau singularity~\eqref{eq:PlanarCrossLandauCurve} crosses the $t=100$ slice in the $(s,t)$ plane. This fact is not surprising, because the leading Landau curves are built into the formalism.

\smallskip

Interestingly, our results exhibit the dynamical emergence of other Landau singularities. 
In Fig.~\ref{fig:subleadingLC}, we display various slices of the double spectral density at fixed $t$. We mark the position of the leading~\eqref{eq:LeadingLC} and the first subleading~\eqref{eq:mm_region} Landau curves, corresponding to the Aks and the double-Aks graphs shown in Fig.~\ref{fig:series} (see first and second graphs from the left in the bottom row respectively). The subleading Landau curve coincides with the boundary of the Mahoux--Martin region~\cite{Mahoux:929517,Correia:2020xtr}.

In contrast to the leading Landau curve, the presence of the subleading Landau curve is less visible at the level of the double spectral density. It becomes more pronounced in the first derivative of the double spectral density with respect to $x$.\footnote{As a toy model for this phenomenon, consider the following threshold behavior $a+b \ \theta(x-x_0)(x-x_0)^{1/2}$.}

\begin{figure}[h!]
\begin{minipage}{\textwidth}
    \centering
    \includegraphics[scale=1]{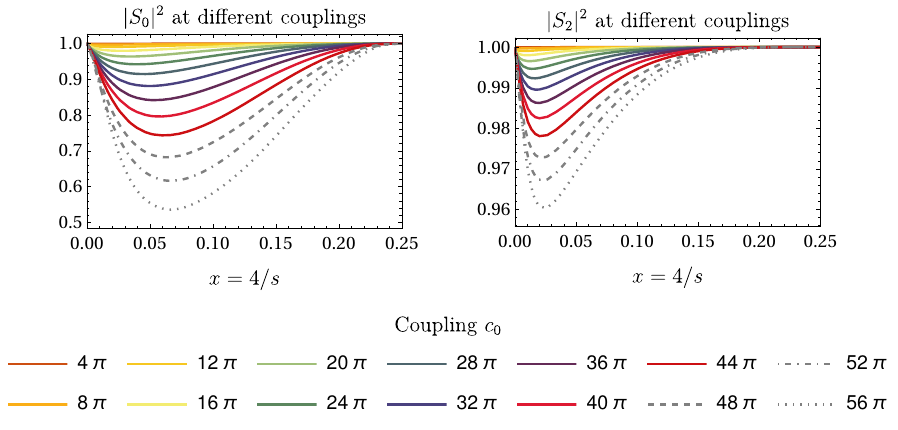}
    \vspace{-25pt}
    \caption{$|S_0|$ and $|S_2|$ at different $c_0$ couplings ranging from $4\pi$ to $56\pi$. The inelasticity grows with the coupling, this observation remains valid for higher spin partial waves. The Reggeized amplitudes ($c_0\leq44\pi$) are displayed as solid curves, while non-Reggeized amplitudes are shown in gray-dashed curves.}
    \label{fig:S0S2coupling}
\end{minipage}
\begin{minipage}{\textwidth}
\vspace{5pt}
    \centering
    \includegraphics[scale=1]{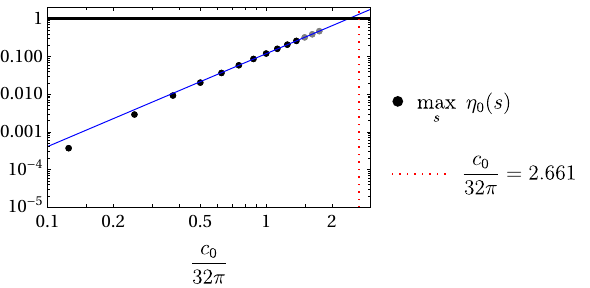}
    \vspace{-10pt}
    \caption{Maximum inelasticity $\max_s\eta_0(s)$ as a function of the coupling $c_0$, shown up to $c_0=56\pi$. The red dotted line indicates the maximal coupling~\cite{Paulos:2017fhb,Guerrieri:2021tak}. The blue curve corresponds to a power-law fit of the form $\alpha\left(\frac{c_0}{32\pi}\right)^p$, with fit parameters $\alpha=0.12$ and $p=2.5$. This fit suggests that $\max_s\eta_0(s)$ would reach unity at $\frac{c_0}{32\pi}=2.384\,$.}
    \label{fig:MaxEtaCouplings}
\end{minipage}
\end{figure}

In Fig.~\ref{fig:Ref0_regge} we display the Regge behavior of $\re\, f_0$ for amplitudes with couplings $c_0 \leq 56\pi$, including both Reggeized and non-Reggeized solutions. Successfully Reggeized amplitudes (solid lines) exhibit decaying tails, with $\re, f_0(s)$ approaching zero from the positive side at large $s$. By contrast, for the non-Reggeized amplitudes ($44\pi < c_0 \leq 56\pi$, gray dashed lines), $\re, f_0(s)$ eventually changes sign at some finite but large value of $s$. This behavior suggests a non-vanishing asymptotic limit, $f_0(s)\to_{s\to\infty}C^{\text{st}}$, indicating that the spin-0 partial wave would no longer be transparent at high energies. At present, we cannot determine whether this behavior is intrinsically related to the failure of Reggeization, or instead reflects limited convergence or finite numerical resolution.

In Fig.~\ref{fig:S0S2coupling} 
%\dnote{Fig.~78 ? and now 85?????} \mnote{solved}
we also show the inelasticities $\eta_0(s)$ and $\eta_2(s)$ for amplitudes with couplings $c_0 \leq 56\pi$, plotted as functions of $x=4/s$ for $s \geq 16$. We observe that, for all amplitudes, the magnitude of $\eta_J(s)$ increases with the coupling. However, the energy scale at which the inelasticity reaches its maximum, $\max_s \eta_J(s)$, appears largely independent of $c_0$.

Particle production is largest in the $J=0$ partial wave. In particular, we find that the maximal inelasticity $\max_s \eta_0(s)$ exhibits a power-law growth with the coupling $c_0$, as shown in Fig.~\ref{fig:MaxEtaCouplings}. A simple power-law fit allows us to estimate a critical value of $c_0$ at which $|S_0|$ vanishes and the scattering becomes purely inelastic. This critical coupling is found to be $c_0/(32\pi) \simeq 2.384$ suggestively close to the extremal value given by bootstrap studies, $c^\text{max}_0/(32\pi) \simeq 2.6613$~\cite{Paulos:2017fhb,Guerrieri:2021tak}.

\section{Multi-particle double discontinuity and Aks screening}
\label{sec:Aks_screening}
In this section we go beyond the 2PRR amplitudes and allow for nonzero multi-particle data: both the single spectral density $\eta_{\rm MP}$, and multi-particle double discontinuity $\rho_{\rm MP}$.

We then use this setup to study the following tension consistently observed in the S-matrix bootstrap literature, starting with~\cite{Paulos:2016fap,Paulos:2016but,Paulos:2017fhb}: on one hand, the  amplitudes that extremize the couplings tend to be fully elastic across all energies. On the other hand, the amplitudes produced by iterations of two-particle unitarity naturally have particle production. Moreover, the Aks theorem~\cite{Aks:1965qga} states that nontrivial scattering in $d>2$ requires particle production, i.e. non-zero inelasticity. Intuitively, this comes from $s\leftrightarrow t$ crossing,\footnote{It trivializes in $d=2$ since $t=0$ always, hence fully elastic S-matrices are possible.} since it generates higher particle cuts in the $s$-channel for any graph that has particle production in the $t$ channel. Importantly, the Aks theorem does not impose a lower bound on the amount of particle production required.

In the language of the present paper, we can restate the phenomenon we want to study as follows: by tuning the multi-particle double discontinuity, can we remove particle production in the low-spin partial waves? We call this phenomenon \emph{Aks screening}, and in this section, we explicitly implement it by adding extra conditions (see Sec.~\ref{sec:TrainingNN}) in our setup that force the following constraints
\be
| S_J(s) |^2 = 1 , ~~~ J \leq J_\text{max} \, . 
\ee
The process outputs non-trivial functions $\eta_{\rm MP}$ and $\rho_{\rm MP}$, as we explain in Section~\ref{sec:aks_w_nn} for $J_{\max}$ up to $2$.
{In Section~\ref{sec:aks_screening}, we 
%study a simplified model to
argue that the screening requires a large $\rho_\MP$, exponentially divergent in $J_\mmax$.} Finally, in Section~\ref{sec:inelastic_sdpb}, we use semidefinite programming methods and study the opposite direction: introducing inelasticity in the extremal amplitudes, following a route similar to~\cite{Antunes:2023irg}.

\subsection{Towards Aks screening with the NN}
\label{sec:aks_w_nn}
The case $J_{\max}=0$ is straightforward in our setup. Enforcing a purely elastic
$S_0(s)$ can be done by setting
\begin{equation}
\begin{aligned}
    \rho_\MP &=0\,,\\
    \eta_0 &= 0 \, ,
    %&\eta_\mathrm{MP}=-\eta_\mathrm{2PRR}\,,
\end{aligned}
\end{equation}
this corresponds to quasi-elastic amplitudes with a nontrivial $\eta_{\MP} \neq 0$.

\begin{figure}[h!]
\begin{minipage}{\textwidth}
    \centering
  \raisebox{1pt}{\subfigure[]{
    \includegraphics[scale=1]{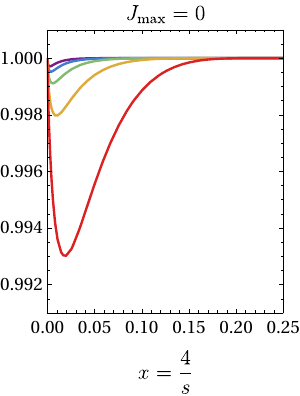}
    \label{fig:AksJ0}
  }}
  \hfill
  \subfigure[]{
    \includegraphics[scale=1]{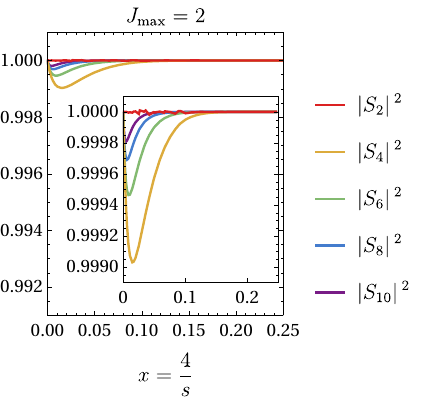}
    \label{fig:AksJ2}
  }
  \vspace{-10pt}
  \caption{\underline{Left:} Quasi-elastic amplitude at $c_0 = 28\pi$, with $\eta_0 = 0$ and $\rho_{\mathrm{MP}} = 0$. \underline{Right:} Amplitude at $c_0 = 28\pi$ with $\eta_0 = 0$. $\rho_{\mathrm{MP}}$ is turned on, and the constraint $|S_2|^2 = 1$ is enforced during training through the loss function. The same vertical scale is used in both panels; a zoomed-in version of the second plot is also provided.}
  \label{fig:Aks_higher_PW}
\end{minipage}
\begin{minipage}{\textwidth}
\vspace{10pt}
    \centering
  \subfigure[]{
   \includegraphics[scale=1]{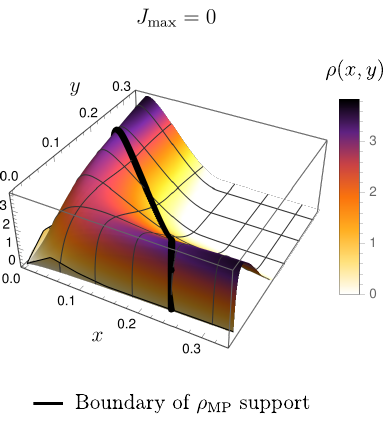}
    \label{fig:3DPlotJmax0}
  }
  \hfill
  \subfigure[]{
    \includegraphics[scale=1]{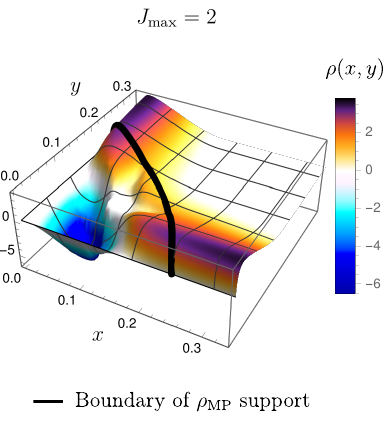}
    \label{fig:3DPlotJmax2}
  }
  \vspace{-10pt}
  \caption{\underline{Left:} 3D plot of the double discontinuity of the quasi-elastic amplitude at $c_0 = 28\pi$. \underline{Right:} Amplitude at $c_0=28\pi$ with $\eta_0=0$ where $\rho_\mathrm{MP}$ has been used to set $|S_2|^2=1$ as visible on the right plot of Fig.~\ref{fig:Aks_higher_PW}. The boundary of the support of $\rho_\mathrm{MP}$ is shown in black.}
  \label{fig:AksDDisc}
\end{minipage}
\end{figure}

We next consider screening inelasticity in spins $J=0$ and $J=2$. Removing the inelasticity $\eta_2$ requires additional freedom, which we introduce
by allowing for a non-zero multi-particle double discontinuity
$\rho_{\mathrm{MP}}(s,t)$.
This is implemented by restricting the Mandelstam equation to the region outside
the support of $\rho_{\mathrm{MP}}$, as explained in
Section~\ref{sec:multi-particle-input}, so that $\rho_{\mathrm{MP}}(s,t)$ can
dynamically evolve during training.

This extension has important consequences: through the Mandelstam equation~\eqref{eq:mandelstam-eqn},
changes in $\rho_{\mathrm{MP}}$ feed back into the elastic double discontinuity
$\rho_{\mathrm{el}}$. As a result, convergence becomes significantly more challenging,
especially at larger values of the coupling.
For this reason, we restrict to an intermediate coupling, $c_0 = 28\pi$, for which stable convergence could be achieved.
The resulting partial waves and double discontinuity for the cases $J_\text{max}=0$ and $J_\text{max}=2$ are shown in Fig.~\ref{fig:Aks_higher_PW}~and Fig.~\ref{fig:AksDDisc} respectively.

These results provide a proof of concept that $\rho_\text{MP}$ can be determined dynamically by enforcing an extra constraint on the amplitude (in our case the no-production constraints on $S_2$). Furthermore, the use of neural network-based solvers made it very easy to relax $\rho_\MP=0$ and allow it to be dynamically generated during training.\footnote{Recall that in iteration-based methods $\rho_\MP$ is a fixed input, and the task of determining it based on certain physical requirements appears to be much more complicated.}

For $J_\mathrm{max} \ge 4$, we were not able to achieve convergence with our setup. In these cases, either the Mandelstam equation or at least one of the constraints $|S_J|^2 = 1$ for $J \leq J_{\max}$ remained poorly satisfied at the end of training. It would be interesting to explore if convergence can be achieved with a more advanced solver or NN architecture.

One particularly interesting quantity to compute for the screened amplitudes is
\begin{equation}
\label{eq:c2def}
    c_2=\frac{1}{4}\frac{\partial^2}{\partial s^2}T(s,t)\Bigr|_{\substack{s=4/3\\t=4/3}}\,.
\end{equation}
The value of $c_2$ for the extremal amplitude that minimizes $c_2$ at fixed $c_0$ is provided by the semidefinite bootstrap~\cite{EliasMiro:2022xaa,Chen:2022nym}.
The value of $c_2$ for the 2PRR amplitude (labeled as $J_\mmax=-1$), as well as for the screened cases $J_\text{max}=0$ and $J_\text{max}=2$ at fixed $c_0=28\pi$, are summarized in Tab.~\ref{tab:c2_screening}. We see that as inelasticity is increasingly screened, $c_2$ approaches its extremal value, consistent with the fact that extremal amplitudes tend to be fully elastic.

\begin{table}[h]
\centering
    \begin{tabular}{c|c}
     $J_\text{max}$ & $c_2$ \\
     \hline
     -1 & 0.7525998 \\
     0 & 0.7509699 \\
     2 & 0.7507403 \\
     $J_\mmax\gg1$& 0.7507396\\
    \end{tabular}
\caption{\textit{Particle-production screening drives the amplitude towards extremality.} Values of $c_2$ defined in Eq.~\eqref{eq:c2def} for amplitudes at $c_0=28\pi$ and different $J_\text{max}$. The case $J_\text{max}=-1$ corresponds to the 2PRR amplitude. The extremal $c_2$ value at $c_0=28\pi$ (labeled as $J_\mmax\gg 1$) comes from semidefinite bootstrap. As progressively more inelasticity is screened, $c_2$ decreases and approaches the extremal value.
}
\label{tab:c2_screening}
\end{table}

Comparing Figs.~\ref{fig:3DPlotJmax0}~and~\ref{fig:3DPlotJmax2} we observe that the screening introduces an oscillatory component to the double discontinuity, which also grows in magnitude. If $\rho_\MP$ can actually be used to screen the inelasticity up to an arbitrary spin $J_\mmax$ our conjecture is that the oscillations and their size get enhanced as $J_\text{max}$ increases. We investigate the growth in the subsection below.

\subsection{Aks screening}
\label{sec:aks_screening}

The property of Aks screening asks whether multi-particle production can be suppressed in a chosen set of partial waves by tuning the multi-particle input
$\eta_{\MP}$ and $\rho_{\MP}$ and what are the consequences of doing so. In the actual Mandelstam equations, $\rho_{\MP}$ backreacts on the elastic double-discontinuity in a fully non-linear manner. In this subsection we study a simplified problem and we ignore (i) backreaction of $\rho_{\MP}$ on $\rho_{\el}$ and (ii) crossing constraints on $\rho_{\MP}$. The goal is not to solve the full coupled system, but to isolate a robust mechanism: screening a growing set of low-spin inelasticities typically requires increasingly oscillatory and large multi-particle input. We call this "rigid screening".

\paragraph{Rigid screening at fixed $s$.}
Using Froissart--Gribov~\eqref{eq:FG} at fixed $s\geq 16m^2$, rigid screening means that we look for a fixed $\rho_{\MP}(s,t)$ such that, given a $\rho_\el(s,t)$ solution to Mandelstam's equation, the inelasticities $\eta_J(s)$ vanish:
\begin{equation}
\label{eq:Aks-screening-rigid-rewrite}
\forall\,J=0,\dots,J_\mmax:\qquad
\eta_J(s)\propto \int_{4}^{\infty} Q_J\!\left(1+\frac{2t}{s-4m^2}\right)\big(\rho_{\el}(t,s)+\rho_{\MP}(t,s)\big)\,dt=0,
\end{equation}
(see definition in Eq.~\ref{eq:eta-J-Q-J}).
% up to an overall non-zero, positive prefactor that we drop for notational simplicity.
At fixed $s$, the elastic and multi-particle contributions start at different
locations $t_{\el}(s)$ and $t_{\MP}(s)$ with $t_{\MP}>t_{\el}$, given by the Landau curves~\eqref{eq:LeadingLC-t} and \eqref{eq:PlanarCrossLandauCurve}. The integration ranges are thus vertical slices in Fig.~\ref{fig:DDiscStructure}) starting at the Landau curves.

\paragraph{Toy model as a truncated moment problem.}
To make the mechanism transparent we compactify the integration domain by
$y=4/t$, and replace $\rho_{\el}\mapsto f(y)$ and $\rho_{\MP}\mapsto -g(y)$.
In addition, we approximate the kernel by a monomial, $Q_J(\cdot)\mapsto y^J$,
which corresponds to the large-$z$ behavior of $Q_J$ and turns the screening constraints into a moment problem. The elastic and MP integration ranges become $y\in[0,y_{\el}]$ and $y\in[0,y_{\MP}]$ with $y_{\MP}<y_{\el}$ (typically $y_{\el}\lesssim 1$ and $y_{\MP}\lesssim 1/4$). The toy screening problem then becomes: given $f$, can one find $g$ such that
\begin{equation}
\label{eq:toy-moments-problem}
m(J)\equiv \int_0^{y_{\el}} y^n f(y)\,dy = \int_0^{y_{\MP}} y^n g(y)\,dy,
\qquad n=0,1,\dots,J_\mmax.
\end{equation}
Inspired by the Mahoux--Martin positivity region in the elastic strip, we assume that there exists a $y_{\rm{MM}}$ in between $y_\MP$ and $y_\el$, $y_\MP<y_{\rm{MM}}<y_\el$  such that $f$ is \textit{strictly positive} in $[y_{\rm{MM}},y_{\el}]$ (we only need strictly non vanishing). The screening function $g$ that solves the moment problem~\eqref{eq:toy-moments-problem} depends on $J_\mmax$ but we drop the dependence for the sake of notational simplicity.

At finite $J_\mmax$ the system \eqref{eq:toy-moments-problem} always admits solutions. However, in the limit $J_\mmax\to\infty$, the equality of all moments forces $f$ and $g$ to be equal\footnote{Their difference $f-g$ would be orthogonal to all polynomials, that are a dense subset of continuous functions in $[0;1]$ by Weierstrass' approximation theorem, hence $f-g\equiv0$.} -- since $g$ vanishes identically in the elastic strip, no such $g$ can screen out $f$'s moments fully.
This can be seen as a weaker version of the Aks theorem.

\paragraph{Bounds from single moment matching.}
Let us now see what happens as demand the equality of a single, arbitrary large moment $J\to\infty$. Since the moments are dominated by their endpoints at large $J$, we expect $\int_0^{y_\MP} y^J g(y) dy\propto  (y_{\MP})^J $ and $\int_0^{y_\el} y^J f(y) dy\propto  (y_{\el})^J $ and thus the screening function must grow as 
\begin{equation}
\label{eq:toy-blowup-single-moment}
\|g\|%_{L^2(0,y_{\MP})}
\ \ge\
%C'\,
%\frac{1}{J^{k+\frac12}}\,
\left(\frac{y_{\el}}{y_{\MP}}\right)^{J},
\end{equation}
where $\|\cdot\|$ is the $L_2$ norm (or any $L_p$ norm), up to fixed $f$-dependent constants and slower-varying powers of $J$, which come from the endpoint behaviour of $f$ near $y=y_\el$. We prove this bound in appendix~\ref{app:cheb-FG}.

The vanishing of not just one but all moments $J=0,\dots,J_\mmax$, generates  a stronger exponential divergence, controlled by 
\begin{equation}
\|g\| \geq {\left(\chi\big(2\frac{y_{\el}}{y_{\MP}}-1\big)\right)^{J_\mmax}}\,,
\end{equation}
where $\chi(x) = x+\sqrt{x^2-1}$, again, up to powers of $J_\mmax$. This enhancement comes from some extremal property of Chebychev polynomials, as explained in appendix~\ref{app:cheb-FG}.

\paragraph{Physical Froissart--Gribov kernel.}
The toy replacement $Q_J\mapsto y^J$ exaggerates the effective separation between elastic and multi-particle Landau curves. For the physical kernel, at fixed $s\geq 16m^2$ it is more natural to use
\begin{equation}
z=1+\frac{2t}{s-4m^2},\qquad \chi(z)=z+\sqrt{z^2-1},\qquad u(z)\equiv \chi(z)^{-1}\in(0,1),
\label{eq:z-chi-u}
\end{equation}
so that for large $J$ and fixed $z>1$, $Q_J(z)$ behaves (up to powers of $J$) like a monomial in $u(z)$.
Repeating the same arguments as above, the equality of a single moment at $J\to\infty$ forces a growth for $\rho_\MP$ as
\begin{equation}
    \|\rho_\MP(s,\cdot)\| \geq R(s)^J\,,\qquad 
    %R=\frac{\chi(z_{\MP}(s))}{\chi(z_{\el}(s))}
    R(s)=\frac{u(z_{\el}(s))}{u(z_{\MP}(s))}>1
    \label{eq:bound-rho-MP-single-moment}
\end{equation}
up to constants (functions of $s$ independent of $J$) and powers of $J$, and where $z_\el,z_\MP$ are just images of $t_\el,t_\MP$ via the map for $z$ defined in Eq.~\eqref{eq:z-chi-u}.
The same trick that leads to the improvement in the toy-case above can be used here too to get an improved bound from matching of all moments $J\leq J_\mmax$, giving
\begin{equation}
\label{eq:bound-rho-MP-all-moments}
\|\rho_{\MP}\| \ \gtrsim\ %\frac{B(s_0)^{J_\mmax}}{J_\mmax^{k+1}}
{B(s)^{J_\mmax}}
\qquad\text{with}\qquad
B(s)\equiv \chi(2\,R(s)-1)>1,
\end{equation}
up to constants (functions of $s$ independent of $J$) and powers of $J_\mmax$. 

\smallskip

\begin{figure}[h]
    \centering
    \includegraphics[scale=1]{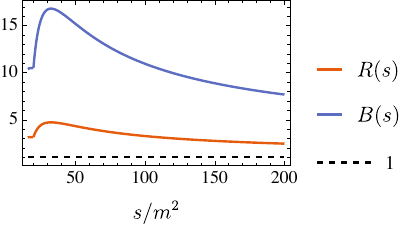}
    \caption{Bounding constants for Aks screening.}
    \label{fig:plRBAksMoments}
\end{figure}

For intermediate $s$, the improvement between \eqref{eq:bound-rho-MP-single-moment} and \eqref{eq:bound-rho-MP-all-moments} is significant, see Fig.~\ref{fig:plRBAksMoments}.
At $s=200m^2$, $B\sim7.7$, and $R\sim2.5$. Asymptotically, $R,B\to1$, as $R\sim_{s\to\infty}1+O(1/s^{1/2})$, while $B$ approaches at slower pace, $B\sim_{s\to\infty} 1+O\left(\frac1{s^{1/4}}\right)$, as can be extracted from the Landau curves~\eqref{eq:LeadingLC-t} and \eqref{eq:PlanarCrossLandauCurve}.

Conversely, these bounds imply that, at fixed $\|\rho_\MP\|$, one can only screen a fixed amount of moments, controlled by
\begin{equation}
    \label{eq:Jmaxbound}
    J_\mmax\lesssim \frac{\log\left(\|\rho_\MP(s,\cdot)\| \right)}{ \log(B(s))}
\end{equation}
From the perspective of $B(s)$ alone, the bound is thus stronger in the IR, where it becomes maximum at $s\simeq 32.2m^2$, where $B(s)\simeq$. Further information would have to come from more constraints on $\rho_\MP$.

\subsection{Injecting inelasticity in semidefinite optimization}
\label{sec:inelastic_sdpb}

We have seen that it is possible in principle to screen inelasticity in partial waves, and that screening drives the amplitude towards extremality.
In this section, we approach the problem from the semidefinite side directly and investigate the effects of \textit{adding} inelasticity. 
Let us emphasize that in this approach, we do not have access to the (converged) double spectral density, and therefore we cannot shed extra light on the problem of screening at higher $J_{\text{max}}$ discussed above.

To that purpose, we turn to the primal S-matrix bootstrap introduced in~\cite{Paulos:2016fap,Paulos:2016but,Paulos:2017fhb} supplemented with some physical inelasticity data as an input. We achieve this by imposing particle production at the level of partial waves, which amounts to imposing new constraints on the standard setup through semidefinite matrices, as already being explored in~\cite{Antunes:2023irg,Guerrieri:2024jkn}.

To be more explicit, we impose %\textit{de-screening} 
\textit{inelasticity} conditions that
\begin{align}
    |S_J(s)|^2 &\leq 1-\eta_J(s), &\quad J \leq J_\text{min} \, ,
    \label{eq:descr1} \\
    |S_J(s)|^2 &\leq 1, &\quad J > J_\text{min} \, ,
    \label{eq:descr2}
\end{align}
for a given set of partial wave inelasticities $\{\eta_J(s)\}$ with $0\leq \eta_J(s) \leq 1$ extracted from the neural network solutions. Unlike the neural network, the inelasticity is not dynamical in the primal setup, it instead becomes an input. To create inelasticity constraints, we extract the 2PRR data of the network at $c_0=40\pi$ as shown in Figure~\ref{fig:SJ_2PRR} up to $J_\text{min}=10$.
We then incrementally increase $J_\text{min}$ to see the particle production effects on certain observables. As usual, we use the SDPB software~\cite{Simmons-Duffin:2015qma,Landry:2019qug} to run our numerics. Details of the numerical ansatz are relegated to Appendix~\ref{app:primal}.

The first thing is to understand how much the allowed range for the subtraction constant $c_0$ changes, given some $\{\eta_J\}$. Adding any amount of non-zero inelasticity prohibits the $S_J$ from the saturating the unity in~\eqref{eq:descr1}. Due to convex nature of partial wave unitarity condition, it is expected that bounds with inelastic input are suboptimal with respect to fully elastic bounds. Extremal $c_0$ values with increasing inelasticity are given in Table~\ref{tab:maxc0}. They confirm that the allowed range shrinks by a small but detectable amount. 

\begin{table}[h]
\centering
    \begin{minipage}{.25\textwidth}
    \begin{tabular}{c|c}
     $J_\text{min}$ & $\max \frac{c_0}{32\pi}$ \\
     \hline
     $-1$ & $2.661336$ \\
     $0$ & $2.657990$ \\
     $2$ & $2.657671$ \\
     $4$ & $2.657641$ \\
     $6$ & $2.657633$ \\
     $8$ & $2.657630$ \\
     $10$ & $2.657629$ \\
    \end{tabular}
    \end{minipage}
    %\hfill
    
\caption{Maximum $c_0$ value allowed by unitarity subject to particle production constraints~\eqref{eq:descr1} obtained from the 2PRR data at $c_0=40\pi$ as a function of $J_\text{min}$. The case $J_\text{min}=-1$ corresponds to fully elastic partial waves. In the limit $J_\text{min} \to \infty$, maximal value for $c_0$ decreases as expected.}
\label{tab:maxc0}
\end{table}

\begin{table}[h]
\centering
    \begin{minipage}{.25\textwidth}
    \begin{tabular}{c|c}
    $J_\text{min}$ & $\min c_2$ at $c_0=28\pi$\\
    \hline
    $-1$ & $0.7507396$ \\
    $0$ & $0.7523833$ \\
    $2$ & $0.7526087$ \\
    $4$ & $0.7526237$ \\
    $6$ & $0.7526396$ \\
    $8$ & $0.7526410$ \\
    $10$ & $0.7526416$
    \end{tabular}
    \end{minipage}
    % \hspace{50pt}
    % \begin{minipage}{.25\textwidth}
    %  \begin{tabular}{c|c}
    % $J_\text{min}$ & $\min c_2$ at $c_0=28\pi$\\
    % \hline
    % $-1$ & $0.7507$ \\
    % $0$ & $0.7524$ \\
    % $2$ & $0.7526$ \\
    % $4$ & $0.7526$ \\
    % $6$ & $0.7526$ \\
    % $8$ & $0.7526$ \\
    % $10$ & $0.7526$
    % \end{tabular}
    % \end{minipage}
\caption{Minimum $c_2$ value at $c_0=28\pi$ subject to particle production constraints~\eqref{eq:descr1} obtained from the 2PRR data at the same coupling. The case $J_\text{min}=-1$ corresponds to the extremal value presented in Figure 2 of~\cite{EliasMiro:2022xaa}. Note that for $J_\text{min} \to \infty$ the primal bootstrap $c_2$ value approaches the value quoted in Tab.~\ref{tab:c2_screening} for the 2PRR amplitude.
}
\label{tab:minc2fixc0}
\end{table}

A quick inspection of various $S_J$ on the solutions to~\eqref{eq:descr1} and \eqref{eq:descr2} maximizing $c_0$, reveals that in general the constraints $1 - |S_J|^2 \geq \eta_J$ are not saturated, meaning that SDPB generates additional inelasticities.
We found that in order to reproduce the 2PRR amplitudes with the desired inelastic profiles one has to minimize $c_2$ while keeping $c_0$ fixed to the value from which the inelasticities $\eta_J$ where extracted.\footnote{This procedure was motivated by the fact that 2PRR amplitudes actually have $c_2$ coefficient close to the extremal amplitude minimizing $c_2$ for every coupling $c_0$.} We show in Table~\ref{tab:minc2fixc0} that as more inelasticity is implemented into this setup, the primal bootstrap $c_2$ approaches the $c_2$ value of the 2PRR amplitude. We now turn to the amplitudes resulting from this minimization at constant $c_0$, again we selected the coupling $c_0=40\pi$ as an example. 

\begin{figure}[h!]
    \centering
    \includegraphics[scale=1]{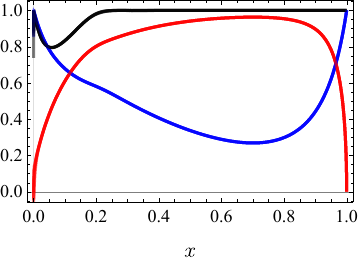}
    \includegraphics[scale=1]{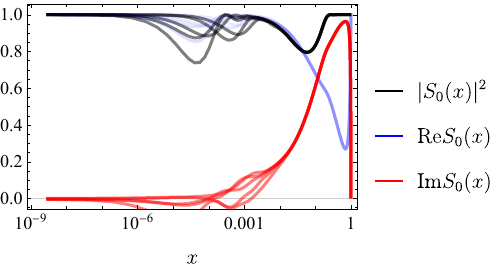}
    \vspace{-10pt}
    \caption{Spin-zero partial wave at $c_0=40\pi$ from the SDPB solution fed with inelasticity data up to $J_\text{min}=10$ obtained from Fig.~\ref{fig:SJ_2PRR}. Shades with increasing opacities correspond to $N_\text{max}=8,9,10,11,12$. Both panels are to be compared with Fig.~\ref{fig:S0_2PRR} -- note the agreement up to $x\simeq10^{-3}$ (or $s\simeq 10^3$) and lack of convergence on the SDPB side for $s>10^3$. The right-most dip of $|S_0(x)|^2$ corresponds to the physical particle production in the $S$-wave, while the subsequent dips are transient.}
    \vspace{10pt}
    \label{fig:S0_sdpb}
\end{figure}
\begin{figure}[h!]
\begin{minipage}{\textwidth}
  \centering
  \includegraphics[scale=1]{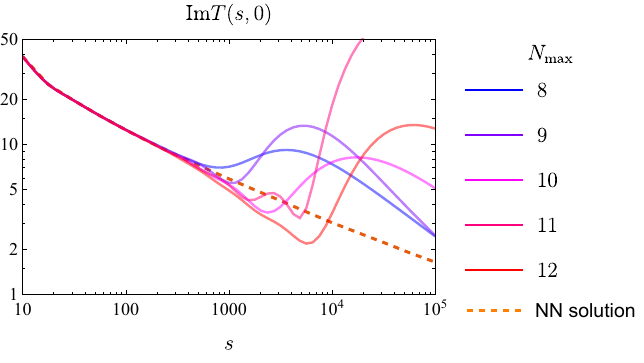}
  \vspace{-10pt}
  \caption{Comparison between the forward amplitudes at $c_0=40\pi$ given by SDPB (solid) and the NN (dashed) solutions as a function of $N_\mmax$ at fixed $J_\text{min}=10$. The farthest wavelet center in the ansatz are $\sigma \in \{ 1191, 1807, 2637, 3726, 5122\}$ respectively for each $N_\mmax$.}
  \vspace{10pt}
  \label{fig:sdpb_ImT}
\end{minipage}

\begin{minipage}{\textwidth}
  \centering
  \includegraphics[scale=1]{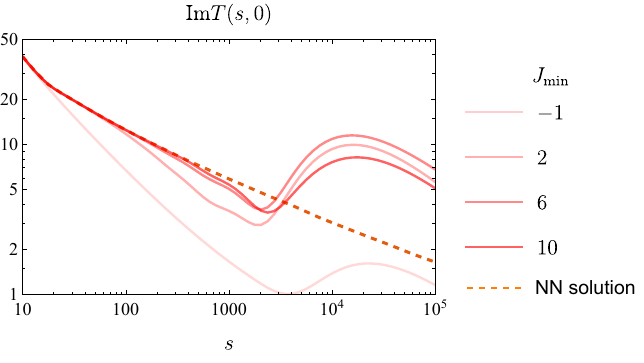}
  \vspace{-10pt}
  \caption{Comparison between the forward amplitudes given by SDPB (solid) and the NN (dashed) solutions as a function of $J_\text{min}$ at fixed $N_\mmax=10$.}
  \label{fig:sdpb_ImT_jmin}
\end{minipage}
\end{figure}

Let us first examine the spin-zero wave given in Figure~\ref{fig:S0_sdpb}, which should be compared with its neural network counterpart in Figure~\ref{fig:S0_2PRR}. We clearly see that SDPB achieves a well-converged solution in the ansatz size $N_\text{max}$ for energies $x > 10^{-3}$, including a perfect saturation of the $\eta_0$ profile. However, in the deep UV region, $x < 10^{-3}$, we observe new and unstable departures from unity at all values of $N_\text{max}$. Due to the lack of convergence, we interpret these extra inelasticities not as genuine particle production, but as transient behavior that would disappear as we introduce further degrees of freedom covering the UV region.
%in the $N_\text{max} \to \infty$ limit.

The reasoning behind this argument is the following: The primal ansatz is comprised of wavelet terms $\rho_\sigma(s)$ labeled by their centers $\sigma$. Each term approaches the constant function away from its center, so any emergent behavior of the amplitude at an energy scale $s_\text{UV}$ needs to be approximated by a sum of wavelets centered near $s_\text{UV}$. The distribution of wavelet centers in our setup is such that they are dense in the IR as $s\to4^+$, and sparse as $s\to\infty$. In particular, the first and last center points in our biggest ansatz $N_\text{max}=12$ are respectively $4+10^{-3}$ and $5 \times 10^3$ -- see Table~\ref{tab:primal_centers} in the appendix for a detailed list. Therefore, we expect that the effective range describing the physical amplitude is restricted to $4 \leq s \lesssim 10^3$. This limitation towards the UV becomes apparent in observables such as the total cross section, or the forward limit of the amplitude displayed in Figure~\ref{fig:sdpb_ImT}. In each SDPB solution at a fixed $N_\mmax$, we observe that the forward amplitude starts to shoot off around the last wavelet center, following a full agreement with the neural network solution across energies of three orders of magnitude.

Yet another interesting aspect is the effect of increasing $J_\text{min}$, which controls the amount of injected inelasticity, as displayed in Figure~\ref{fig:sdpb_ImT_jmin} for the forward amplitude in the well-converged window of energies $s<10^3$. When no particle production is imposed, the large energy behavior shows departure from the 2PRR solution already at $O(10)$ energies. As $J_\text{min}$ is increased, an agreement with the neural network solution builds up gradually. Perhaps the disagreement is not surprising, since for the fully elastic amplitude, the total cross section misses the inelastic contribution which becomes non-zero after four-particle threshold $s=16$ and increases as $s \to \infty$.

Finally, as typically observed in this setup, the double spectral density did not exhibit a converging behavior at any value of $N_\mmax$ or $J_\text{min}$.

\section{Conclusions and future directions}
\label{sec:conclusion}

The S-matrix bootstrap (in its current form) does not constitute a closed set of equations: $2 \to 2$ scattering amplitudes are the only ones simple enough to analyze quantitatively, yet particle production brings in higher-point amplitudes in the bootstrap (some of these contributions are nevertheless controlled by two-particle unitarity through crossing).\footnote{By placing the theory in an AdS ``box'' \cite{Paulos:2017fhb,vanRees:2023,Lauria:2024,Antunes:2021}, we obtain a discrete spectrum of energy levels, which correspond to local boundary CFT operators. All of these operators enter the crossing equations on an equal footing. Consequently, unlike the S-matrix bootstrap, the conformal bootstrap yields a closed system of equations: any intermediate operator can also be treated as an external operator without modifying the basic axioms, see e.g. \cite{Aprile:2026uxe}.}
It is therefore important to \emph{separate} as much as possible the unknown multi-particle data and the two-to-two scattering dynamics.

\subsection{Separation of the multi-particle data}

The essence of the Atkinson–Mandelstam approach to the S-matrix bootstrap adopted in this paper is to separate the multi-particle data from the two-to-two scattering dynamics by exploiting \emph{the detailed analytic structure} of the $2\to 2$ scattering amplitude in gapped theories. {This separation is achieved by (a) enforcing the correct support of the single- and double-spectral densities in the Mandelstam representation of the scattering amplitude, so that they reproduce the expected pattern of normal thresholds and Landau curves; and (b) imposing two-particle unitarity via the Mandelstam equation.}

\begin{figure}[h]
    \centering
    \includegraphics[scale=1.4]{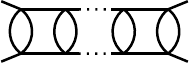}
    \caption{Iterated $s$-channel two-particle unitarity graph (the “Aks graph”) with a $2n$-particle cut in the $t$-channel.}
    \label{fig:multi-Aks}
\end{figure}

One particularly interesting aspect of this formulation is that it captures part of the multi-particle physics from first principles: combining crossing symmetry with two-particle unitarity forces the generation of an infinite series of elementary graphs with the 2PRR property. While these graphs have a two-particle cut in one channel, the crossed channel typically involves many particles; see Fig.~\ref{fig:multi-Aks}.
The rest of multi-particle processes, which in terms of graphs involve more than 4-particle cuts in all channels, are repackaged into an unknown set of functions $\rho_\MP(s,t)$ (supported above the leading multi-particle Landau curve) and $\eta_\MP(s)$ (supported above the leading multi-particle normal threshold).

In this way, \emph{the multi-particle data} $\left(\eta_\MP(s), \rho_\MP(s,t) \right)$ provide the variational degrees of freedom that parametrize the space of consistent scattering amplitudes in the S-matrix bootstrap. This is the approach that we adopted in the present paper. The main \textit{conceptual} novelty compared to~\cite{Tourkine:2023xtu} lies in this variational aspect. 

\subsection{Space of solutions}
In the approach described above, the problem of the S-matrix bootstrap at the level of $2 \to 2$ scattering amplitude is the one of characterizing the space of solutions to the Atkinson-Mandelstam problem. At present very little is known about this space.

In the first part of the paper, we have numerically constructed and analyzed a set of solutions with $\eta_\MP(s) = \rho_\MP(s,t)=0$ characterized by the subtraction constant $1.4 \gtrsim {c_0 \over 32 \pi} >0$. In the second part of the paper, we have constructed numerical solutions with non-zero multi-particle data which minimizes particle production in low-spin partial waves. From rigorous dual bootstrap~\cite{Guerrieri:2020kcs,Guerrieri:2021tak} bounds, we know that $-8\lesssim \frac{c_0}{32\pi} \leq 2.66$.

The primal bootstrap constructions based on semidefinite optimization and the $\rho$-ansatz fill most of this space~\cite{Paulos:2017fhb,Chen:2022nym,EliasMiro:2022xaa}, however they do not solve the Atkinson-Mandelstam problem as described above as they do not respect the detailed analytic structure of the scattering amplitude. An obvious open question in this context is: what is the relationship between the solutions to the $\rho$-based primal bootstrap and the solutions to the Atkinson-Mandelstam bootstrap?

Based on this work we propose the following \emph{microscopic interpretation of the extremal amplitudes}: they are essentially solutions to the Atkinson-Mandelstam problem, where the multi-particle data has been tuned to minimize particle production. It would be very interesting to explore and test this idea further.

\subsection{Reggeization and UV-completeness}
In our work, we have found that an important aspect of solving the Atkinson-Mandelstam problem is checking its UV completeness. By ``UV completeness" in this context we mean that we are able to reliably characterize the amplitude at arbitrary high energies. A particularly important high-energy limit is the Regge limit. In particular, this limit is what determined the maximal value ${c_0 \over 32 \pi} \simeq 1.4$ that we reported in the paper. Above this value, we were not able to reliably demonstrate the Reggeization of the amplitude.

In practice, we have solved the unitarity equations with a UV cutoff in energy. A crucial step of our analysis was the check that as we increase the UV cutoff the Regge behavior of the amplitude does not change. In our analysis of the 2PRR amplitudes, for ${c_0 \over 32 \pi} \lesssim 1.4$ we observed slow logarithmic decay of the spectral densities.

We also observed that the inclusion of inelasticity of 2PRR amplitudes barely affects the values of low-energy coupling constants of the amplitude in the primal bootstrap~\cite{Paulos:2017fhb}, while it is crucial to align with the Regge behavior of the 2PRR amplitudes (to an extent limited by the ansatz size). 

To make further progress in understanding the solution space of the Atkinson–Mandelstam bootstrap, it is important to better characterize solutions in the Regge limit and understand how they interplay with the finite-energy properties of the amplitude.

\subsection{Bounding particle production}

When the detailed analytic structure of the amplitude is taken into account there is an interesting tension---we can call it \emph{the Aks tension}---between the following statements:
\begin{itemize}
\item two-particle unitarity and crossing symmetry;
\item multi-particle unitarity as the statement about the support of $\rho_{\text{MP}}$;
\item boundedness of $\rho_{\text{MP}}$;
\item small (or even vanishing) particle production.
\end{itemize}
As emphasized by Aks~\cite{Aks:1965qga}, and demonstrated explicitly in the present paper two-particle unitarity and crossing naturally lead to particle production. If we insist on small or vanishing particle production, we can use the multi-particle data to reduce it, a phenomenon we called \emph{the Aks screening} in the present paper. However, due to multi-particle unitarity and the support of $\rho_{\text{MP}}(s,t)$ above the leading multi-particle Landau curve, the screening requires that $\rho_{\text{MP}}(s,t)$ takes larger and larger values. Conversely, we estimated in Eq.~\eqref{eq:Jmaxbound} the number of partial waves that can be screened entirely is restricted by $\log(\|\rho_\MP\|)$.

\subsection{Multi-particle data}

\begin{figure}[h]
    \centering
    \includegraphics[scale=1]{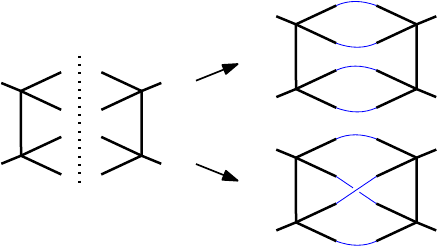}
    \caption{Bose symmetry of the four-particle final state relates the 2PRR Aks graph to the MP non-planar box graph.}
    \label{fig:bose}
\end{figure}

\begin{figure}[h]
    \centering
    \includegraphics[scale=1]{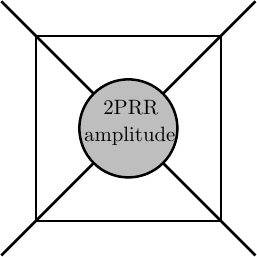}
    \caption{An example of an MP graph that follows from four-particle unitarity, where the 2PRR amplitude is a sub-process of a larger diagram.}
    \label{fig:mroom}
\end{figure}

At this point the reader might wonder: what is the physical principle that would rule out the 2PRR model considered in the present paper? It is very easy to understand by considering a four-particle cut of the 2PRR amplitude. The result is shown in Fig. \ref{fig:bose}. It is then clear that \emph{the Bose symmetry of the four-particle state} requires including the nonplanar box diagram. We can also consider a process, where the two-to-two scattering is effectively embedded inside the higher-point amplitude, see Fig. \ref{fig:mroom}. It is clearly required by multi-particle unitarity but not included in the 2PRR model.

It is also interesting to ask: how can one extract $\rho_{\text{MP}}(s,t)$ in a scattering experiment? We can imagine the following: first, we  consider a low-energy experiment ($16m^2>s \geq 4 m^2$) and read off $f_J(s)$ for all $J$. From this we can calculate $\rho_{el}(s,t)$. Consider then inelastic cross-section in the impact parameter space $\sigma_{\text{inel}}(s,b) = {\rm Im}f(s,b)-|f(s,b)|^2$. Since we have already measured $\rho_{el}(s,t)$, we can calculate the contribution of $\rho_{el}(t,s)$ to $\sigma_{\text{inel}}(s,b)$, let us call it $\sigma_{\text{inel}}^{Aks}(s,b)$. We can then measure $\rho_{MP}(s,t)$ from $\sigma_{\text{MP}}(s,b)=\sigma_{\text{inel}}(s,b)-\sigma_{\text{inel}}^{Aks}(s,b)$. It would be very interesting to explore $\rho_{\text{MP}}(s,t)$ in physical theories, such as QCD.

\subsection{Future directions}

Here we list some of the most pressing future directions:
\begin{itemize}
\item \textbf{Two subtractions, singular two-particle threshold, resonances.} An obvious, but important extension of the present work would be to allow for two subtractions. We expect that this would allow in principle to explore the negative $c_0$ part of the parameter space, that is expected to contain amplitudes that Reggeize through QCD-like exchanges of infinite towers of resonances~\cite{EliasMiro:2022xaa,Chen:2022nym}
and display Froissart-growth~\cite{Correia:2025uvc}, which we would find particularly interesting to study with our method.
Further extensions include incorporating the singular threshold behavior in Eq.~\eqref{eq:singular}, which would open up a new class of solutions. More generally, additional physical features can be imposed directly at the level of the loss function. For example, one may constrain the position of resonances by enforcing the presence of zeros of $|S_J|$ at specified locations in the complex plane.
Together, these extensions would help characterize the space of solutions to the Atkinson--Mandelstam problem.

\item \textbf{Analytic solution in the Regge limit.} For all amplitudes in Section~\ref{sec:2PRR_Regge}, we observed transparency and logarithmically decaying observables such as spectral densities, and the total cross-section in the $s \to \infty$ limit. It would be very interesting to derive the large-$s$ expansion starting from elastic unitarity and Mandelstam equation~\eqref{eq:mandelstam-eqn} analytically. The fact that the amplitude vanishes at large-$s$ up to a real constant suggests that the leading asymptotic is given by the $J=0$ Regge pole. It would be very interesting to see if the methods of Regge Field Theory REFs can be used to find the analytic solution for the 2PRR amplitudes in the Regge limit.
\item \textbf{Aks screening at higher spins.} We used $\rho_\text{MP}$ as an additional degree of freedom to screen particle production in individual partial waves in Sec.~\ref{sec:aks_w_nn}. 
Our current methods could not solve the complex numerical problem of Aks screening beyond $J_\mmax=2$. It would be important to overcome this problem and find a way to push $J_\mmax$ to higher values, to see if expectations from our argument in Sec.~\ref{sec:aks_screening} survive the back-reaction of $\rho_\MP$ and full crossing symmetry. Understanding how $\rho_\text{MP}$ changes as a function of $J_\mmax$ would be a very interesting refinement of the Aks theorem.
\item \textbf{Experimental input.} The gradient descent solver strategy used in this article allows further interesting applications. For instance, if we aim to describe pion scattering~\cite{Guerrieri:2018uew,Guerrieri:2020bto,Guerrieri:2024jkn,He:2023lyy,He:2025gws,Bose:2020cod,Bose:2020shm}, we would want to be able to include for instance chiral-perturbation theory data, i.e. Adler zeros, $\rho$-meson resonance, scattering lengths etc. This kind of data can be included straightforwardly in our setup, by adding new terms to the loss function. Such constraints on partial waves of the form $S_J(s_*)=x$ can be simply added to the global loss function $\mathcal{L}\to\mathcal{L}+w |S_J(s_*)-x|$. 
Of course, this simple option comes at a cost: the more terms we have in the loss, the more the chances that losses interfere with each other and screen each other out, rendering convergence more cumbersome to reach in practice.
One element is however particularly simple to add: inelastic profiles. 

In~\cite{Guerrieri:2024jkn}, a gradient-free optimization algorithm (particle-swarm optimization) was used to do a scan in the space of amplitudes constrained by non-convex conditions, alongside the semidefinite bootstrap setup, where the physical particle production was imposed as a convex condition (such as~\eqref{eq:descr1}) above the $K\bar K$ production threshold in the pion amplitude~\cite{Garcia-Martin:2011iqs}. In our approach, it is possible to enforce the inelasticity conditions exactly, rather than an upper bound, along with further non-convex conditions in the loss function. Particle production in the S-wave is immediate to enforce, for instance, as $\eta_\MP$ being the input. In higher waves, it has to be added via the loss, as described above.

\item \textbf{Neural network architecture.}
Alternative architectures could be considered for parametrizing the spectral densities. In particular, Kolmogorov–Arnold Networks (KANs)~\cite{liu2025kankolmogorovarnoldnetworks}. Unlike standard architectures such as the one used here, KANs use learnable activation functions, typically implemented as splines with trainable parameters. This structure may offer a more compact representation of our smooth spectral densities. 
\item \textbf{Large-$N$ models.} Theories with flavor symmetry often arise as good microscopic descriptions of nonperturbative amplitudes~\cite{Paulos:2018fym,Cordova:2018uop,Cordova:2019lot,He:2018uxa,EliasMiro:2019kyf,Henning:2022xlj,Cordova:2023wjp,Cordova:2025bah}.
A natural question is whether the 2PRR amplitudes can be realized as large-$N$ limits of some microscopic models. While we do not yet have a concrete proposal, we note that the large-$N$ counting in the tensor models \cite{Klebanov:2016xxf,Giombi:2017dtl,Klebanov:2018fzb} naturally produces a large-$N$ enhancement for a sub-class of 2PRR graphs. We leave a systematic study of four-point scattering in large-$N$ tensor models to future work. Let us also notice that a sub-class of 2PRR graphs appears in the large $N$ limit of the fishnet theory \cite{Korchemsky:2018hnb}. 

\item \textbf{Higher dimensions.} Recently in \cite{Gumus:2025hwq}, higher dimensional bounds on S-matrices were studied. Most notably, evidence for distinct families of UV completions was observed, suggesting the onset of some form of non-locality. As this study is based on the semidefinite bootstrap, it is subject to the usual caution that bounds could be affected by inelasticity. In four dimensions, our present article suggests that Aks screening is a mechanism solid enough to protect the bounds, but it would be interesting to see if this holds in higher dimensions. More generally, since our present numerical strategy provides UV completions to arbitrary energies, it would be particularly interesting to study the Reggeization mechanisms in higher dimensions and their possible relation to local vs non-local QFT completions. Similarly, it would be interesting to study $\phi^4$ theory in lower dimensions, $d=3$, with our methods.
\end{itemize}

\paragraph{Acknowledgements.}
We acknowledge discussions with Benjamin Basso, Stefano De Angelis, Andrea Guerrieri, Simon Metayer, Jesse Thaler.
%, \snote{add} \pinote{add}
This work has received funding from Agence Nationale de la Recherche (ANR), project ANR-22-CE31-0017. This project has received funding from the European Research Council (ERC) under the European Union’s
Horizon 2020 research and innovation program (grant agreement number 949077). We thank the Yukawa Institute, Yu Nakayama and the organizers of the 2025 conference ``Progress of Theoretical Bootstrap'' for hospitality, where parts of this work were finalized.

\appendix
\addtocontents{toc}{\protect\setcounter{tocdepth}{1}} % Suppress entries below 'section' % no longer in table of content !

\section{Definitions}
\label{app:definitions}
We follow the definitions of \cite{Correia:2020xtr} and \cite{Tourkine:2023xtu}.
\paragraph{Inputs and outputs.}
Throughout this paper we work with a once--subtracted Mandelstam representation
\eqref{eq:mandelstamRep1} and treat the amplitude as specified by:
(i) the subtraction constant $c_0 \equiv T(s_0,t_0)$ at the crossing--symmetric point,
(ii) the multi--particle double spectral density $\rho_{\rm MP}(s,t)$ (when allowed),
and (iii) a possible extra $S$--wave inelasticity input $\eta_{\rm MP}(s)$.
Given these inputs, the unknowns are the single spectral density $\rho(s)$ and the
elastic double discontinuity $\rho_{\rm el}(s,t)$, which we solve for by imposing
two--particle unitarity and analyticity.
All observables (partial waves, inelasticities, Regge limits) are computed a posteriori
from the resulting spectral densities. 

\paragraph{Coupling convention.}
To connect the subtraction constant $c_0$ in~\eqref{eq:mandelstamRep1} with the ${\lambda \over 4!} \phi^4$ perturbation theory, let us first note that the amplitude is a constant for all energies at the leading order, $T(s,t) = -\lambda + O(\lambda^2)$. The next-to-leading order corrections need a renormalization condition that we then choose to be
\begin{equation}
 T(4/3,4/3) = c_0 = -\lambda\, .
\end{equation}
Under this condition, the usual ${\lambda \over 4!} \phi^4$  theory with a repulsive potential that is bounded from below corresponds to $\lambda > 0$ and  $c_0 < 0$. In this regime, the theory is known to possess a Landau pole at large energies, at exponentially large energies predicted by the beta function~\eqref{eq:phi4running} for $\lambda >0$.

On the other hand, the regime $c_0>0$ that we study in this work, the quartic coupling $\lambda$ is negative and hence describes an attractive force between the external particles. Note that the potential of this theory must receive corrections at $O(\phi^6)$ to render the potential bounded. The theory is asymptotically-free which can be seen by~\eqref{eq:phi4running} for $\lambda<0$.

\paragraph{Partial waves.}

We have also used the notation for partial waves of the amplitude defined as follows
\begin{subequations}
\be
T(s,t) &=16 \pi \sum_{\substack{J=0,\\ J - \text{even}}}^{\infty} (2J+1) f_J(s)  P_J \Big( 1+{2 t \over s - 4 m^2} \Big), \\
S_J(s) &\equiv 1 + i {\sqrt{s - 4 m^2} \over \sqrt{s} } f_J(s) , \label{eq:pwave}\ee
where $f_J(s)$ are partial wave coefficients and $P_J(z)$ are the Legendre polynomials. Partial wave coefficients can be obtained by doing the projection
\begin{equation}
f_J(s)=\frac{1}{32\pi}\int_{-1}^{1} \!\! dz \,P_{J}(z)\,T(s,t(s,z))\,,\qquad t(s,z)=-\frac{1}{2}(s-4)(1-z) \, .
\label{eq:PW}
\end{equation}
\label{eq:partialwave_expansion}
\end{subequations}
In particular, Eq.~\eqref{eq:PW} applied for $J=0$ and the Mandelstam representation~\eqref{eq:mandelstamRep1} for the amplitude $T(s,t)$ leads to
\begin{equation}
    \begin{aligned}
        f_0(s) = &\dfrac{1}{16\pi}\left(c_0+\int_{4m^2}^\infty\dfrac{\mathrm{d}s'}{\pi}\dfrac{(s-s_0)\rho(s')}{(s-s)(s'-s_0)}+2\int_{4m^2}^\infty\dfrac{\mathrm{d}t'}{\pi}\rho(t')\!\left[\dfrac{\log\left(\frac{s-4m^2-t'}{t'}\right)}{s-4m^2}-\dfrac{1}{t'-t_0}\right]\right.\\
    &+2\int_{4m^2}^\infty\dfrac{\mathrm{d}s'\mathrm{d}t'}{\pi}\dfrac{(s-s_0)\rho(s',t')}{(s'-s)(s'-s_0)}\!\left[\dfrac{\log\left(\frac{s-4m^2-t'}{t'}\right)}{s-4m^2}-\dfrac{1}{t'-t_0}\right]\\
    &\left.+\int_{4m^2}^\infty\dfrac{\mathrm{d}t'\mathrm{d}u'}{\pi}\left[1-\dfrac{A_2^{(4)}(s,t',u')}{s+t'+u'-4m^2}\right]\right)\,,\\
    A_2^{(4)}(s,t',u') =\, &\dfrac{(t_0-t')(s_0+t_0-s-t')\tanh^{-1}\!\left(\frac{s-4m^2}{\sqrt{s+2t'-4m^2}}\right)}{s+2t'-4m^2}\\
    &+\dfrac{(u_0-u')(s_0+u_0-s-u')\tanh^{-1}\!\left(\frac{s-4m^2}{\sqrt{s+2u'-4m^2}}\right)}{s+2u'-4m^2}\,.
    \end{aligned}
    \label{eq:f0def}
\end{equation}

The Froissart-Gribov projection allows to compute the partial wave projection suitably analytically continued in spin. It reads:
\begin{equation}
\label{eq:FG}
f_J(s) = \frac{1}{8\pi^2} \int^\infty_4 \frac{2\,dt}{s-4} Q^{(4)}_J\left(1+\frac{2t}{s-4}\right) T_t(s,t) \ ,
\end{equation}
where $Q_J$ are the Legendre Q-functions.

Low energy coefficients have the dispersive formulas
\begin{equation}
\label{eq:dispersive_taylor}
\begin{aligned}
c_2 &= \int_4^\infty \frac{ds'}{\pi} \frac{\im T(s',\frac{4}{3})}{(s-4/3)^3} \ , \\    
c_3 &= \int_4^\infty \frac{ds'}{\pi} \frac{3 \, \im T(s',\frac{4}{3})}{(s-4/3)^4} - \int_4^\infty \frac{ds'}{\pi} \frac{4 \, \partial_t \im T(s',t)|_{t=4/3}}{(s-4/3)^3} \ , \\
c_4 &= \frac{1}{2} \int_4^\infty \frac{ds'}{\pi} \frac{\im T(s',\frac{4}{3})}{(s-4/3)^5} \ .
\end{aligned}
\end{equation}

\section{Semidefinite bootstrap}
\label{app:primal}
In this appendix, we give the full details for the semidefinite bootstrap setup we used in Section~\ref{sec:inelastic_sdpb}. We use the primal ansatz 
\be
T(s,t) = \sum_{\Sigma^2(N_\mmax)} \alpha_{\sigma,\tau} \left( 
\rho_\sigma(s) \rho_\tau(t) + \text{crossed} \ 
%+ \rho_\sigma(s) \rho_\tau(4{-}s{-}t) + \rho_\sigma(t) \rho_\tau(4{-}s{-}t) 
\right) \ ,
\ee
where the wavelet term $\rho_\sigma(s)$ centered at $\sigma$ is given by
\be
\rho_\sigma(s) = \frac{\sqrt{4-\sigma}-\sqrt{4-s}}{\sqrt{4-\sigma}+\sqrt{4-s}} \, .
\ee
Note that a wavelet function goes to a constant $\rho_\sigma(s) \to -1$ as $s \to \pm \infty$. Sum over the pair of wavelet centers $\{\sigma,\tau\}$ is fixed by the ansatz size $N_\mmax$ and runs through the square of center grid, $\Sigma(N_\mmax) \times \Sigma(N_\mmax)$, as will be described below.

Primal bootstrap solves the semidefinite programming problem for the unknown coefficients $\{\alpha_{\sigma,\tau}\}$ by imposing partial wave unitarity, 
\be
|S_J(s)|^2 \leq 1 \quad , \quad J<J_\text{max} \ , 
\ee
on a number of grid points in $s$ (that we call \emph{unitarity grid}) and for a finite number spins $J_\text{max}$. The unitarity grid is constructed using the inverse map of the mother wavelet
\be
\rho_{20/3}: (4,\infty) \mapsto \{ \, e^{i\theta} \, | \, 0 < \theta < \pi \, \} \ 
\ee
by choosing different Chebyshev nodes on the boundary of the unit disk given by the set $\theta_N=\{\frac{\pi}{2} \left( 1+\cos( \pi \frac{k}{N+1} ) \right) \, | \, 1 \leq k \leq N\}$. In all our runs, we fix $N=300$ and $J_\text{max}=16$ yielding sufficiently many constraints, such that our results did not depend on adding further ones.

The wavelet centers at a fixed ansatz size are distributed on the following nodes
\be
\Sigma(N_\mmax) = \bigcup_{k=1}^{N_\mmax} \theta_k
\ee
that we call \emph{center grid}. Below we give a table summarizing important properties of this grid for various $N_\mmax$ values, including the wavelet centers.

\begin{table}[H]
\begin{minipage}{\textwidth}
\centering
\begin{tabular}{c|c|c|c}
$N_\mmax$ & \text{Ansatz size} & \text{First 3 wavelets} & \text{Last 3 wavelets} \\
\hline
8 & 135 &
$\begin{array}{c} 4.00599 \\ 4.00955 \\ 4.0162 \\ \end{array}$
& $\begin{array}{c} 443.027 \\ 748.304 \\ 1190.86 \\ \end{array}$
\\
\hline
9 & 173 &
$\begin{array}{c} 4.00394 \\ 4.00599 \\ 4.00955 \\ \end{array}$
& $\begin{array}{c} 748.304 \\ 1190.86 \\ 1806.9 \\ \end{array}$
\\
\hline
10 & 238 & 
$\begin{array}{c} 4.0027 \\ 4.00394 \\ 4.00599 \\ \end{array}$
& $\begin{array}{c} 1190.86 \\ 1806.9 \\ 2636.9 \\ \end{array}$
\\
\hline
11 & 288 & 
$\begin{array}{c} 4.00191 \\ 4.0027 \\ 4.00394 \\ \end{array}$
& $\begin{array}{c} 1806.9 \\ 2636.9 \\ 3725.61 \\ \end{array}$
\\
\hline
12 & 378 & 
$\begin{array}{c} 4.00139 \\ 4.00191 \\ 4.0027 \\ \end{array}$
& $\begin{array}{c} 2636.9 \\ 3725.61 \\ 5122.01 \\ \end{array}$
\\
\end{tabular}
\end{minipage}

\caption{The number of wavelet centers and their positions with respect to various values of $N_\mmax$. }
\label{tab:primal_centers}
\end{table}

\section{Aks-screening lemmas}
\label{app:cheb-FG}

Let us first demonstrate Eq.~\eqref{eq:toy-blowup-single-moment}.

\paragraph{Single-moment screening.}

What controls the behavior of $g$ is the asymptotics of the last-moment.
Starting from Eq.~\eqref{eq:toy-moments-problem}, by Cauchy-Schwarz inequality, we have that
\begin{equation}
|m(J)|\equiv \left|\int_0^{y_\MP} y^J g(y)\,dy\right|
\le \|g\| \left(\int_0^{y_\MP} y^{2J}\,dy\right)^{1/2},
\end{equation}
where 
\begin{equation}
    \|g\| = \left(\int_0^{y_\MP} g(y)^{2}\,dy\right)^{1/2}\,.
\end{equation}
We then have a first bound on $g$, relative to $m(J)$:
\begin{equation}
\label{eq:g-CS}
\|g\| \ \ge\ |m(J)|\,(y_\MP)^{-J-1/2}\,(J+1/2).
\end{equation}
The bound is maximal for $J=J_\mmax$. We now have to find a estimate for $m(J)$ in terms of the norm of $f$.

Here it is crucial to use that $f$ is not vanishing in the elastic range. By Mahoux--Martin positivity, we even have the stronger condition that $f$ is strictly positive in a neighbourhood of $y_\el$, bounded by $y_{\rm MM}$, thus there exists $\varepsilon>0,c>0$  such that 
\begin{equation}
\label{eq:f-endpoint-lowerbound-L2}
f(y)\ \ge\ c\,(y_{\el}-y)^k\qquad \forall\,y\in[y_{\el}-\varepsilon,y_{\el}] .
\end{equation}
where $k>-1$ for the function to be integrable in Froissart-Gribov, as is the case in $d=4$ for the double-discontinuity (see~\cite{Correia:2020xtr}).

Let us split the integral that defines $m(J)$ into an endpoint piece and a bulk piece:
\begin{equation}
m_J=\int_{y_{\el}-\varepsilon}^{y_{\el}} y^{J} f(y)\,dy
+\int_{0}^{y_{\el}-\varepsilon} y^{J} f(y)\,dy.
\end{equation}
The sign of $f$ can change in the bulk, but $\int_{0}^{y_{\el}-\varepsilon} y^{J} f(y)\,dy\geq - \int_{0}^{y_{\el}-\varepsilon} y^{J} |f(y)|\,dy$, and using \eqref{eq:f-endpoint-lowerbound-L2} we have:
\begin{equation}
\label{eq:mJ-split-L2}
m_J\ \ge\
c\int_{y_{\el}-\varepsilon}^{y_{\el}} y^{J} (y_{\el}-y)^k\,dy
-\int_{0}^{y_{\el}-\varepsilon} y^{J} |f(y)|\,dy.
\end{equation}
The bulk term is bounded by the $L_1$ norm of $f$:
\begin{equation}
\label{eq:bulk-bound-L2}
\int_{0}^{y_{\el}-\varepsilon} y^{J} |f(y)|\,dy
\le (y_{\el}-\varepsilon)^{J}\int_{0}^{y_{\el}-\varepsilon} |f(y)|\,dy
= (y_{\el}-\varepsilon)^{J}\,\|f\|_{L^1(0,y_{\el})}.
\end{equation}
For the endpoint term, set $y=y_{\el}-\delta$:
\begin{equation}
\int_{y_{\el}-\varepsilon}^{y_{\el}} y^{J} (y_{\el}-y)^k\,dy
=\int_{0}^{\varepsilon} (y_{\el}-\delta)^{J}\,\delta^k\,d\delta.
\end{equation}
Choose $\varepsilon$ large enough so that  $J\ge \frac{y_{\el}}{2\varepsilon}$ so that $\frac{y_{\el}}{2J}\le\varepsilon$, and
restrict to $\delta\in[0,y_{\el}/(2J)]$:
\begin{equation}
\int_{0}^{\varepsilon} (y_{\el}-\delta)^{J}\,\delta^k\,d\delta
\ge
\int_{0}^{y_{\el}/(2J)} (y_{\el}-\delta)^{J}\,\delta^k\,d\delta.
\end{equation}
Then, use that $(y_{\el}-\delta)^J\ge e^{-1}y_{\el}^J$ for $\delta\le y_{\el}/(2J)$\footnote{Use $\log(1-x)\ge -x/(1-x)$ and $1/(1-x)\le 2$ for
$x\le 1/2$, one gets $J\log(1-x)\ge -1$.} and obtain
\begin{equation}
\int_{0}^{y_{\el}/(2J)} (y_{\el}-\delta)^{J}\,\delta^k\,d\delta
\ge
e^{-1}y_{\el}^J\int_{0}^{y_{\el}/(2J)} \delta^k\,d\delta
=
e^{-1}y_{\el}^J\frac{1}{k+1}\left(\frac{y_{\el}}{2J}\right)^{k+1}.
\end{equation}
Inserting this and \eqref{eq:bulk-bound-L2} into \eqref{eq:mJ-split-L2} yields, for
$J\ge \frac{y_{\el}}{2\varepsilon}$,
\begin{equation}
\label{eq:mJ-explicit-L2}
m_J\ \geq\ 
\frac{c\,e^{-1}}{k+1}\,\frac{y_{\el}^{J+k+1}}{(2J)^{k+1}}
-\|f\|_{L^1(0,y_{\el})}\,(y_{\el}-\varepsilon)^{J}.
\end{equation}
Since $(y_{\el}-\varepsilon)^J/y_{\el}^J=(1-\varepsilon/y_{\el})^J$ is exponentially small, there exists
$J_0$ such that for all $J\ge J_0$ the RHS is at least half the first term. This shows that there exist constants $J_0\in\mathbb{N}$ and $C>0$ (depending only on $f,y_{\el},\varepsilon,c,k$)
such that for all $J\ge J_0$,
\begin{equation}
\label{eq:mJ-lowerbound-L2}
|m_J|\ \ge\ C\,\frac{y_{\el}^{J+k+1}}{J^{k+1}} .
\end{equation}
Inserting \eqref{eq:mJ-lowerbound-L2} into \eqref{eq:g-CS} gives \eqref{eq:toy-blowup-single-moment}.\qed

\paragraph{All-moments bound via an extremal polynomial.}
The bound above obtained for equality of one moment is enough in theory as it discards the possibility of asymptotic screening with finite-norm screening function. It can be greatly improved to the case of all screened moments with a standard trick.
Since \eqref{eq:toy-moments-problem} holds for all $n\le J_\mmax$, it also holds for any polynomial
$P(y)=\sum_{n=0}^{J_\mmax} c_n y^n$ of degree at most $J_\mmax$:
\begin{equation}
\int_0^{y_\el} P(y) f(y)\,dy = \int_0^{y_\MP} P(y) g(y)\,dy.
\end{equation}
The CS inequality gives again
\begin{equation}
\label{eq:dual-L1}
\|g\|\ \ge\ \frac{\left|\int_0^{y_\el} P(y) f(y)\,dy\right|}{\|P\|}.
\end{equation}
where the norm $\| \cdot\|$ is given by an integral over $[0;y_\MP]$ by definition. The polynomial $P$ is however defined beyond this range (like the monomials $y^J$), and we need  now to choose $P$ so as to maximize growth outside $[0,y_\MP]$ and thus the numerator, while keeping $\|P\|\le 1$.
The extremal choice is given in terms of Chebyshev polynomials. Those have the property that they extremize polynomial growth outside of $[-1;1]$. More precisely,  for any other polynomial $p(x)$ of the degree $n$ such that $\forall x\in[-1;1],\,|p(x)|\leq1$, then (see e.g.\cite[section 2.7]{rivlin2020chebyshev})
\begin{equation}
    |p(x)|\leq |T_N(x)|\,\qquad x\in \mathbb{R} \backslash [-1;1]
\end{equation}
We thus use the shifted Chebyshev polynomials
\begin{equation}
P(y)=T_{J_\mmax}\!\left(2\frac{y}{y_\MP}-1\right),
\qquad
\|P\|\le 1.
\end{equation}
Outside of the interval $[-1;1]$, the growth of Chebychev polynomials is given by
\begin{equation}
T_N(x)=\cosh\!\big(N\,\arccosh(x)\big)\sim \frac12\,\chi(x)^N,\qquad \chi(x)=x+\sqrt{x^2-1}.
\end{equation}
Therefore the integral
\begin{equation}
I_{J_\mmax}\equiv \int_0^{y_\el} T_{J_\mmax}\!\left(2\frac{y}{y_\MP}-1\right)\, f(y)\,dy
\end{equation}
is dominated near the endpoint $y=y_\el$, and the exact same calculation as above yields a similar exponential bound but where $y_\el^J$ is replaced with $\chi\left(2\frac{y_\el}{y_\MP}-1\right)^J$.

The improvement is significant: for $y_\el\simeq 1$ and $y_\MP\simeq 1/4$ one has $y_\el/y_\MP \simeq 4$ and $\chi(7)\approx 13.9$.

\bibliographystyle{JHEP}
\bibliography{bib-papers}

\appendix

\end{document}